\documentclass[aip,reprint]{revtex4-1}

\usepackage[utf8]{inputenc}
\usepackage[T1]{fontenc}
\usepackage{graphicx}
\usepackage{dcolumn}
\usepackage{bm}
\usepackage{etoolbox}
\usepackage{amsmath}
\usepackage{amssymb}
\usepackage{url}
\usepackage{multirow}
\usepackage{xr}
\usepackage{filecontents}

\newcommand{\Qscaled}{\tilde{Q}}
\newcommand{\Jscaled}{\delta \tilde{J}}
\newcommand{\Iscaled}{\delta \tilde{I}}
\newcommand{\epotscaled}{\tilde{\psi}_0}
\newcommand{\Ib}{I^{(0)}}
\newcommand{\Jb}{J^{(0)}}
\newcommand{\delJ}{\delta J}
\newcommand{\delI}{\delta I}
\newcommand{\lDinf}{\lambda_{\mathrm{D}}^\infty}
\newcommand{\lD}{\lambda_{\mathrm{D}}^\infty}
\newcommand{\kBT}{k_{\mathrm{B}} T} 
\newcommand{\eps}{\epsilon \epsilon _0}
\newcommand{\Dellnc}{\Delta \ln c}
\newcommand{\Delc}{\Delta c}
\newcommand{\Dp}{D_+ + D_-}
\newcommand{\Dm}{D_+ - D_-}

\newcommand{\cinf}{c_\infty}

\newcommand{\lDu}{l_{\mathrm{Du}}^\infty}

\graphicspath{{./figures_paper/}}

\makeatletter
\def\@email#1#2{%
 \endgroup
 \patchcmd{\titleblock@produce}
  {\frontmatter@RRAPformat}
  {\frontmatter@RRAPformat{\produce@RRAP{*#1\href{mailto:#2}{#2}}}\frontmatter@RRAPformat}
  {}{}
}%
\makeatother
\begin{document}


\title{Scaling laws for concentration-gradient-driven electrolyte transport through a 2D membrane} 


\author{Holly C. M. Baldock}
\author{David M. Huang}
\email{david.huang@adelaide.edu.au}
\affiliation{Department of Chemistry, School of Physics, Chemistry and Earth Sciences, The University of Adelaide, Adelaide, SA 5005, Australia}



\begin{abstract}
Two-dimensional (2D) nanomaterials exhibit unique properties that are promising for diverse applications, including those relevant to concentration-gradient-driven transport of electrolyte solutions through porous membranes made from these materials, such as water desalination, osmotic power, and iontronics. Here we derive general equations, and determine scaling laws in the thick and thin electric-double-layer limits, that quantify the variation of the concentration-gradient-driven flow rate, solute flux and electric current with the pore radius, surface charge density and Debye screening length for the transport of a dilute electrolyte solution through a circular aperture in an infinitesimally thin planar membrane. We also determine scaling laws for the electric-field-driven flow rate in the thin electric-double-layer limit in the same geometry. We show that these scaling laws accurately capture the scaling relationships from finite-element numerical simulations within the Debye--H\"uckel regime, and extend the theory to obtain scaling laws in the thin electric-double-layer limit that hold even when the electric potential energy is large compared with the thermal energy. These scaling laws indicate unusual behavior for concentration-gradient-driven flow in a 2D membrane that is not seen in thicker membranes, which has broad implications for liquid transport through membranes whose thickness comparable to, or smaller than, their pore size.
\end{abstract}


\maketitle

\section{Introduction}

Electrolyte transport through porous membranes is central to many applications,\cite{siwyNanporesDNASequencingFiltration2023, marbachMolecularInsights2019} such as water desalination,\cite{duChanoisMembranesSelectiveIonSeparations2021} energy harvesting from salinity gradients (osmotic power or "blue energy"), \cite{duChanoisMembranesSelectiveIonSeparations2021, zhangNanofluidicsOsmoticEnergy2021a, siriaNewAvenuesLargescale2017, rankinEffectHydrodynamicSlip2016} energy storage, \cite{aluruConfinementSingleDigitNanpore2023, pomerantsevaEnergyStorageFuture2019} nanopore-based sensing,\cite{siwyNanporesDNASequencingFiltration2023, yingBeyondDNASequencing2022} and biomimetic ion-based information processing and storage (iontronics).\cite{robinModellingSpikingIonicTransport2021, robinNanofluidicsCrossroads2023, kamsmaBrainIontronicNanochannels2024, gkoupidenisOrganicBioinspiredElectrotronics2024} Membranes made from two-dimensional (2D) nanomaterials, including graphene and molybdenum disulfide (MoS$_2$), exhibit unique properties compared with conventional membranes \cite{sahuColloquiumIonicPhenomena2019} that are of practical interest for applications such as desalination,\cite{zhangUltrathinMembranesSeparations2022, pakulskiMembranesWaterPurificationEnergyHarvesting2020, surwadeDesalinationSingleLayerGraphene2015} osmotic power, \cite{pakulskiMembranesWaterPurificationEnergyHarvesting2020, macha2DMaterialsEmerging2019, fengSinglelayerMoS2Nanopores2016} sensing,\cite{venkatesanNanoporeSensorsNucleic2011} and iontronics.\cite{robinMemorySynapseDynamicsChannels2023} For example, exceptionally high osmotic power densities observed in single-layer MoS$_2$ membranes\cite{fengSinglelayerMoS2Nanopores2016} and a 2D nanoporous polymer membrane\cite{chengGateableOsmoticPowerPolymer2023} have been partially attributed to the large concentration gradients that occur across these ultrathin membranes, which are up to several orders of magnitude larger than those for conventional membranes. 

Despite the great promise of 2D membranes, significant knowledge gaps exist about the fluid-transport processes that govern applications. Most of the applications mentioned above, such as desalination and osmotic power, are underpinned by transport processes driven or modulated by electrolyte concentration gradients. Furthermore, in nanopore-based sensing, use of a salt gradient along with an applied electric field has been shown to improve measurement sensitivity significantly over an electric field alone.\cite{wanuunuElectrostaticDNASaltGradient2010} Nevertheless, although equations have been derived previously to quantify the pressure-driven flow of a pure liquid,\cite{sampson1891, heiranianRevisitingSampsonTheory2020} the electric-field-driven solution flux \cite{maoElectroosmoticFlowNanopore2014} and electric current of an electrolyte,\cite{hallAccessResistance1975, leeLargeElectricSizeSurfaceCondudction2012} and the concentration-gradient-driven solution and solute fluxes of a solution containing a neutral solute through a 2D membrane,\cite{rankinEntranceEffectsConcentrationgradientdriven2019} no theory to date has quantified the relationships between the concentration-gradient-driven fluid fluxes of an electrolyte solution through a 2D membrane and relevant membrane and electrolyte properties. The development of such a theory would aid the predictive design of 2D membranes that are optimized for applications. 

Of particular interest is understanding the parameters that control the process of diffusioosmosis, the net flow of a solution with respect to a surface due to a concentration gradient. Diffusioosmosis is responsible for the exceptionally high osmotic power densities measured in boron nitride nanotubes,\cite{siriaGiantOsmoticEnergy2013} and has been invoked to explain even higher osmotic power densities measured for 2D MoS$_2$ membranes,\cite{fengSinglelayerMoS2Nanopores2016} as well as hitherto unexplained discrepancies between the ionic conductance driven by concentration gradients and electric fields for graphene and MoS$_2$ membranes.\cite{macha2DMaterialsEmerging2019}

A better understanding of fluid transport through 2D membranes has significance that extends well beyond 2D membranes to thicker membranes, since fluid transport into (and out of) the pores of a thick membrane from (and to) the fluid outside strongly resembles transport through a 2D membrane (Fig. \ref{schematic}) and thus can be described by similar theories. For pressure-induced fluid flow \cite{sisanEndNanochannels2011} and electric-field-induced ionic currents \cite{leeLargeElectricSizeSurfaceCondudction2012} through a pore in a finite-thickness membrane, the flow resistance due to the pore ends has been shown to be well described by the resistance of a 2D membrane. The influence of the pore ends to fluid transport through membranes, known as "entrance effects", becomes dominant as the membrane thickness approaches the pore size,\cite{maoElectroosmoticFlowNanopore2014, melnikovElectrosmoticFlowNanopore2017} which occurs for nanoporous membranes at thicknesses significantly above the atomic scale, or when the flow resistance of the membrane is small, which is the case for carbon nanotube membranes due to their high interfacial slip.\cite{sisanEndNanochannels2011, rankinEffectHydrodynamicSlip2016} To date, many theoretical models of fluid transport involving concentration gradients in porous media have only considered the flow inside the pores and have neglected entrance effects.\cite{fairReverseElectrodialysis1971, prieveMigrationColloidConcentrationGradient1982, LaviTheoreticalOsmoticPower2024, bonthuisFlowCarbonNanotubes2011} 

In this work, we derive general equations that quantify the variation of the concentration-gradient-driven flow rate, solute flux and electric current as functions of the pore radius, surface charge density and Debye screening length for transport of a dilute electrolyte solution through a circular aperture in an infinitesimally thin planar membrane, as well as the scaling laws of the electric-field-driven flow rate in the thin electric-double-layer limit in the same geometry. From these novel equations, whose accuracy we verify by comparison with finite-element numerical simulations in the Debye--H\"uckel regime, we determine the underlying scaling laws that govern the concentration-gradient-driven fluxes in the thick and thin electric-double-layer limits. We also determine the parameters that dictate the scaling of the fluxes with surface charge density and electrolyte concentration in simulations both inside and outside the Debye--H\"uckel regime when the width of the electric double layer in smaller than the pore radius. Using these relationships, we extend the analytical theory to outside the Debye--H\"uckel regime and determine general scaling laws for the fluxes as a function of surface charge density, pore size and electrolyte concentration in the thin electric-double-layer limit.

\section{Theory}

This derivation takes a similar approach to that of Ref.~\citenum{rankinEntranceEffectsConcentrationgradientdriven2019} for the concentration-gradient-driven transport of a solution containing a neutral solute through an ultra-thin membrane; however, a more rigorous consideration of a perturbation expansion is considered here to decouple the contributions due to the applied concentration gradient and induced electric field gradient in the case of an electrolyte solution. For low-Reynolds-number steady-state transport of a dilute electrolyte solution and assuming that the system can be described by continuum hydrodynamics, the governing equations are\cite{probstein1994}
\begin{align}
        \label{poisson}
         \epsilon \epsilon_0 \nabla ^2 \psi + \rho &= 0, \\
        \label{stokes}
        -\nabla p - \rho \nabla \psi + \eta \nabla ^2 \bm{u} &= 0,\\ 
         \label{nernst-planck}
        \nabla \cdot \bm{j} _i = \nabla \cdot \left[ c ^{(i)} \bm{u} - D_{i} \nabla c^{(i)} - Z_i e \mu _i c^{(i)} \nabla \psi \right] &= 0,  \\
        \label{newt}
      \nabla \cdot \bm{u} &= 0,
\end{align}
where $\psi$ is the electric potential, $\rho= e \sum_i Z_i c^{(i)}$ is the total ionic charge density, $\bm{u}$ is the solution velocity, $p$ is the pressure, $\eta$ is the solution shear viscosity, and $\bm{j}_i$, $c ^{(i)}$, $Z_i$, $D_i$ and $\mu _i$ are the flux density, concentration, valence, diffusivity and mobility, respectively, of species~$i$. Moreover, $e$ is the elementary charge, $\epsilon$ is the dielectric constant of the medium (taken to be water here) and $\epsilon _0$ is the vacuum permittivity. At the membrane surface, we assume that the solution velocity and ion fluxes satisfy the no-slip ($\bm{u}=0$) and zero flux ($\bm{\hat{n}} \cdot \bm{j}_i = 0$) boundary conditions, where $\bm{\hat{n}}$ is the unit vector normal to the surface. 

\begin{figure}[tb!]
    \centering
    \includegraphics{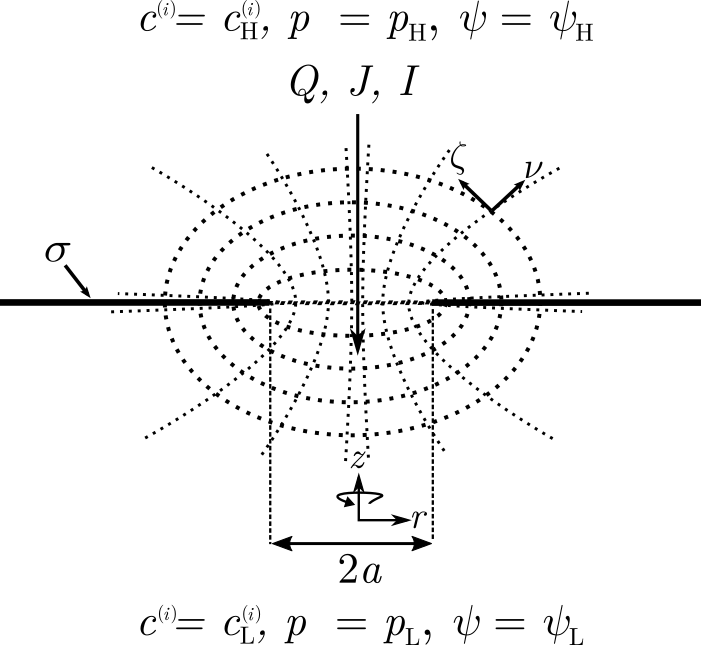}
    \caption{Schematic of flow of a solution through a circular aperture of radius $a$ in an infinitesimally thin planar wall with surface charge density $\sigma$. $c_{\mathrm{H}}^{(i)}$ is the concentration of species~$i$, $p_{\mathrm{H}}$ is the pressure and $\psi_{\mathrm{H}}$ is electric potential far from the membrane in the higher solute-concentration reservoir, and $c_{\mathrm{L}}^{(i)}$ is the concentration of species~$i$, $p_{\mathrm{L}}$ is the pressure and $\psi_{\mathrm{L}}$ is electric potential far from the membrane in the lower solute-concentration reservoir. $Q$, $J$ and $I$ are the flow rate, total solute flux and electric current, respectively. $\zeta$ and $\nu$ are oblate–spheroidal coordinates; contours of constant $\zeta$ and $\nu$ are shown as dashed lines and unit vectors are shown at one point in space.}
    \label{schematic}
\end{figure}

We consider fluid flow through a circular aperture of radius $a$ in an infinitesimally thin planar wall, as illustrated in Fig.~\ref{schematic}, induced by a concentration difference $\Delc^{(i)} = c _{\mathrm{H}}^{(i)} - c_{\mathrm{L}}^{(i)}$ between the two sides of the membrane, where charge neutrality in the bulk solution in each reservoir requires that $\sum_i Z_i c^{(i)}_{\mathrm{H/L}}  = 0$. The pressure far from the pore is the same on both sides, i.e., $p_{\mathrm{H}} = p_{\mathrm{L}}$. Analogous to Ref.~\citenum{rankinEntranceEffectsConcentrationgradientdriven2019}, we assume without loss of generality that the ion concentration in the presence of a concentration gradient for an oblate-spheroidal ($\zeta$, $\nu$, $\phi$) coordinate system has the form 
\begin{equation}
        \label{Boltzmann_dist}
        c^{(i)}(\zeta, \nu) = c_{\mathrm{s}}^{(i)}(\zeta, \nu) \exp{\left(-\frac{Z_i e \psi(\zeta, \nu)}{\kBT} \right)}, 
\end{equation}
where $c_{\mathrm{s}}^{(i)}(\zeta, \nu)$ is to be determined. We can relate $\zeta$ and $\nu$ to a cylindrical coordinate system by $r = a\sqrt{(1 + \nu^2)(1 - \zeta^2)}$ in the radial direction and $z = a \nu \zeta$ in the axial direction, for which $0 \leq \zeta \leq 1$, $-\infty < \nu < \infty$, and $0 \leq \phi < 2 \pi$.\cite{rankinEntranceEffectsConcentrationgradientdriven2019, morse1953a} 

We assume that $D_i$ and $\mu _i$ are related by the Einstein relation \cite{probstein1994} $\mu _i = \frac{D_i}{\kBT}$, where $k_{\mathrm{B}}$ is the Boltzmann constant and $T$ is the temperature. Non-dimensionalizing the velocity, electric potential, concentrations, spatial gradients, ion diffusivities and ion flux densities according to $\bm{u} = \frac{\eps}{\eta a} ( \frac{\kBT}{e} ) ^2 \hat{\bm{u}}$, $c^{(i)} = c_{\infty} \hat{c}^{(i)}$, $\psi = \frac{\kBT}{e} \hat{\psi}$, $\nabla = \frac{\hat{\nabla}}{a}$, $\hat{D}_i = D_0 \hat{D}_i$ and $\bm{j}_i = \frac{D_0 c_{\infty}^{(i)} \hat{\bm{j}}_i}{a}$, respectively, we can write the ion flux density in Eq.~\eqref{nernst-planck} as
\begin{equation}
        \label{nernst-planck_hat}
        \hat{\bm{j}} _i = \mathrm{Pe} \ c^{(i)} \bm{u} - \hat{D}_i \left[ \hat{\nabla} c^{(i)} + Z_i c^{(i)} \hat{\nabla} \hat{\psi} \right],
\end{equation}
where $\mathrm{Pe} = \frac{\eps}{\eta D_0} ( \frac{\kBT}{e} )^2$ is the Péclet number, \cite{maoElectroosmoticFlowNanopore2014} $D_0$ is a characteristic diffusivity, $\cinf = \frac{1}{N} \sum _i\cinf^{(i)}$ is the average bulk concentration of the electrolyte solution containing $N$ species, and $\cinf ^{(i)} = (c_{\mathrm{H}}^{(i)} + c_{\mathrm{L}}^{(i)})/2$ is the average bulk concentration of species~$i$. Assuming that advection of the solute is negligible compared with diffusion ($\mathrm{Pe} \ll 1$), Eq.~\eqref{nernst-planck_hat} reduces to
\begin{equation}
    \label{diff-advec}
    \hat{\bm{j}} _i = - \hat{D}_i \left[ \hat{\nabla} \hat{c}^{(i)} + Z_i \hat{c}^{(i)} \hat{\nabla} \hat{\psi} \right].
\end{equation}
From Eqs.~\eqref{Boltzmann_dist} and \eqref{diff-advec},
\begin{equation}
        \label{diff-advec2}
        \hat{\bm{j}} _i = - \hat{D}_i e^{-Z_i \hat{\psi}} \hat{\nabla} \hat{c}_{\mathrm{s}} ^{(i)},
\end{equation}
where $c_{\mathrm{s}} ^{(i)} = c_{\infty} \hat{c}_{\mathrm{s}} ^{(i)}$. Inserting Eq.~\eqref{diff-advec2} into Eq.~\eqref{nernst-planck} and using the Debye--H\"uckel approximation, $|Z_i \hat{\psi}| \ll 1$, gives 
\begin{equation}
    \label{diff-advec3}
    \hat{\nabla} \cdot \hat{\bm{j}}_i =  - \hat{D}_i \hat{\nabla} \cdot \left[ \left(1 - Z_i \hat{\psi} \right) \hat{\nabla} \hat{c}_{\mathrm{s}}^{(i)} \right] = 0.
\end{equation}

We assume that each system variable can be represented by a perturbation expansion with respect to the equilibrium value, where $\beta \ll 1$ is a dimensionless quantity that characterizes the perturbation due to the applied concentration difference. Hence, to first order
\begin{align}
    \label{velocity_perturb}
    \bm{u} &= \beta \bm{u}_1, \\
    \label{conc_perturb}
    c_{\mathrm{s}}^{(i)}  &= c _{\mathrm{s}_0}^{(i)} + \beta c_{\mathrm{s}_1}^{(i)}, \\
    \label{potential_perturb}
    \psi &= \psi _0 + \beta \psi _1, \\
    \label{pressure_perturb}
    p &= p_0 + \beta p_1,
\end{align}
where $\bm{u}_0=\bm{0}$, since there is no net flow at equilibrium, and $c _{\mathrm{s}_0}^{(i)} = \cinf^{(i)}$. Expanding to $O(\beta)$ and assuming negligible contributions due to $Z_i \hat{\psi} _0$ (i.e., a small equilibrium electric potential), Eq.~\eqref{diff-advec3} reduces to
\begin{equation}
        \label{laplace}
        \hat{\nabla}^2 \hat{c}_{\mathrm{s} _1}^{(i)} = 0 ,   
\end{equation}
where $\bm{\hat{n}} \cdot \hat{\nabla} \hat{c} _{\mathrm{s} _1} ^{(i)} = 0$ on the membrane surface. The boundary conditions far from the membrane are $c^{(i)} \rightarrow c_{\mathrm{H}}^{(i)} = \cinf^{(i)} + \frac{\Delta c_{i}}{2}$ and $c^{(i)} \rightarrow c_{\mathrm{L}}^{(i)} = \cinf^{(i)} - \frac{\Delta c_{i}}{2}$ in the upper and lower half planes, respectively, and $\psi \rightarrow 0$; hence, $\beta \hat{c}_{\mathrm{s} _1}^{(i)} \rightarrow \pm \frac{\Delta \hat{c}_{i}}{2}$ far from the membrane surface in the upper and lower half planes, respectively. Solving Eq.~\eqref{laplace} subject to the boundary conditions on $\hat{c}_{\mathrm{s}_1}^{(i)}$ gives \cite{morse1953b}
\begin{equation}
        \label{cs_solution}
        \beta \hat{c}_{\mathrm{s} _1}^{(i)}(\zeta, \nu) = \frac{\Delta \hat{c}_i}{\pi}\tan^{-1}{\nu},
\end{equation}
which is analogous to the result for a neutral solute.\cite{rankinEntranceEffectsConcentrationgradientdriven2019}

Summing Eq.~\eqref{diff-advec2} over $i$ for a binary $Z$:$Z$ electrolyte, where $Z_+ = -Z_- = Z$, $\Delc_+ = \Delc_- = \Delc$, $\cinf^+ = \cinf^- = \cinf$ and $c^+ _{\mathrm{s}} = c^- _{\mathrm{s}} = c _{\mathrm{s}}$, we can write the re-dimensionalized total ion flux density $\bm{j} = \bm{j}_+ + \bm{j}_-$ and electric current density $\bm{j}_{\mathrm{e}} = e (Z_+ \bm{j}_+ + Z_- \bm{j}_-)$, expanded up to $O(\beta)$ for $Ze|\psi| \ll \kBT$, as
\begin{align}
        \label{solute_flux_density}
        \bm{j} = - \beta \nabla c_{\mathrm{s}_1} & \left\{ (\Dp)\left[1 + \frac{1}{2} \left(\frac{Ze \psi _0}{\kBT} \right) ^2 \right] \right.  \nonumber \\
        &  -  \left. (\Dm)\left( \frac{Ze\psi _0 }{\kBT}\right) \right\}  
\end{align}
and
\begin{equation}
        \label{electric-current-density2}
        \bm{j}_{\mathrm{e}} = \beta Ze \nabla c_{\mathrm{s}_1}  \left[ (\Dp) \left( \frac{Ze\psi _0}{\kBT} \right) - (\Dm) \right],
\end{equation}
respectively. The total solute flux $J$ and electric current $I$ induced by the concentration difference across the membrane can be obtained from the surface integrals of the total ion flux density and electric current density, respectively, over the pore aperture;\cite{rankinEntranceEffectsConcentrationgradientdriven2019} i.e.,
\begin{align}
        \label{solute_flux1}
        J &= \iint _S \mathrm{d} S \,  \bm{\hat{n}} \cdot \bm{j}, \\
        \label{electric_current1}
        I &= \iint _S \mathrm{d} S \,  \bm{\hat{n}} \cdot \bm{j}_{\mathrm{e}},
\end{align}
where $\bm{\hat{n}} = \bm{\hat{\nu}}$ is the unit vector normal to the pore mouth. Inserting Eq.~\eqref{solute_flux_density} into Eq.~\eqref{solute_flux1} and evaluating the integral at the pore mouth $(\nu=0)$ gives
\begin{align}
        \label{solute_flux3}
        J = - a (\Dp) &\left\{ 2 \Delc + \frac{\eps \Dellnc}{(\lDinf)^2} \int ^1 _0 \mathrm{d} \zeta \, \left [ \frac{ ( \psi_0|_{\nu=0} )^2}{2\kBT} \right. \right.  \nonumber \\
         & -  \left. \left. \left( \frac{\Dm}{\Dp} \right) \frac{\psi_0|_{\nu=0}}{Ze} \right] \right\}, 
\end{align}
where $\Delc/\cinf = (c_{\mathrm{H}} - c_{\mathrm{L}})/\cinf \approx \Dellnc$ in the surface contributions to the fluxes at small $\Delc$, and
\begin{equation}
        \label{Debye_length}
        \lDinf = \sqrt{\frac{\eps \kBT}{2(Ze)^2\cinf}}
\end{equation}
is the equilibrium (i.e., in the absence of a concentration difference $\Delc$) Debye screening length of a $Z$:$Z$ electrolyte. \cite{probstein1994} The first term in Eq.~\eqref{solute_flux3} is the bulk contribution to the total solute flux, $\Jb$. Inserting Eq.~\eqref{electric-current-density2} into Eq.~\eqref{electric_current1} and integrating over the pore mouth gives
\begin{align}
        \label{electric_current3}
        I = &-2a (\Dm) ze \Delta c \nonumber \\ 
        &+ a (\Dp) \frac{\eps \Dellnc}{(\lDinf)^2} \int ^1 _0 \mathrm{d} \zeta \, \psi_0 |_{\nu=0},
\end{align}
where the first term is the bulk contribution to the electric current, $\Ib$. The $O(\psi_0 |_{\nu=0})$ term in Eq.~\eqref{solute_flux3} and the bulk contribution to Eq.~\eqref{electric_current3} are negligible when the electric field in the pore mouth induced by the difference in ion diffusivities is small. Since the bulk and surface terms in the total solute flux and electric current have different dependencies on the concentration difference, it is not possible to define a single transport coefficient for the total flux; hence, we define the surface contributions $\delJ = J-\Jb$ and $\delI = I-\Ib$.

For the related problem of electroosmosis through a circular aperture, it has previously been shown using the reciprocal theorem \cite{happel1983a, maoElectroosmoticFlowNanopore2014} that
\begin{equation}
    \label{flow rate integral}
    Q = -\frac{1}{\Delta p} \iiint _V \mathrm{d} V \, \bm{\bar{u}}\cdot \bm{F} ,
\end{equation}
where $\bm{\bar{u}}$ is the fluid velocity induced by a pressure difference $\Delta p = p_{\mathrm{H}} - p_{\mathrm{L}}$ in the same system geometry for $\Delc=0$, $\bm{F}$ is the electric body force in the Stokes equation (Eq.~\eqref{stokes}), and the integral is over the volume occupied by the fluid. The analytical solution for the pressure-driven flow velocity in this geometry is \cite{rankinEntranceEffectsConcentrationgradientdriven2019}
\begin{equation}
    \label{stream-function}
        \bm{\bar{u}}= -\frac{a \zeta ^2 \Delta p}{2 \pi \eta \sqrt{(1 + \nu ^2)(\nu^2 + \zeta ^2)}} \bm{\hat{\nu}} .
\end{equation}
For a binary electrolyte, the electric body force $\bm{F}$ is
\begin{equation}
    \label{stokes-eterm1}
    \bm{F} = -\rho \nabla \psi = -e(Z_+ c^+ + Z_- c^-) \nabla \psi.
\end{equation}
Inserting Eq.~\eqref{Boltzmann_dist} for $c^\pm$ into Eq.~\eqref{stokes-eterm1} and assuming $Ze|\psi| \ll \kBT$ and a $Z$:$Z$ electrolyte gives
\begin{equation}
    \label{stokes-eterm2}
   -\rho \nabla \psi  = \frac{(Ze)^2}{\kBT} c_{\mathrm{s}} \nabla \psi^2 ,
\end{equation}
where $Ze \sinh(\frac{Ze\psi}{\kBT})\nabla \psi = \kBT \nabla \cosh (\frac{Ze\psi}{\kBT}) \approx \frac{(Ze)^2}{2\kBT} \nabla \psi^2$. Since there is zero fluid velocity at equilibrium, expanding Eq.~\eqref{stokes-eterm2} to $O(\beta)$ gives
\begin{align}
        \label{body-force1}
         -\rho_1 \nabla \psi_0 - \rho_0 \nabla \psi_1 &= \frac{(Ze)^2}{\kBT} \left[c_{\mathrm{s}_1} \nabla \psi_0^2 + 2 \cinf \nabla (\psi _0 \psi _1) \right] \nonumber \\
         &= \nabla \left(\frac{\eps}{(\lDinf)^2} \psi _0 \psi _1 \right) 
           + \frac{\eps}{2 (\lDinf)^2} \frac{c_{\mathrm{s} _1}}{\cinf}\nabla \psi_0^2  .
\end{align}
In the absence of an applied electric field, $\psi_1 \rightarrow 0$ far from the membrane surface. Hence, there is no net contribution from $\nabla \left(\frac{\eps}{(\lDinf)^2} \psi _0 \psi _1 \right)$ to the total volumetric flow rate in Eq.~\eqref{flow rate integral}. By inserting Eq.~\eqref{cs_solution} for $c_{\mathrm{s}_1}$ into Eq.~\eqref{body-force1}, we can write the  electric body force as
\begin{equation}
    \label{body-force}
    \bm{F} = \frac{\eps}{2 (\lDinf)^2} \frac{\Dellnc}{\pi} \tan^{-1}\nu \nabla \psi_0^2 .
\end{equation}
Inserting Eqs.~\eqref{body-force} and \eqref{stream-function} into Eq.~\eqref{flow rate integral} gives
\begin{align}
    \label{flow_rate1}
    Q &= \frac{a^3 \eps \Dellnc}{2 \pi \eta ( \lDinf )^2} \int ^1 _0 \mathrm{d} \zeta \, \zeta ^2 \int ^\mathrm{\infty} _{-\mathrm{\infty}} \mathrm{d} \nu \,\tan^{-1}\nu \frac{\partial (\psi _0 ^2)}{\partial \nu}  \nonumber\\
    &= -\frac{a^3 \eps \Dellnc}{\pi \eta ( \lDinf ) ^2} \int ^1 _0 \mathrm{d} \zeta \, \zeta ^2 \int ^\mathrm{\infty} _0 \mathrm{d} \nu \, \frac{\psi _0^2}{1 + \nu^2}
\end{align}
as the total flow rate, by integrating by parts in the second step.

\subsection{Limiting cases and scaling behavior}
Letting $\psi _0 = \frac{a \sigma}{\eps} \epotscaled$, the flow rate and surface contributions to the solute flux and electric current with respect to the dimensionless variable $\epotscaled$ are
\begin{equation}
    \label{flow_rate4}
    Q = -\frac{a^5 \sigma ^2 \Dellnc}{\pi \eta \eps (\lDinf)^2} \int ^1 _0 \mathrm{d} \zeta \, \zeta ^2 \int ^\mathrm{\infty} _0 \mathrm{d} \nu \, \frac{ \epotscaled^2}{1 + \nu^2} ,  
\end{equation}
\begin{align}
    \label{solute_flux4}
    \delJ = &- \frac{(\Dp) a^2 |\sigma| \Dellnc}{Ze(\lDinf)^2} \left[\frac{a}{l _{\mathrm{GC}}}  \int ^1 _0 \mathrm{d} \zeta \ ( \epotscaled |_{\nu=0} )^ 2 \right. \nonumber \\
    &- \left. \text{sgn} (\sigma) \left( \frac{\Dm}{\Dp} \right) \int ^1 _0 \mathrm{d} \zeta \, \epotscaled|_{\nu=0} \right ]
\end{align}
and
\begin{equation}
    \label{electric_current4}
    \delI = \frac{(\Dp) a^2 \sigma \Dellnc}{(\lDinf)^2} \int ^1 _0 \mathrm{d} \zeta \, \epotscaled|_{\nu=0} ,
\end{equation}
where 
\begin{equation}
    l_{\mathrm{GC}} = \frac{2 \eps \kBT}{Ze|\sigma|}
\end{equation}
is the Gouy--Chapman length for a $Z$:$Z$ electrolyte.\cite{probstein1994} The analytical solution for the dimensionless equilibrium electric potential for $Ze|\psi| \ll \kBT$ in this geometry\cite{maoElectroosmoticFlowNanopore2014} is
\begin{align}
    \label{epot_in}
    \epotscaled = &- \int ^{\mathrm{\infty}} _0 \mathrm{d} s \, \frac{J_1(s) J_0(\hat{r} s)}{\sqrt{(a/\lDinf)^2 + s^2}} e^{-\hat{z}\sqrt{(a/\lDinf)^2 + s^2}} \nonumber\\ 
    &+ \frac{e^{-(a/\lDinf) \hat{z}}}{(a/\lDinf)} ,
\end{align}
where $\hat{z} = z/a $ and $\hat{r} = r/a $ are the dimensionless cylindrical coordinates, and $J_0$ and $J_1$ are Bessel functions of the first kind. The surface and volume integrals in Eqs.~\eqref{flow_rate4}--\eqref{electric_current4} have thus been reduced to dimensionless functions that depend on $a/ \lDinf$ only. 

\subsubsection{Thick electric double layer}
When the Debye screening length is much larger than the pore radius ($\lDinf \gg a$), the potential at the membrane surface is approximately constant and equal to the potential at a planar surface everywhere. Inserting the expression $\epotscaled \approx \lDinf / a$ obtained from Eq.~\eqref{epot_in} in this limit into Eqs.~\eqref{flow_rate4}--\eqref{electric_current4} gives the scaling laws
\begin{align}
    \label{Q_propto}
    Q &\propto a^3 \sigma ^2 \Dellnc , \\
    \label{J_propto}
    \delJ &\propto a \sigma ^2 \Dellnc , \\
    \label{I_propto}
    \delI &\propto \frac{a \sigma }{\lDinf} \Dellnc,
\end{align}
where we have assumed that the $O(\epotscaled)$ term in Eq.~\eqref{solute_flux4} is negligibly small. We show that this assumption is valid for $\lDinf/l_{\mathrm{GC}} \gg \frac{|D_+ - D_-|}{\Dp}$ in Sec.~\ref{sec:delJ} of the supplementary material. For simulations of potassium chloride (KCl) with average bulk concentrations of $0.3$, $3$ and $30$~mol~m$^{-3}$ (see Sec.~\ref{simulations}), this condition holds for $|\sigma| \gg 0.04$, $0.12$ and $0.36$~mC~m$^{-2}$, respectively.  

\subsubsection{Thin electric double layer}

Let us assume that the equilibrium electric potential can be approximated as decaying exponentially with the distance $d$ from the membrane surface over the equilibrium Debye screening length $\lDinf$. When the Debye screening length is much smaller than the pore radius ($\lDinf \ll a$), the potential also decays rapidly in space. Thus, we assume for simplicity that the equilibrium electric potential is approximately equal to the potential at the surface within a distance of $\lDinf$ from the surface and is zero elsewhere.

The full analytical expression for the flow rate considers the strength of the electric potential everywhere, but for thin electric double layers, it is reasonable to ignore contributions to the flow rate from ion interactions with the pore edge ($r\leq a$). For $r > a$, where $d \approx a\nu\zeta$ and $\nu > \zeta$ when $d \leq \lDinf \ll a$,\cite{rankinEntranceEffectsConcentrationgradientdriven2019} we assume that the dimensionless surface potential is approximately that at a planar surface, i.e. $\epotscaled(d) \approx \lDinf/a$ for $d < \lDinf$. Since the rate of exponential decay is doubled when $\epotscaled$ is squared, we assume that $(\epotscaled(d))^2 \approx 0$ when $d \geq \lDinf/2$. Thus, the integral in Eq.~\eqref{flow_rate4} can be approximated in the $\lDinf \ll a$ limit as
\begin{align}
    \int ^1 _0 \mathrm{d} \zeta \, & \zeta ^2  \int ^\mathrm{\infty} _0 \mathrm{d} \nu \, \frac{ (\epotscaled ) ^2}{1 + \nu^2}  \nonumber\\
    &\approx \left(\frac{\lDinf}{a} \right) ^2 \int ^1 _0 \mathrm{d} \zeta \, \zeta ^2 \int ^\mathrm{\infty} _\zeta \mathrm{d} \nu \, \frac{ H(\lDinf/2 - a\nu\zeta)}{1 + \nu^2} \nonumber \\
    \label{epot5}
    &\approx \frac{1}{8} \left (\frac{\lDinf}{a} \right )^4 ,
\end{align}
where $H(x)$ is the Heaviside function and we have used the solution to an analogous integral derived in Ref. \citenum{rankinEntranceEffectsConcentrationgradientdriven2019}. Inserting Eq.~\eqref{epot5} into Eq.~\eqref{flow_rate4}, we find in this limit that
\begin{equation}
    \label{Q_propto2}
    Q \propto a (\lDinf )^2 \sigma ^2 \Dellnc .
\end{equation}

For the total solute flux and electric current, we only need to consider the region inside the pore mouth $(r<a$, $\nu=0)$, where $d \approx a\zeta^2/2$ is the distance from the pore edge when $\lDinf \ll a$ and we take the surface potential to be the potential at the pore edge in this limit, i.e. $\epotscaled(d) \approx \lDinf/(2a)$ for $d < \lDinf$ (see Sec.~\ref{sec:epot} of the supplementary material for details of the derivation). The surface integral of $\epotscaled$ over the pore aperture can thus be approximated as
\begin{align}
      \int ^1 _0 \mathrm{d} \zeta \, \epotscaled|_{\nu=0} &\approx \frac{\lDinf}{2a} \int ^1 _0 \mathrm{d} \zeta \, H\left( \sqrt{\frac{2\lDinf}{a}} - \zeta \right) \nonumber \\
      \label{epot_integral_shortDebye2}
      &= \frac{1}{\sqrt{2}}\left( \frac{ \lDinf}{a} \right) ^{\frac{3}{2}} .
\end{align}
On the other hand, assuming $(\epotscaled(d))^2 \approx 0$ when $d\geq \lDinf/2$ since the rate of decay is doubled when $\epotscaled$ is squared,
\begin{align}
      \int ^1 _0 \mathrm{d} \zeta \, (\epotscaled|_{\nu=0})^2 &\approx \left( \frac{\lDinf}{2a} \right)^2 \int ^1 _0 \mathrm{d} \zeta \, H\left( \sqrt{\frac{\lDinf}{a}} - \zeta \right) \nonumber \\
      \label{epot_integral_shortDebye3}
      &= \frac{1}{4} \left( \frac{ \lDinf}{a} \right) ^{\frac{5}{2}}.
\end{align}
Inserting Eqs.~\eqref{epot_integral_shortDebye2} and \eqref{epot_integral_shortDebye3} into Eqs.~\eqref{solute_flux4} and \eqref{electric_current4}, respectively, gives the scaling relationships
\begin{align}
    \label{J_propto2}
        \delJ &\propto \left( a \lDinf \right) ^\frac{1}{2}  \sigma ^2  \Dellnc \\
    \label{I_propto2}
        \delI &\propto \left( \frac{a}{\lDinf} \right) ^{\frac{1}{2}} \sigma \Dellnc 
\end{align}
in this limit, where we verify that Eqs.~\eqref{epot5}, \eqref{epot_integral_shortDebye2} and \eqref{epot_integral_shortDebye3} are suitable approximations for the full expressions of the fluxes for thin electric double layers in Sec.~\ref{sec:epot} of the supplementary material. Note that we have neglected the $O(\epotscaled)$ term in Eq.~\eqref{solute_flux4} in this limiting case, which is a valid for $\lDinf/l_{\mathrm{GC}} \gg 2\sqrt{2} \frac{|D_+ - D_-|}{\Dp}$ (see Sec.~\ref{sec:delJ} of the supplementary material for details of derivation). For simulations of KCl where $\cinf = 0.3$, $3$ and $30$ mol m$^{-3}$, this condition holds for $|\sigma| \gg 0.11$, $0.34$ and $1.0$~mC~m$^{-2}$, respectively; thus, we can neglect the contribution to $\delJ$ due to the difference in ion diffusivities when referring to simulations of sufficiently high surface charge. As the bulk contribution to the total solute flux dominates at low surface charge, the contribution to the total solute flux due to the difference in ion diffusivities can be ignored in all simulations of KCl. For an equilibrium salt concentration of 250 mol m$^{-3}$ (i.e. the average concentration of seawater and fresh water), which generally corresponds to the thin electric-double-layer regime for nanoscale pores, this condition requires that $|\sigma| \gg 3$~mC~m$^{-2}$ for KCl and $|\sigma| \gg 35$~mC~m$^{-2}$ for sodium chloride (NaCl) \cite{grayAIPHandbook1972} to ignore the contribution to $\delJ$ due to the difference in ion diffusivities. As surface charge density magnitudes of $24$--$88$ mC m$^{-2}$ have been estimated for MoS$_2$ (which could be increased to $300$--$800$ mC m$^{-2}$ at higher pH)\cite{fengSinglelayerMoS2Nanopores2016} and 600 mC m$^{-2}$ for graphene, \cite{rollingsIonSelectivityGraphene2016} this assumption should be reasonably accurate under most conditions of interest for applications such as water desalination and osmotic power. 

\subsubsection{Heuristic derivation for high surface charge and non-overlapping electric double layers}
\label{scaling-heuristic}

To go beyond the Debye--H\"uckel approximation, we make use of analytical equations that have previously been derived for electrolyte solutions in a planar channel for non-overlapping electric double layers and arbitrary strengths of the electric potential to obtain scaling relationships for the fluxes in a 2D membrane outside the Debye--H\"uckel regime (at fixed $\lDinf/a$). Further details of the derivation of these relationships, which are particularly useful for quantifying the variation of the fluxes with the surface charge density when the surface potential is large, are given in Sec.~\ref{sec:sigma} of the supplementary material. We assume that the width of the electric double layer is smaller than the pore radius such that the scaling relationships of the surface contributions to the fluxes can be represented as the product of a function of the potential at a planar surface and some power-law scaling of $\lDinf/a$ that describes the interaction range of the potential in a 2D membrane geometry. When $\lDinf/a$ is fixed, Eq.~\eqref{flow_rate4} gives the scaling relationships 
\begin{equation}
    \label{Q_scaling_inside_DH}
    Q \propto \frac{\eps a}{\eta} \left(\frac{\sigma \lDinf}{\epsilon \epsilon_0} \right)^2 \Dellnc 
\end{equation}
where $\sigma \lDinf/(\epsilon \epsilon_0)$ is the potential at a planar surface in the Debye--H\"uckel regime. Exchanging $(\sigma \lDinf/(\epsilon \epsilon_0))^2$ in Eq.~\eqref{Q_scaling_inside_DH} for the factor in the equation for the concentration-gradient-driven fluid velocity outside the electric double layer for a planar channel that reduces to $(\sigma\lDinf/(\epsilon\epsilon_0))^2$ in the limit of small potentials\cite{prieveMigrationColloidConcentrationGradient1982} gives the scaling for the flow rate at fixed $\lDinf/a$ and arbitrary surface potentials as
\begin{equation}
    \label{other_Scaling_Q}
    Q \propto \left( \frac{\kBT}{Ze} \right)^2  \frac{\epsilon \epsilon _0 a}{\eta} \ln{\left( 1 + \frac{\lDu}{4\lDinf} \right)} \Dellnc,
\end{equation}
where
\begin{equation}
    \label{dukhin}
    \lDu = \frac{|\sigma|}{2 \cinf Ze} \left[-\frac{l_{\mathrm{GC}}}{\lDinf} + \sqrt{\left( \frac{l_{\mathrm{GC}}}{\lDinf} \right)^2 + 1} \right]
\end{equation}
is the Dukhin length (near a charged planar wall) of a $Z$:$Z$ electrolyte.\cite{leeLargeElectricSizeSurfaceCondudction2012} From Eq.~\eqref{other_Scaling_Q}, we see that $Q$ is proportional to $\sigma^2$ when $l_{\mathrm{GC}} \gg \lDinf$, which does not always correspond to the Debye--H\"uckel regime. When $l_{\mathrm{GC}} \ll \lDinf$, which occurs at high surface charge for a thin electric double layer, Eq.~\eqref{other_Scaling_Q} reduces to
\begin{equation}
    \label{other_Scaling_Q2}
    \ln{\left( 1 + \frac{\lDu}{4\lDinf} \right)} \approx \ln{ \left(1 + \frac{\lDinf}{l_{\mathrm{GC}}} \right)} + \ln{\left(\frac{1}{2}\right)} .
\end{equation}
Thus, Eq.~\eqref{other_Scaling_Q} predicts scaling of the flow rate with $\ln{|\sigma|}$ for 2D membranes with large surface charge densities and non-overlapping electric double layers when all other parameters are fixed. 
 
When ions transported by an applied concentration gradient interact electrostatically with a charged surface, $\kBT/(Z_ie) \Dellnc$ plays an analogous role to an electric field applied on species~$i$. As $\bm{j} = \bm{j}_+ + \bm{j}_-$ is the total ion flux density and $\bm{j}_e = e(Z_+\bm{j}_+ + Z_-\bm{j}_-)$ is the electric current density of a $Z$:$Z$ electrolyte, we assume that the concentration-gradient-driven total solute flux and electric-field-driven electric current exhibit the same scaling with the potential at a planar surface when the width of the electric double layer is smaller than the pore radius. When the contribution due to the difference in ion diffusivities can be ignored, we can use an analogous equation in Ref.~\citenum{leeLargeElectricSizeSurfaceCondudction2012} for the electric-field-driven electric current near a planar surface to write, at fixed $\lDinf/a$ and arbitrary surface potentials, that
\begin{equation}
\label{Jpropto3}
    \delJ \propto \frac{\kBT}{(Ze)^2} \kappa ^{\infty} _\mathrm{s} \Dellnc ,
\end{equation}
where 
\begin{equation}
    \label{conductance}
    \kappa_\mathrm{s} ^{\infty} = \frac{Ze(\Dp) |\sigma|}{2 \kBT} \left[-\frac{l_{\mathrm{GC}}}{\lDinf} + \sqrt{\left( \frac{l_{\mathrm{GC}}}{\lDinf} \right)^2 + 1} \right] 
\end{equation}
is the surface conductivity (near a planar wall) of a $Z$:$Z$ electrolyte within the Poisson--Nernst--Planck framework (Eqs.~\eqref{poisson} and \eqref{nernst-planck} with $\bm{u} = 0$ and analogous boundary conditions to those applied here but for a planar wall).\cite{leeLargeElectricSizeSurfaceCondudction2012}
We can also represent Eq.~\eqref{conductance} in terms of hyperbolic functions and use the relationship between the total solute flux density and the electric current density for a $Z$:$Z$ electrolyte (see supplementary material, Sec.~\ref{sec:sigma}) to show that
\begin{equation}
\label{Ipropto3}
    \delI \propto \sigma (D_+ + D_-) \Dellnc .
\end{equation}
When the width of the electric double layer is smaller than the pore radius, Eq.~\eqref{Jpropto3} predicts that $\delJ$ is proportional to $\sigma^2$ when $l_{\mathrm{GC}} \gg \lDinf$, whereas $\delJ$ is proportional to $\sigma$ when $l_{\mathrm{GC}} \ll \lDinf$. By contrast, Eq.~\eqref{Ipropto3} predicts that $\delI$ is proportional to $\sigma$ in this regime (independent of $\lDinf/l_{\mathrm{GC}}$). We have used the forms of the equations for the ion flux and electric current density (see Eqs.~\eqref{total_ion_flux_density} and \eqref{total_electric_current_density} in the supplementary material) to infer that the contribution from the difference in ion diffusivities is proportional to $(D_+ - D_-) \sigma \Dellnc / (Ze)$ at fixed $\lDinf/a$ and that of $\delI$ is proportional to $\kBT\kappa ^{\infty} _\mathrm{s} \Dellnc/(Ze)$. These relationships indicate that these contributions can generally be ignored in $\delJ,$ but could become significant in $\delI$ at sufficiently high surface charge, depending on the diffusion coefficients of the ions.

\section{Results and discussion}

\begin{table}
\caption{\label{scaling-flux} Scaling relationships for the total flow rate, $Q$, surface contribution to the solute flux, $\delJ$, and surface contribution to the electric current, $\delI$, with the pore radius $a$, equilibrium Debye screening length $\lDinf$ and pore length $L$ (where relevant) in the Debye--H\"uckel regime for various systems, applied gradients and limiting regimes. The entry for the gradient indicates the transport process ($\Dellnc$ and $\Delc$ for concentration-gradient-driven flow and $\Delta \psi$ for electric-field-driven flow) and the linear scaling of the flux with the applied gradient. The membrane--solute interaction range $\lambda$ for the neutral solute plays an analogous role to $\lDinf$ in the electrolyte.}
\begin{ruledtabular}
\renewcommand{\arraystretch}{1.3}
\begin{tabular}{c c c c c}
Flux & System & Gradient & Limit & Scaling \\
\hline 
\multirow{8}{*}{$Q$} & circular aperture,  & \multirow{2}{*}{$\Dellnc$} & $\lDinf \gg a$ & $a^3$  \\ 
& $Z$:$Z$ electrolyte &                 & $\lDinf \ll a$ & $a (\lDinf)^2$ \\ 
\cline{2-5}
& circular aperture, & \multirow{2}{*}{$\Delc$} & $\lambda \gg a$ & $a^3$ \\
& neutral solute \cite{rankinEntranceEffectsConcentrationgradientdriven2019}    &            & $\lambda \ll a$ & $a \lambda ^2$ \\
\cline{2-5} 
& circular aperture,  & \multirow{2}{*}{$\Delta \psi$} & $\lD \gg a$ & $a^3(\lD)^{-1}$  \\ 
& $Z$:$Z$ electrolyte\footnotemark[1] &                & $\lD \ll a$ & $a \lD$ \\
\cline{2-5}
& cylinder,           & \multirow{2}{*}{$\Dellnc$} & $\lDinf \gg a$ & $a^4 L^{-1}$ \\
& $Z$:$Z$ electrolyte &                 & $\lDinf \ll a$ & $a^2 (\lDinf)^2  L^{-1}$ \\
\hline
\multirow{6}{*}{$\delJ$} & circular aperture,  & \multirow{2}{*}{$\Dellnc$} & $\lDinf \gg a$ & $a$ \\ 
& $Z$:$Z$ electrolyte &            & $\lDinf \ll a$ & $(a \lDinf)^{\frac{1}{2}}$ \\ 
\cline{2-5} 
& circular aperture, & \multirow{2}{*}{$\Delc$} & $\lambda \gg a$ & $a$ \\
& neutral solute \cite{rankinEntranceEffectsConcentrationgradientdriven2019}     &            & $\lambda \ll a$ & $(a \lambda)^{\frac{1}{2}}$ \\
\cline{2-5} 
& cylinder,            & \multirow{2}{*}{$\Dellnc$} & $\lDinf \gg a$ & $a^2 L^{-1}$ \\
& $Z$:$Z$ electrolyte &             & $\lDinf \ll a$ & $a \lDinf L^{-1}$ \\
\hline 
\multirow{4}{*}{$\delI$} & circular aperture,  & \multirow{2}{*}{$\Dellnc$} & $\lDinf \gg a$ & $a (\lDinf)^{-1}$  \\ 
& $Z$:$Z$ electrolyte &            & $\lDinf \ll a$ & $(a/ \lDinf) ^{\frac{1}{2}}$ \\ 
\cline{2-5} 
& cylinder,            & \multirow{2}{*}{$\Dellnc$} & $\lDinf \gg a$ & $a^2 (\lDinf)^{-1} L^{-1}$ \\
& $Z$:$Z$ electrolyte  &           & $\lDinf \ll a$ & $a L^{-1}$ \\
\end{tabular}
\end{ruledtabular}
\footnotetext[1]{Scaling in the $\lD \gg a$ regime derived in Ref.~\citenum{maoElectroosmoticFlowNanopore2014}.}
\end{table}

To validate the derived scaling laws, finite-element method (FEM) simulations of concentration-gradient-driven transport of KCl solutions through $0.2$~nm thick membranes containing pores of various radii and surface charge densities, with various upper and lower reservoir concentrations, were carried out with COMSOL Multiphysics (version 4.3a).\cite{comsol4.3a} Details of these simulations are given in Sec.~\ref{sec:FEM} of the supplementary material, which we verified were all within the low Péclet-number regime. Simulations of electrolyte transport with zero electric potential were carried out to calculate $\Jb$ and $\Ib$ (see Sec~\ref{sec:bulk} in the supplementary material), where $\delJ$ and $\delI$ were recovered by calculating $J-\Jb$ and $I-\Ib$, respectively. The theory predicts that the flow rate $Q$ and the surface contributions to the total solute flux, $\delJ$, and electric current, $\delI$, are linearly related to the logarithm of the concentration difference $\Dellnc$. We have verified this scaling for the range of concentration differences $1/100 \cinf \leq \Delc \leq 18/11 \cinf$ studied in the numerical simulations (see Figs.~\ref{fig:Delc1}--\ref{fig:Delc3} in the supplementary material). The linear scaling of $Q$, $\delJ$ and $\delI$ with $\Dellnc$ in the linear-response regime has been shown for concentration-gradient-driven electrolyte transport in cylindrical pores in this regime.\cite{rankinEffectHydrodynamicSlip2016} These scaling relationships are expected for electrolyte transport and differ to the linear scaling of fluxes with $\Delc$ for the neutral solute case in Ref.~\citenum{rankinEntranceEffectsConcentrationgradientdriven2019} as a result of the concentration dependence of the interaction range (Debye screening length) for electrolytes. 

\subsection{Scaling of fluxes}
\label{simulations}

The case of a circular aperture in an infinitesimally thin wall is relevant beyond the case of a 2D membrane, since the resistance to flow in this geometry is useful for approximating the resistance of the pore ends of a finite-length pore, with the total flow resistance often accurately approximated as the sum of the resistances due to the pore ends and pore interior.\cite{rankinEffectHydrodynamicSlip2016, leeLargeElectricSizeSurfaceCondudction2012} We thus highlight the differences between our results and the scaling behaviors for a long cylinder, which are useful for modeling the interior of a finite-length pore\cite{rankinEffectHydrodynamicSlip2016, leeLargeElectricSizeSurfaceCondudction2012} and transport in nanotubes \cite{borgMultiscaleSimulationNanotube2018} (see Sec.~\ref{sec:cylinder} of the supplementary material for the derivation of the scaling laws in the Debye--H\"uckel regime for concentration-gradient-driven electrolyte transport in a long cylinder). As our results can also be useful for understanding flow driven by both an applied electric field and applied concentration gradient, which has relevance to osmotic power generation,\cite{duChanoisMembranesSelectiveIonSeparations2021, zhangNanofluidicsOsmoticEnergy2021a, rankinEffectHydrodynamicSlip2016} we also compare our results with electric-field-driven flow in this geometry\cite{maoElectroosmoticFlowNanopore2014, leeLargeElectricSizeSurfaceCondudction2012} and derive the scaling of electroosmosis with $\lD/a$ in the thin electric-double-layer limit for the first time in Sec.~\ref{sec:EDF} of the supplementary material. Moreover, we compare our results with those of concentration-gradient-driven transport of a neutral solute in the same geometry \cite{rankinEntranceEffectsConcentrationgradientdriven2019} to highlight the implications of extending this theory to electrolytes. The scaling laws derived in this work and for comparable cases in the Debye--H\"uckel regime are shown in Table~\ref{scaling-flux}. 

\begin{figure}[tb!] 
\includegraphics{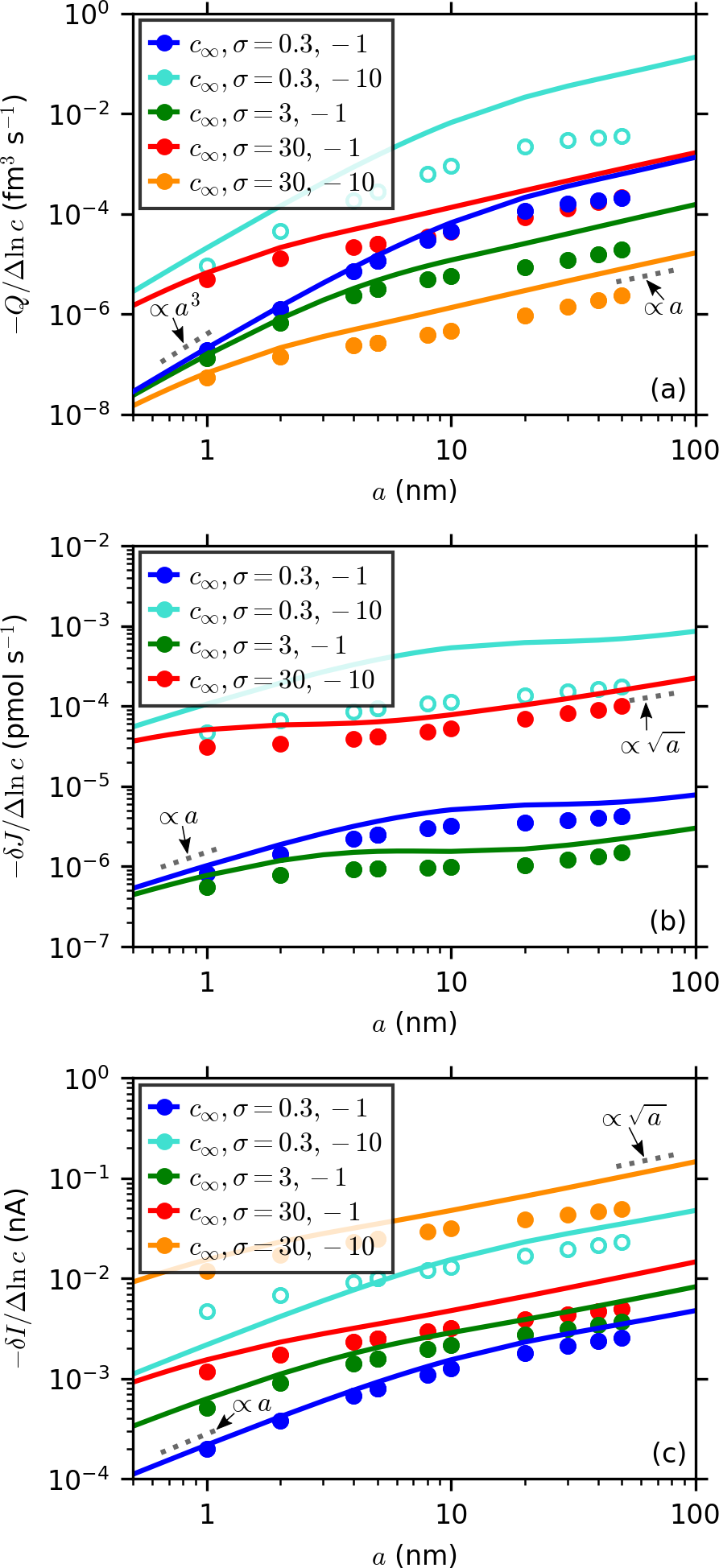}
   \caption{\label{apore_scaling} (a) Flow rate $Q$, (b) surface contribution to the solute flux, $\delJ$, and (c) surface contribution to the electric current, $\delI$, over $\Dellnc$ vs pore radius $a$ from FEM simulations (symbols) and theory (solid lines) for surface charge densities $\sigma = -1$ and $-10$~mC~m$^{-2}$ and a range of equilibrium electrolyte concentrations $\cinf$ (mol~m$^{-3}$). FEM simulations are shown for surface potential energies that are $< \kBT$ (filled symbols) and $> \kBT$ (empty symbols) far from the pore mouth, respectively, and the scaling is shown for various powers of $a$ (dotted lines).}
\end{figure}

\begin{figure}[tb!] 
\includegraphics{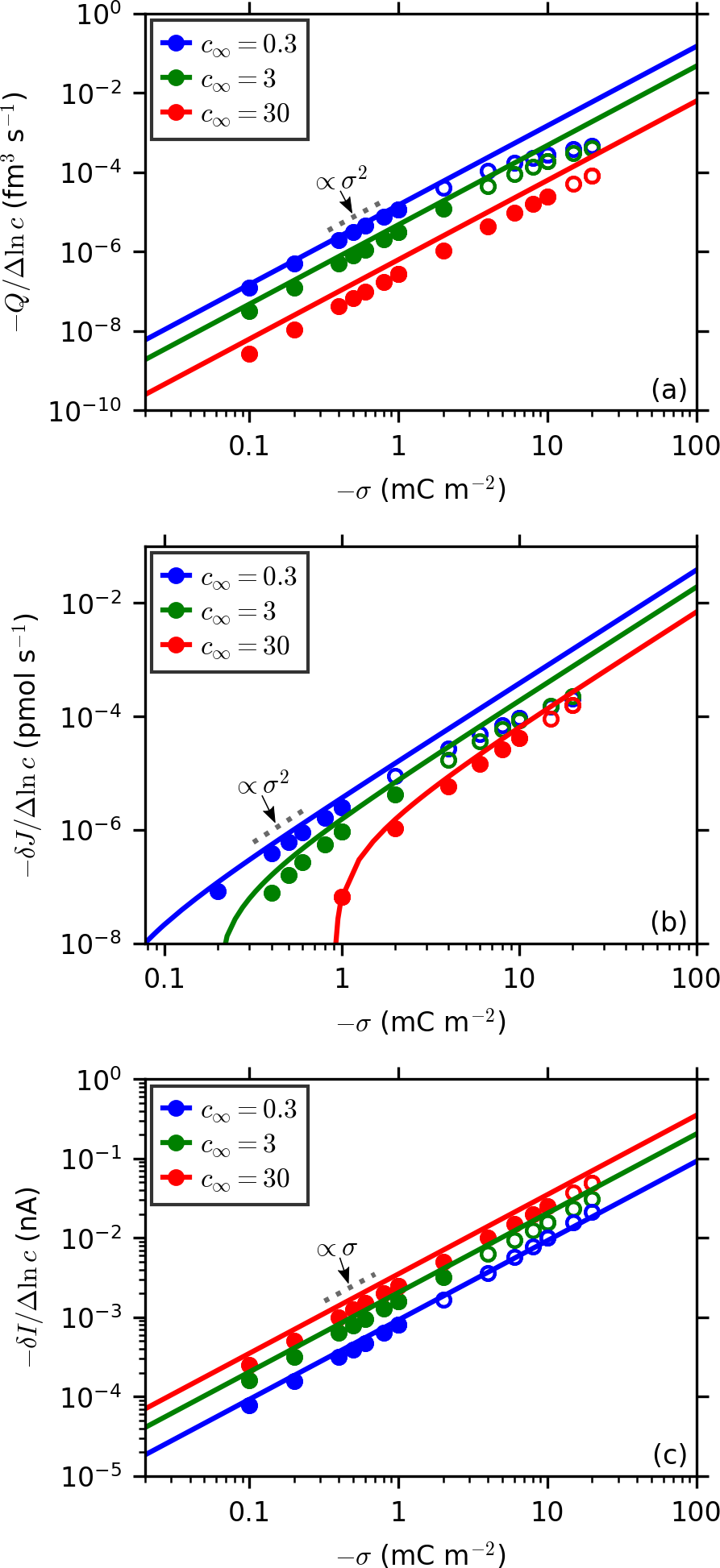}
   \caption{\label{sigma_scaling} (a) Flow rate $Q$, (b) surface contribution to the solute flux, $\delJ$, and (c) surface contribution to the electric current, $\delI$, over $\Dellnc$ vs surface charge density $\sigma$ from FEM simulations (symbols) and theory (solid lines) for pore radius $a = 5$ nm and a range of equilibrium electrolyte concentrations $\cinf$ (mol~m$^{-3}$). FEM simulations are shown for surface potential energies that are $< \kBT$ (filled symbols) and $> \kBT$ (empty symbols) far from the pore mouth, respectively, and the scaling is shown for various powers of $\sigma$ (dotted lines).}
\end{figure}

Figures~\ref{apore_scaling} and \ref{sigma_scaling} compare the simulation results for the concentration-gradient-driven flow rate $Q$ and the surface contributions to the solute flux, $\delJ$, and electric current, $\delI$, with the scaling with the pore radius $a$ and surface charge density $\sigma$, respectively, predicted by the theory in the Debye--H\"uckel regime for various average bulk concentrations $\cinf$. When the equilibrium Debye screening length $\lDinf$ is much larger than the pore radius ($\lDinf \gg a$), the theory predicts that $Q$ is proportional to $a^3$, and that $\delJ$ and $\delI$ are proportional to $a$. On the other hand, it predicts that $Q$ is proportional to $a$ and that $\delJ$ is proportional to $\sqrt{a}$ when the equilibrium Debye screening length is much smaller than the pore radius ($\lDinf \ll a$). These scaling relationships can be seen for the simulation data in Fig.~\ref{apore_scaling} to hold in the Debye--H\"uckel regime when $a$ and $\cinf$ are small ($\lDinf \gg a$) and large ($\lDinf \ll a$), respectively. As shown in Table~\ref{scaling-flux}, the scaling of $Q$ and $\delJ$ with the pore radius in both the thick and thin electric-double-layer limits are identical to those for concentration-driven transport of a solution containing a neutral solute in a 2D membrane; \cite{rankinEntranceEffectsConcentrationgradientdriven2019} these scaling laws for $Q$ are also analogous to those of the electric-field-driven flow rate in a 2D membrane.\cite{maoElectroosmoticFlowNanopore2014} By contrast, the scaling relationships of the concentration-gradient-driven fluxes with $a$ in a 2D membrane differ from those in a long cylinder, where Table~\ref{scaling-flux} indicates that all of the fluxes have a weaker dependence on the pore radius in a 2D membrane than in thick membranes in the Debye--H\"uckel regime, and thus are less strongly limited as the pore size decreases.

Figure~\ref{apore_scaling} shows good quantitative agreement between the theory and simulations for $Q$, $\delJ$ and $\delI$ in the Debye--H\"uckel regime for thick electric double layers. In the thin electric-double-layer limit, the predicted scaling of $Q$ and $\delJ$ with the pore radius from the theory agrees with the simulations, but the fluxes from the theory are shifted by roughly a constant factor relative to the simulations. The theory also predicts that $\delI$ is proportional to $\sqrt{a}$ in the thin electric-double-layer limit; however, the simulations show a slightly weaker dependence of $\delI$ on $a$ than predicted by the theory when $a$ and $\cinf$ are both large, which we address further in Sec.~\ref{sec:universal_scaling}. Nevertheless, the bulk contributions to the solute flux and electric current (in KCl), which are accurately predicted by the theory (see Fig.~\ref{fig:bulk} in the supplementary material), dominate in the Debye--H\"uckel regime and thin electric-double-layer limit. Thus, the theory predicts the total solute flux $J$ and total electric current $I$ reasonably well under these conditions. We would like to note that since $\delJ$ is the difference between the total solute flux and the bulk contribution to the total solute flux, which were calculated from separate simulations, $\delJ$ is sensitive to numerical error for large electrolyte concentrations and large pore radii ($\lDinf \ll a$) in the Debye--H\"uckel regime when the bulk contribution dominates. The scaling relationships of the fluxes with $a$ for thick electric double layers do not hold outside the Debye--H\"uckel regime where the theory is expected to break down, as shown by the empty symbols in Fig.~\ref{apore_scaling}. Note that the scaling relationships of the fluxes with the equilibrium Debye screening length, $\lDinf$, are discussed in Sec~\ref{sec:universal_scaling}.

The analytical theory predicts that $Q$ and $\delJ$ (when the contribution to $\delJ$ from the difference in ion diffusivities can be neglected) are proportional to $\sigma ^2$, while $\delI$ is proportional to $\sigma$, which is evident for all $\cinf$ shown in Fig.~\ref{sigma_scaling} in the Debye--H\"uckel regime. These scaling relationships are the same as for thick membranes (see Sec.~\ref{sec:cylinder} of the supplementary material) and differ from those observed for electric-field-driven flow, for which the electroosmotic flow rate and electric current are proportional to $\sigma$ and $\sigma^2$, respectively, in the Debye--H\"uckel regime.\cite{maoElectroosmoticFlowNanopore2014, leeLargeElectricSizeSurfaceCondudction2012} Figures~\ref{Q_largepsi1} and \ref{fig:J-large_sigma}(a) in the supplementary material indicate that the scaling predicted by the heuristic theory for non-overlapping electric double layers and arbitrary strengths of the electric potential is approximately true for $Q$ when $\cinf \geq 1$~mol~m$^{-3}$ and $\delJ$ for all $\cinf$ in the simulations (where $a=5$~nm), which includes concentrations ($\cinf < 3$~mol~m$^{-3}$) for which the theory might be expected to break down, since $\lDinf > a$. Figures~\ref{sigma_scaling}(c) and \ref{fig:J-large_sigma}(b) in the supplementary material indicate that the linear scaling of $\delI$ with $\sigma$ holds for all $\sigma$ (i.e. outside the Debye--H\"uckel regime) in simulations with non-overlapping electric double layers ($\cinf \geq 3$~mol~m$^{-3}$ for $a=5$ nm), as predicted by the heuristic theory. 

\subsection{Universal scaling laws}
\label{sec:universal_scaling}

\begin{figure}[tb!]
\includegraphics{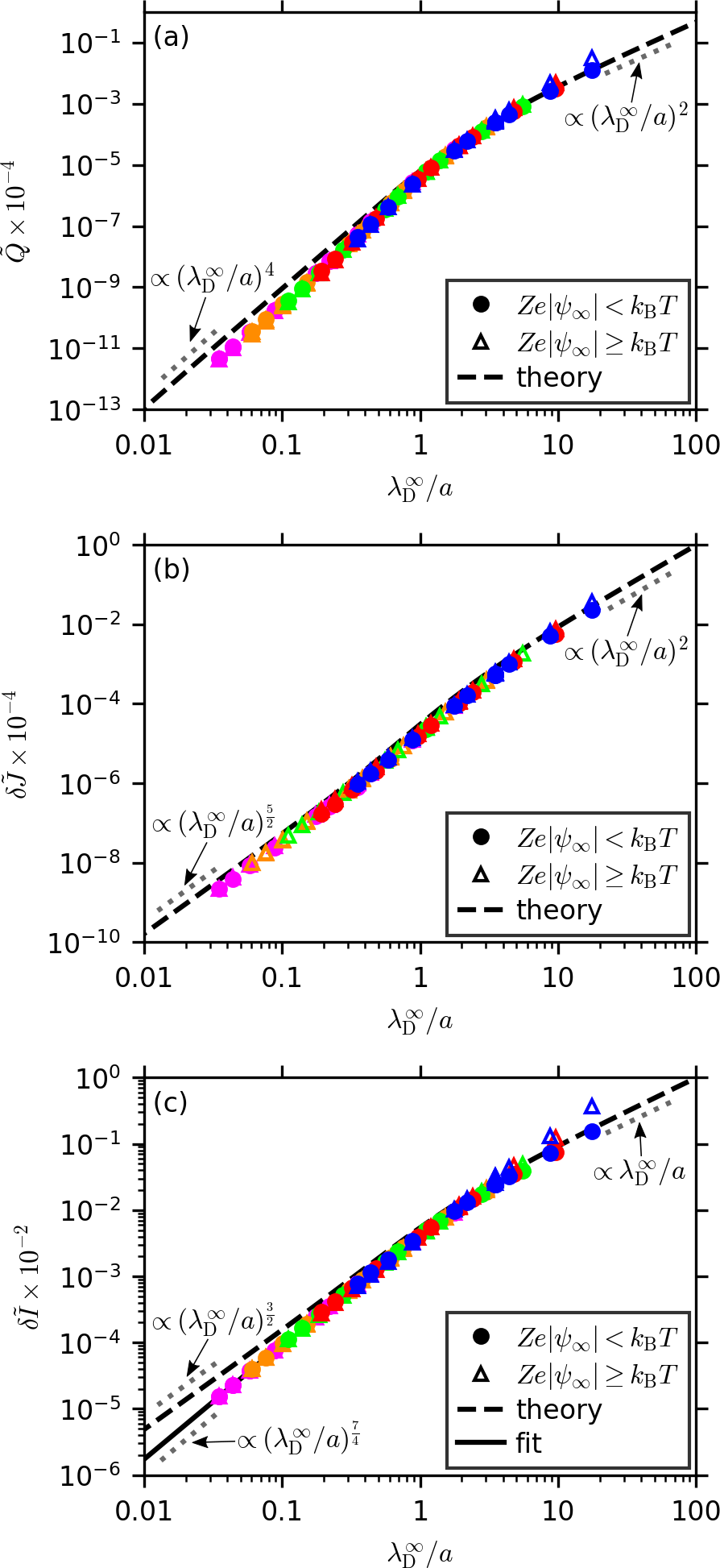}
\caption{\label{kappaa_scaling} Dimensionless (a) flow rate $\Tilde{Q}$, (b) surface contribution to the total solute flux, $\Jscaled$, and (c) surface contribution to the electric curent, $\Iscaled$, vs $\lDinf / a$ from FEM simulations for surface potential energies that are $< \kBT$ (filled circles) and $\geq \kBT$ (unfilled triangles) far from the pore mouth and equilibrium bulk concentrations $\cinf$ of $0.3$ (blue), $1$ (red), $3$ (green), $10$ (orange) and $30$ (magenta) mol~m$^3$ and from theory (dashed lines). The solid line in (c) is a power-law fit to the simulation data for $\lDinf/a \leq 0.1$ in the Debye--H\"uckel regime. The scaling is shown for various powers of $\lDinf / a$ in the $\lDinf \ll a$ and $\lDinf \gg a$ regimes (dotted lines).}
\end{figure}

We have assumed that the fluxes inside and outside the Debye--H\"uckel regime can be non-dimensionalized to only depend on $\lDinf/a$, and thus subjected the surface contributions to the fluxes to the non-dimensionalization
\begin{align}
    \label{Qtilde2}
    Q &= - \frac{16}{\pi} \left( \frac{\kBT}{Ze} \right) ^2 \frac{\eps a^5 \Dellnc}{\eta (\lDinf)^4} \ln{\left(1 + \frac{\lDu}{4 \lDinf} \right)} \, \Qscaled  , \\
    \label{Jtilde2}
    \delJ &= - \frac{\kBT}{(Ze)^2} \frac{a^3  \kappa_\mathrm{s}^{\infty} \Dellnc }{(\lDinf)^3} \, \Jscaled , \\ 
    \label{Itilde2}
    \delI &= \frac{a^2 \sigma (\Dp)  \Dellnc}{(\lDinf)^2} \, \Iscaled  .
\end{align}
From the results in Sec.~\ref{simulations} and Sec.~\ref{sec:sigma} of the supplementary material, it can be shown that $\Qscaled$, $\Jscaled$ and $\Iscaled$ are equal to the dimensionless integrals in Eq.~\eqref{flow_rate4}, the first term in Eq.~\eqref{solute_flux4}, and Eq.~\eqref{electric_current4}, respectively, in the theory in the Debye--H\"uckel regime when the difference in ion diffusivities can be ignored, i.e.
\begin{align}
    \label{Qtilde_pre}
    \Qscaled &= \int ^1 _0 \mathrm{d} \zeta \, \zeta ^2 \int ^{\infty} _0  \mathrm{d} \nu \frac{(\epotscaled)^2}{1 + \nu^2}, \\
    \label{Jtilde_pre}
    \Jscaled &= \int ^1 _0 \mathrm{d} \zeta \, (\epotscaled |_{\nu = 0})^2, \\ 
    \label{Itilde_pre}
    \Iscaled &= \int ^1 _0 \mathrm{d} \zeta \, \epotscaled |_{\nu = 0},
\end{align}
and the dimensional prefactors in Eqs.~\eqref{Qtilde2}--\eqref{Itilde2} reduce to the prefactors in Eqs.~\eqref{flow_rate4}--\eqref{electric_current4}, respectively, in this regime. Figure~\ref{kappaa_scaling} depicts the variation of $\Qscaled$, $\Jscaled$ and $\Iscaled$ with $\lDinf / a$ in the theory and the simulations (at fixed $\Dellnc$); we have excluded data in which $\lDinf/l_{\mathrm{GC}}$ is less than $10|\Dm|/(\Dp)$ from Fig.~\ref{kappaa_scaling}(b) such that we can ignore the fluxes induced by the difference in ion diffusivities. Almost all the simulation data of the dimensionless fluxes both inside and outside of the Debye--H\"uckel regime collapse on to a single universal curve as a function of $\lDinf/a$, with some deviation when the electric double layers overlap. This universal scaling is captured reasonably well by the theory given by Eqs.~\eqref{Qtilde_pre}--\eqref{Itilde_pre} with the dimensionless potential $\epotscaled$ for a 2D membrane in the Debye--H\"uckel regime.

As $\epotscaled \approx \lDinf / a$ in the Debye--H\"uckel regime for thick electric double layers, the scaling relationships for the dimensionless fluxes are $\Qscaled, \Jscaled \propto (\lDinf/a)^2$ and $\Iscaled \propto (\lDinf/a)$ in this limit, which agree with the simulations (Fig.~\ref{kappaa_scaling}). This scaling indicates that the flow rate, $Q$, and surface contribution to the solute flux, $\delJ$, do not depend on the equilibrium Debye screening length, $\lDinf$, and that the surface contribution to the electric current, $\delI$, is inversely proportional to $\lDinf$ (i.e. $\propto (\cinf)^{\frac{1}{2}}$) in the Debye--H\"uckel regime for thick electric double layers, which agrees with the scaling depicted in Table~\ref{scaling-flux}. Figure~\ref{kappaa_scaling} shows that the power-law scaling of the fluxes with $\lDinf/a$ outside the Debye--H\"uckel regime is weaker than the predicted scaling relationships, where the theory is expected to break down. On the other hand, the theory gives the scaling $\Tilde{Q} \propto \left( \lDinf /a \right) ^4$, $\Jscaled \propto \left( \lDinf /a\right)^{\frac{5}{2}}$ and $\Iscaled \propto \left( \lDinf /a\right)^{\frac{3}{2}}$ in the Debye--H\"uckel regime for thin electric double layers. Figure~\ref{kappaa_scaling} shows that the scaling of $\Qscaled$ and $\Jscaled$ with $\lDinf/a$ in the theory holds for all surface potentials in the simulations, whereas there is a discrepancy between the scaling of $\Iscaled$ in the theory and the simulations. This predicted scaling indicates that $Q$ is proportional to $(\lDinf)^2$  (i.e. $\propto 1/\cinf$) and $\delJ$ is proportional to $(\lDinf)^{\frac{1}{2}}$ (i.e. $\propto (\cinf)^{-\frac{1}{4}}$) in the thin electric-double-layer limit, which agrees with the scaling depicted in Table~\ref{scaling-flux} and holds for arbitrary strengths of the electric potential. Figure~\ref{kappaa_scaling}(c) shows that the scaling relationship $\Iscaled \propto (\lDinf/a)^{\frac{7}{4}}$, which predicts that $\delI$ is proportional to $(a/\lDinf)^{\frac{1}{4}}$ (i.e. $\propto (\cinf)^{\frac{1}{8}}$), instead describes the scaling in the simulations for thin electric double layers.

Table~\ref{scaling-flux} indicates that the scaling of $Q$ with $\lDinf$ (or $\lambda$ for a neutral solute) in the limiting regimes of $\lDinf$ is analogous to that for concentration-gradient-driven flow of a neutral solute in a 2D membrane and concentration-gradient-driven flow of an electrolyte solution in thick membranes. By contrast, the electroosmotic flow rate in the same geometry scales inversely with $\lD$ for thick electric double layers \cite{maoElectroosmoticFlowNanopore2014} and varies linearly with $\lD$ for thin electric double layers (Table~\ref{scaling-flux}). The theory also predicts that, when the contribution due to the difference in ion diffusivities can be ignored, $\delJ$ does not depend on $\lDinf$ for thick electric double layers, which is the same for all cases in Table~\ref{scaling-flux}. Similarly, the proportionality of $\delI$ with $(\lDinf)^{-1}$ is seen in both thick and 2D membranes. Table~\ref{scaling-flux} indicates that the proportionality of $\delJ$ with $(\lDinf)^{\frac{1}{2}}$ for thin electric double layers is analogous to that for a neutral solute in the same geometry but not for an electrolyte with a thick membrane, in which $\delJ$ is proportional to $\lDinf$. Moreover, $\delI$ is independent of $\lDinf$ for concentration-gradient-driven electrolyte transport for thick membranes but appears to exhibit fractional power-law scaling with $\lDinf$ for 2D membranes. The theory for the electric-field-driven electric current in the same geometry in Ref.~\citenum{leeLargeElectricSizeSurfaceCondudction2012} does not predict such fractional power-law scaling with $\lDinf/a$ despite the similarities between these cases. However, a dependence on fractional powers of $\lDinf/a$ due to the pore geometry could be contained within the numerical constants in this theory.

\subsection{Experimental implications}

Substituting the results in Sec.~\ref{sec:universal_scaling} for a thin electric double layer into the heuristic theory in Sec.~\ref{scaling-heuristic}, the scaling relationships of the surface contributions to the fluxes for thin electric double layers and arbitrary strengths of the electric potential with all parameters are

\begin{align}
    \label{Q-propto4}
    Q &\propto \left( \frac{\kBT}{Ze} \right)^2  \frac{\epsilon \epsilon _0 a}{\eta} \ln{\left( 1 + \frac{\lDu}{4\lDinf} \right)} \Dellnc, \\
    \label{J-propto4}
    \delJ &\propto \frac{\kBT}{(Ze)^2} \left( \frac{a}{\lDinf} \right)^{\frac{1}{2}} \kappa ^{\infty} _\mathrm{s} \Dellnc, \\
    \label{I-propto4}
    \delI &\propto \left( \frac{a}{\lDinf} \right)^{\frac{1}{4}} \sigma (D_+ + D_-) \Dellnc ,
\end{align}
where Eq.~\eqref{J-propto4} assumes that the contribution due to the difference in ion diffusivites is negligibly small. We fit the simulation data in Fig.~\ref{kappaa_scaling} to obtain approximate equations that predict the fluxes for arbitrary surface potentials and non-overlapping electric double layers when $D_+ \approx D_-$ in Sec.~\ref{sec:kappa*a2} of the supplementary material. When the width of the electric double layer is much smaller than the pore radius, the theory predicts fractional power-law scaling of the surface contributions to the solute flux and electric current with the equilibrium Debye screening length and pore radius that differs markedly from that seen in thick membranes. For example, the electric-field-driven electric current in a long channel does not depend of electrolyte concentration when the surface contribution dominates. \cite{bocquetNanofluidicsBulkInterfaces2010} By contrast, the scaling relationships in Eq.~\eqref{I-propto4} indicate a weak but non-zero dependence of the concentration-gradient-driven electric current on $(\cinf)^{\frac{1}{8}}$ when the bulk contribution can be ignored.

Scaling of the (electric-field-driven) ionic conductance with fractional powers of the electrolyte concentration have been observed in the low-salt regime in carbon nanotubes\cite{secchiScalingTransportCarbonNanotubes2016} ($\propto (\cinf)^{\frac{1}{3}}$) and boron nitride nanotubes\cite{zhongwuIonTransportBNNT2024} ($\propto (\cinf)^{\frac{1}{4}}$) in experiments (with KCl) and rationalized in terms of a salinity-dependent surface charge.\cite{secchiScalingTransportCarbonNanotubes2016, zhongwuIonTransportBNNT2024} These scaling relationships versus salt concentration in turn imply fractional scaling of the ionic conductance with the Debye length, specifically with $(\lambda_{\mathrm{D}}^{\infty})^{-\frac{2}{3}}$ and $(\lambda_{\mathrm{D}}^{\infty})^{-\frac{1}{2}}$, respectively. As the 2D membrane geometry results in power-law scaling of the surface contribution to the solute flux and electric current with the equilibrium Debye screening length and pore radius for concentration-gradient-driven transport, entrance effects could explain, or contribute to, unusual fractional power-law scaling of the ionic conductance with salt concentration observed experimentally. In particular, we have shown that the concentration-gradient-driven solute flux exhibits analogous scaling with the pore radius, equilibrium bulk electrolyte concentration and surface charge density to the electric-field-driven electric current, where the theory predicts that the concentration-gradient-driven solute flux is proportional to $(\lDinf)^{-\frac{1}{2}}$ when the bulk contribution can be neglected (i.e. the Gouy--Chapman length is much smaller than the Debye length) and the electric double layer is thin relative to pore size (see Sec.~\ref{sec:kappa*a2} of the supplementary material for details of the derivation of this scaling relationship). 

\section{Conclusions}

We have, for the first time, derived general equations and scaling laws for the concentration-gradient-driven fluid fluxes, and scaling laws of the electric-field-driven flow rate in the thin electric-double-layer limit, of an electrolyte solution through a circular aperture in an infinitesimally thin planar membrane, which we have verified by comparison with finite-element simulations. Although these equations are not fully quantitative, they uncover the fundamental scaling relationships for these fluxes as functions of the pore radius, surface charge density and the Debye screening length, which constitute the main results of this work. We have shown, by comparing with simulations both inside and outside the Debye--H\"uckel regime, that these scaling relationships accurately quantify the variation of all the fluid fluxes in the Debye--H\"uckel regime for thick electric double layers relative to the pore radius, and that of the fluid flow rate and solute flux irrespective of the magnitude of the electric potential when the electric double layer is thin relative to the pore radius. The scaling laws determined in this work indicate unusual fractional power-law scaling with the pore radius and the equilibrium Debye screening length for the surface contributions to the total solute flux and electric current, which is not seen in other geometries. These results are important for understanding concentration-gradient-driven transport of an electroyte solution in membranes made from two-dimensional nanomaterials, such as graphene and molybdenum disulfide, as well as entrance effects in thicker membranes, particularly those with low-friction internal surfaces such as carbon nanotubes. Thus, these results have significant implications for applications involving concentration-gradient-driven transport of an electrolyte solution in a porous membrane, including desalination, salinity-gradient power conversion for energy generation and sensing.

\section*{Supplementary material}
The supplementary material contains further details on the derivation of scaling relationships for concentration-gradient-driven flow through a circular aperture in a infinitesimally thin membrane; general expressions for the concentration-gradient-driven fluxes for thin electric double layers; derivations of scaling relationships for concentration-gradient-driven flow of an electrolyte solution in a cylindrical pore and electroosmosis through a circular aperture in a infinitesimally thin membrane in the limit that the pore radius is much larger than the Debye screening length; and further details and supplementary results of finite-element numerical simulations.

\section*{Acknowledgments}
This work was supported by the Australian Research Council under the Discovery Projects funding scheme (Grant No. DP210102155). H.C.M.B. acknowledges the support of an Australian Government Research Training Program Scholarship.

\section*{Author declarations}
\subsection*{Conflict of interest}
The authors declare no conflict of interest.

\subsection*{Author Contributions}

\textbf{Holly C. M. Baldock}: Methodology (equal); Investigation (lead); Data curation (supporting); Formal Analysis (lead); Writing -- original draft (lead); Writing -- review \& editing (supporting) \textbf{David M. Huang}: Conceptualization (lead); Methodology (equal); Investigation (supporting); Data curation (lead); Formal Analysis (supporting); Funding acquisition (lead); Project administration (lead); Supervision (lead); Writing -- original draft (supporting); Writing -- review \& editing (lead).

\section*{References}
\bibliography{2D_membrane_CDF-references}
\makeatletter\@input{2D_membrane_CDF-sm-aux.tex}\makeatother

\end{document}



\title{Supplementary Material: Scaling laws for concentration-gradient-driven electrolyte transport through a 2D membrane} 



\author{Holly C. M. Baldock}
\author{David M. Huang}
\email[Electronic mail: ]{david.huang@adelaide.edu.au}
\affiliation{Department of Chemistry, School of Physics, Chemistry and Earth Sciences, The University of Adelaide, Adelaide, SA 5005, Australia}



\pacs{}

\maketitle 

\section{Derivation of scaling relationships for concentration-gradient-driven flow through a circular aperture in a 2D membrane} \label{sec:CDF}

\subsection{Electric potential for a thin electric double layer (relative to the pore radius)} \label{sec:epot}

Assuming that the equilibrium electric potential can be approximated as decaying exponentially with the distance $d$ from the surface over the equilibrium Debye screening length, we can assume the potential near a planar surface (Eq.~\eqref{epot_in} in the main paper with $r \rightarrow \infty$) to write the dimensionless equilibrium potential $\tilde{\psi} _0$ for thin electric double layers and $r>a$ (in the Debye--H\"uckel regime) as
\begin{align}
\label{epot3}
    \tilde{\psi} _0(d)|_{r>a} & \approx \frac{\lDinf}{a} \exp{\left (-\frac{d}{\lDinf} \right )}.
\end{align}

Using Eq.~\eqref{epot_in} in the main paper, the full analytical expression for the potential at the pore edge in a 2D membrane (in the Debye--H\"uckel regime) is
\begin{align}
    \label{epot_in_edge}
    \epotscaled|_{\hat{r} = 1, \hat{z} = 0} = & - \int ^{\mathrm{\infty}} _0 \mathrm{d} s \, \frac{J_1(s) J_0(s)}{\sqrt{(a/\lDinf)^2 + s^2}} + \frac{\lDinf}{a} .
\end{align}
As the derivative of $J_1(s)$ with respect to $s$ is $-J_0(s)$, where $\sqrt{(a/\lDinf)^2 + s^2} \rightarrow a/\lDinf$ in the limit $a/
\lDinf \rightarrow \infty$, Eq.~\eqref{epot_in_edge} reduces to
\begin{align}
    \label{epot_in_edge2}
    \epotscaled|_{\hat{r} = 1, \hat{z} = 0} & \approx \frac{\lDinf}{a} \left[ \int ^{\mathrm{\infty}} _0 \mathrm{d} s \, J_0(s) \frac{\mathrm{d}}{\mathrm{d}s}(J_0(s)) + 1 \right] = \frac{\lDinf}{2a} ,
\end{align}
which is the approximate potential at the pore edge of a 2D membrane in the Debye--H\"uckel regime and thin electric-double-layer limit. Assuming that the potential inside of the pore mouth ($\nu=0$, i.e. $z = 0$ for $r \leq a$) decays exponentially with the distance $d$ from the pore edge over $\lDinf$, we can write the potential inside of the pore mouth as
\begin{align}
\label{epot3_edge}
    \tilde{\psi} _0(d)|_{\nu=0} \approx \frac{\lDinf}{2a} \exp{\left (-\frac{d}{\lDinf} \right )} .
\end{align}

In cylindrical coordinates $(r,z)$, the distance from the membrane surface is
\begin{align}
    d(r,z)= 
    \begin{cases} 
        \sqrt{(a-r)^2 + z^2}, & r\leq a \\
        z, & r>a
    \end{cases} \mathrm{.}
\end{align}
Using the relationship between cylindrical and oblate-spheroidal $(\zeta, \nu)$ coordinates, where $r = a\sqrt{(1+v^2)(1-\zeta^2)}$ and $z = a\nu\zeta$, the distance $d$ from the membrane surface in oblate-spheroidal coordinates is
\begin{align}
\label{dist}
    d(\zeta, \nu)= 
    \begin{cases} 
        a\sqrt{2 + \nu^2 - \zeta^2 - 2\sqrt{(1 + \nu^2)(1 - \zeta^2)}}, & \sqrt{(1 + \nu^2)(1 - \zeta^2)} \leq 1 \\
        a \nu \zeta, & \sqrt{(1 + \nu^2)(1 - \zeta^2)} > 1
    \end{cases} \mathrm{.}
\end{align}

The integral in Eq.~\eqref{flow_rate4} of the main paper for the flow rate with $\epotscaled$ given by Eq.~\eqref{epot3} cannot be solved in closed form. Thus, we assume for simplicity that the exponential decay of the dimensionless equilibrium electric potential with $d$ over the length scale $\lDinf$ for a thin electric double layer can be represented by the step function
\begin{equation}
    \label{epot_step}
    \exp{\left (-\frac{d}{\lDinf} \right )} \approx 
    \begin{cases} 
        1, & d < \lDinf \\
        0, & d \geq \lDinf 
    \end{cases}
\end{equation}
when using Eqs.~\eqref{epot3} to derive the scaling relationships for the flow rate in the thin electric-double-layer limit in the main paper. For simplicity and consistency, we also use Eq.~\eqref{epot_step} to approximate Eq.~\eqref{epot3_edge} and thus derive the scaling relationships for the solute flux and electric current in the thin electric-double-layer limit in the main paper.

\begin{figure}[!h]
    \centering
    \includegraphics[scale = 1, trim={0.175cm 0.0875cm 0.175cm 0.0875cm},clip]{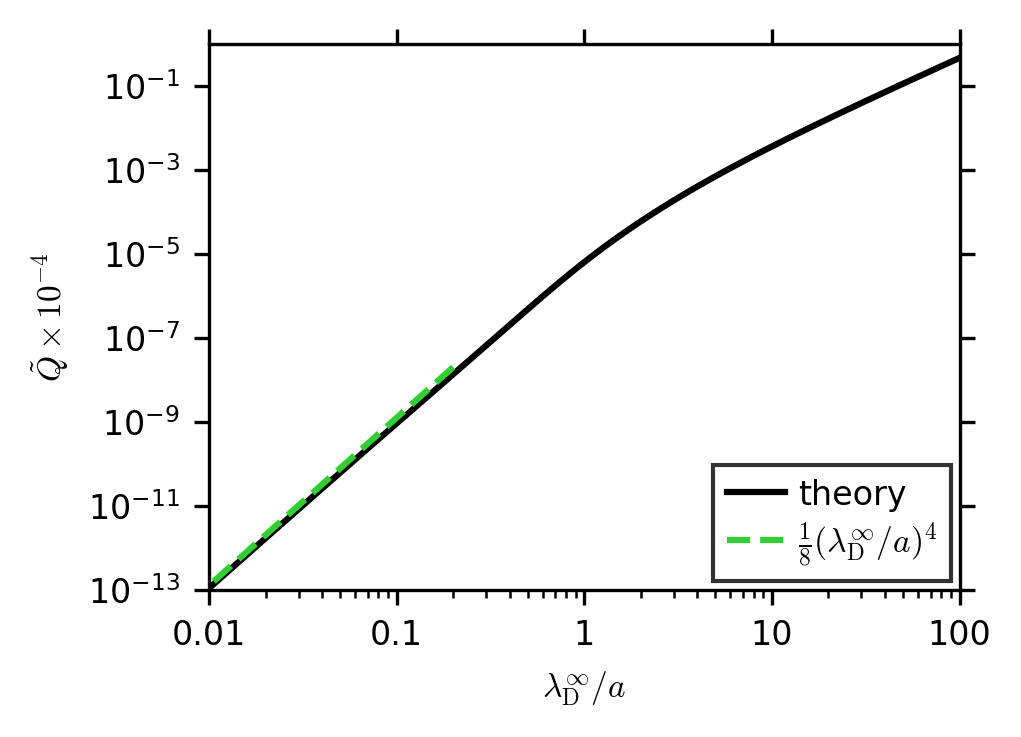}
    \caption{Dimensionless flow rate $\Qscaled$ defined by Eq.~\eqref{Qtilde2} of the main paper vs the equilibrium Debye screening length $\lDinf$ over the pore radius $a$ using the exact expression for the dimensionless equilibrium electric potential $\tilde{\psi}_0$ in Eq.~\eqref{epot_in} of the main paper (solid line) and the approximation to $\tilde{\psi}_0$ for the thin electric-double-layer limit in Eq.~\eqref{epot3} with Eq.~\eqref{epot_step} (dashed line).}
    \label{Q-kappaa-large}
\end{figure}

\begin{figure}[!h]
 \centering
 \begin{minipage}{.5\textwidth}
\begin{flushleft}
    \includegraphics[scale = 1, trim={0.175cm 0.0875cm 0.175cm 0.0875cm},clip]{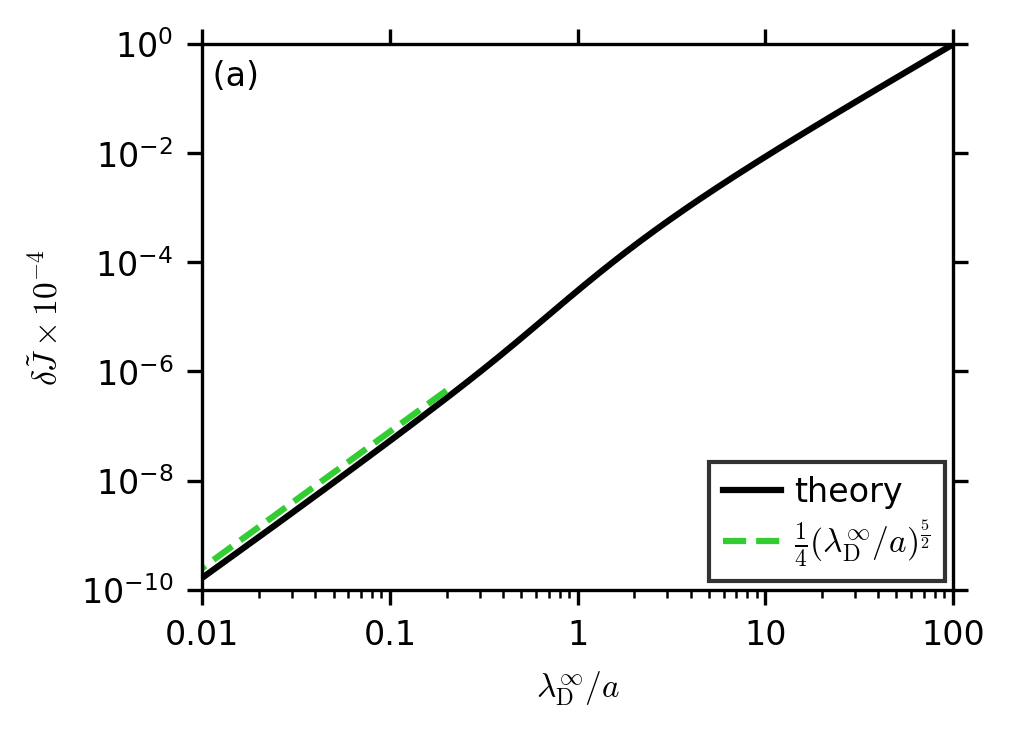}
\end{flushleft}    
  \end{minipage}%
 \begin{minipage}{.5\textwidth}
\begin{flushright}
\includegraphics[scale = 1, trim={0.125cm 0.0875cm 0.125cm 0.0875cm},clip]{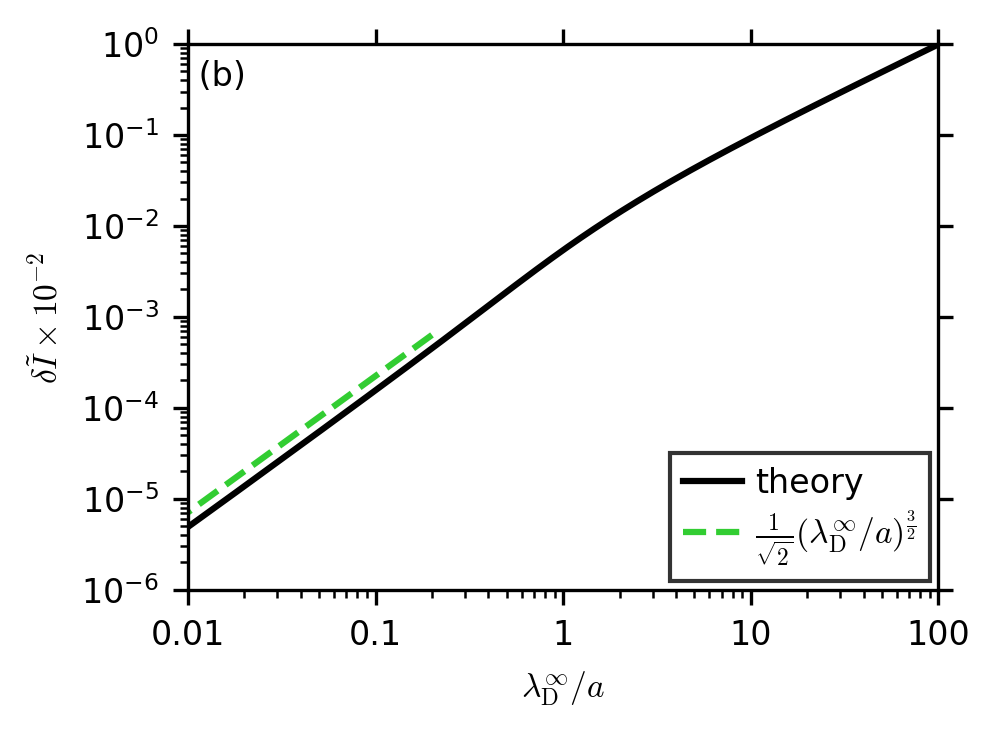}
\end{flushright}
  \end{minipage}
    \caption{\label{J-I-kappaa-large} Dimensionless surface contribution to the (a) solute flux, $\Jscaled$, defined by Eq.~\eqref{Jtilde2} of the main paper and (b) electric current, $\Iscaled$, in Eq.~\eqref{Itilde2} of the main paper vs the equilibrium Debye screening length $\lDinf$ over the pore radius $a$ using the exact expression for the dimensionless equilibrium electric potential $\tilde{\psi}_0$ in Eq.~\eqref{epot_in} of the main paper (solid line) and the approximation to $\tilde{\psi_0}$ for the thin electric-double-layer limit in Eq.~\eqref{epot3_edge} with Eq.~\eqref{epot_step} (dashed line).}
\end{figure}

We compare the quantitative agreement between the integrals given in Eqs.~\eqref{flow_rate4}--\eqref{electric_current4} in the main paper in the thin electric-double-layer limit using the approximate equilibrium potential (Eq.~\eqref{epot3} or \eqref{epot3_edge} with Eq.~\eqref{epot_step}) and the exact equation in the Debye--H\"uckel regime (Eq.~\eqref{epot_in} of the main paper). As shown in Fig.~\ref{Q-kappaa-large}, the approximation (Eq.~\eqref{epot5}) shows favorable quantitative agreement with the full theory for the dimensionless flow rate in Eq.~\eqref{flow_rate4}. Figure~\ref{J-I-kappaa-large} shows good quantitative agreement when approximating the surface integral of the equilibrium potential inside the pore mouth as given in Eqs.~\eqref{epot_integral_shortDebye2} and \eqref{epot_integral_shortDebye3}, respectively. As the approximate surface integrals of the electric potential and the squared electric potential over the pore aperture show similar discrepancies compared with the full analytical theory in the thin electric-double-layer limit ($\approx 1.5$), we assume that the condition to ignore the contribution to $\delJ$ from the difference in ion diffusivities, $\lDinf/l_{\mathrm{GC}} \gg 2\sqrt{2} \frac{|D_+ - D_-|}{\Dp}$, also holds for the exact potential.

\clearpage

\subsection{Conditions for neglecting effects of the difference in ion diffusivities for the concentration-gradient-driven solute flux} \label{sec:delJ}

Substituting $\Tilde{\psi_0} = \lDinf/a$ into Eq.~\eqref{solute_flux4} of the main paper, which holds for $\lDinf \gg a$, gives
\begin{align}
\notag
    \delJ & = - \frac{(\Dp) a^2 |\sigma| \Dellnc}{Ze(\lDinf)^2} \left[\frac{a}{l _{\mathrm{GC}}}  \int ^1 _0 \mathrm{d} \zeta \, \left(\frac{\lDinf}{a} \right)^ 2 - \text{sgn} (\sigma) \left( \frac{\Dm}{\Dp} \right) \int ^1 _0 \mathrm{d} \zeta \, \left(\frac{\lDinf}{a} \right) \right ]     \\ 
\label{delJ_conc}
          & = - \frac{(\Dp) a |\sigma| \Dellnc}{Ze\lDinf} \left[\frac{\lDinf}{l _{\mathrm{GC}}} - \text{sgn} (\sigma) \left( \frac{\Dm}{\Dp} \right) \right ] ,
\end{align}
where the second term in Eq.~\eqref{delJ_conc} is negligible when $\lDinf/l_{\mathrm{GC}} \gg \frac{|D_+ - D_-|}{\Dp}$. 

Substituting Eqs.~\eqref{epot_integral_shortDebye2} and \eqref{epot_integral_shortDebye3} from the main paper, which is approximately true for $\lDinf \ll a$, into Eq.~\eqref{solute_flux4} of the main paper gives
\begin{align}
\notag
    \delJ & = - \frac{(\Dp) a^2 |\sigma| \Dellnc}{Ze(\lDinf)^2} \left[\frac{a}{l _{\mathrm{GC}}} \left(\frac{\lDinf}{2a} \right)^ 2 \int ^1 _0 \mathrm{d} \zeta \, H \left(\sqrt{\lDa} - \zeta \right) \right. \\
    \notag
     & \ \ \ \ - \left. \text{sgn} (\sigma) \left( \frac{\Dm}{\Dp} \right) \left(\frac{\lDinf}{2a} \right) \int ^1 _0 \mathrm{d} \zeta \, H \left( \sqrt{\frac{2 \lDinf}{a}} - \zeta \right) \right] \\
\label{delJ_conc2}
& \approx - \frac{(\Dp) |\sigma| \Dellnc}{4 Ze} \left [ \frac{\lDinf}{l _{\mathrm{GC}}}  - 2\sqrt{2} \text{sgn} (\sigma) \left( \frac{\Dm}{\Dp} \right) \right ]  \sqrt{\frac{a}{\lDinf}} ,
\end{align}
where the second term in Eq.~\eqref{delJ_conc2} is negligible when $\lDinf/l_{\mathrm{GC}} \gg 2\sqrt{2} \frac{|D_+ - D_-|}{\Dp}$. 

\clearpage

\subsection{Heuristic derivation outside of the Debye--H\"uckel regime for non-overlapping electric double layers} \label{sec:sigma}

We can rearrange the full analytical theory for the concentration-gradient-driven flow rate in the Debye--H\"uckel regime (Eq.~\eqref{flow_rate4} in the main paper) to give
\begin{equation}
    \label{Q_scaling_derive1}
    Q = -\frac{\eps a}{\pi \eta} \left( \frac{\kBT}{Ze} \right) ^2 \left( \frac{Ze \sigma \lDinf}{\kBT \eps}\right) ^2 \left[ \left( \frac{a}{\lDinf} \right )^4 \int ^1 _0 \mathrm{d} \zeta \, \zeta ^2 \int ^\mathrm{\infty} _0 \mathrm{d} \nu \, \frac{ (\epotscaled ) ^2}{1 + \nu^2} \right]  \Dellnc.
\end{equation}
At fixed $\lDinf/a$, Eq.~\eqref{Q_scaling_derive1} gives the scaling
\begin{equation}
    \label{Q_scaling_derive2}
    Q \propto \frac{\eps a}{\pi \eta} \left( \frac{4 \kBT}{Ze} \right) ^2 \left( \frac{Ze \sigma \lDinf}{4 \kBT \eps}\right) ^2  \Dellnc.
\end{equation}
For the theory of diffusioosmosis in a planar channel for non-overlapping double layers and arbitrary strengths of the electric potential,\cite{prieveMigrationColloidConcentrationGradient1982} it was shown that the the concentration-gradient-driven fluid velocity outside the electric double layer is proportional to $\ln (1-\gamma^2)$, where
\begin{align}
    \gamma = \tanh{ \left( \frac{Ze \psiinf }{4 \kBT} \right)}
\end{align}
and $|\psiinf| = 2 \kBT \sinh^{-1}{\left(\lDinf / l_{\mathrm{GC}} \right)}/Ze$ is the magnitude at a planar surface for a $Z$:$Z$ electrolyte. As given in Ref.~\citenum{prieveMigrationColloidConcentrationGradient1982} for the limit $Ze|\psiinf| \ll 4\kBT$,
\begin{equation}
    \ln{(1 - \gamma^2)} \approx -\left( \frac{Ze \psiinf}{4 \kBT} \right) ^2.
\end{equation}
Thus, in the Debye--H\"uckel regime, 
\begin{equation}
\label{Q_scale2}
    \ln{(1 - \gamma ^2)} \approx -\left( \frac{Ze \sigma \lDinf}{4 \kBT \epsilon \epsilon_0} \right) ^2.
\end{equation}
We can write $\ln{(1 - \gamma ^2)}$ in terms of the ratio of the Dukhin length near a planar wall, $\lDu$, to the equilibrium Debye screening length, $\lDinf$, for arbitrary surface potentials such that
\begin{align}
    \label{Qproptoref2}
    \ln{(1 - \gamma^2)} = -\ln{\left( 1 + \frac{\lDu}{4 \lDinf} \right)} ,
\end{align}
where $\lDinf$ and $\lDu$ are given by Eqs.~\eqref{Debye_length} and \eqref{dukhin} in the main paper, respectively. We assume that the factor on the right-hand side of Eq.~\eqref{Q_scale2} can be replaced by the right-hand side of Eq.~\eqref{Qproptoref2} in Eq.~\eqref{Q_scaling_derive2} to extend the validity of the flow rate equation outside the Debye--H\"uckel regime, which gives the scaling relationships for the flow rate in a 2D membrane at fixed $\lDinf/a$ and arbitrary electric potentials when the width of the electric double layer is smaller than the pore radius (Eq.~\eqref{other_Scaling_Q} in the main paper). When $l_{\mathrm{GC}} \gg \lDinf$, $|\psiinf| \approx 2 \kBT \lDinf / (Ze l_{\mathrm{GC}}) = |\sigma| \lDinf /(\eps)$, such that Eq.~\eqref{Q_scale2} holds within a regime that is less restrictive on the surface charge density than the Debye--H\"uckel regime for short Debye lengths.  When $l_{\mathrm{GC}} \ll \lDinf$, $\sqrt{(l_{\mathrm{GC}}/\lDinf)^2 +1} \approx 1$ and the Duhkin length near a plane wall can be given as
\begin{equation}
    \label{dukhin_approx}
    \lDu \approx 2\lDinf \left(\frac{\lDinf}{l_{\mathrm{GC}}} -1 \right). 
\end{equation}
Substituting Eq.~\eqref{dukhin_approx} into Eq.~\eqref{Q_scale2} thus gives Eq.~(\ref{other_Scaling_Q2}) in the main paper in this regime. As $\lDinf/l_{\mathrm{GC}} \rightarrow \infty$, Eq.~(\ref{other_Scaling_Q2}) in the main paper reduces to
\begin{align}
    \label{other_Scaling_Q3}
    \ln{ \left(1 + \frac{\lDinf}{l_{\mathrm{GC}}} \right)} + \ln{\left(\frac{1}{2}\right)} \approx \ln{ \left(\frac{\lDinf}{2l_{\mathrm{GC}}} \right)} & = \ln{\left(\frac{Ze|\sigma| \lDinf}{4\kBT\eps} \right)} \nonumber \\
    & = \ln{\left(\frac{Ze \lDinf}{4\kBT \eps} \right)} + \ln{|\sigma|}.
\end{align}
Thus, the heuristic theory predicts that the flow rate varies with the logarithm of $|\sigma|$ at high surface charge densities when the Debye length is smaller than the pore radius. 
 
\begin{figure}[!h]
 \centering
 \begin{minipage}{.5\textwidth}
\begin{flushleft}
    \includegraphics[scale = 1, trim={0.175cm 0.0875cm 0.175cm 0.0875cm},clip]{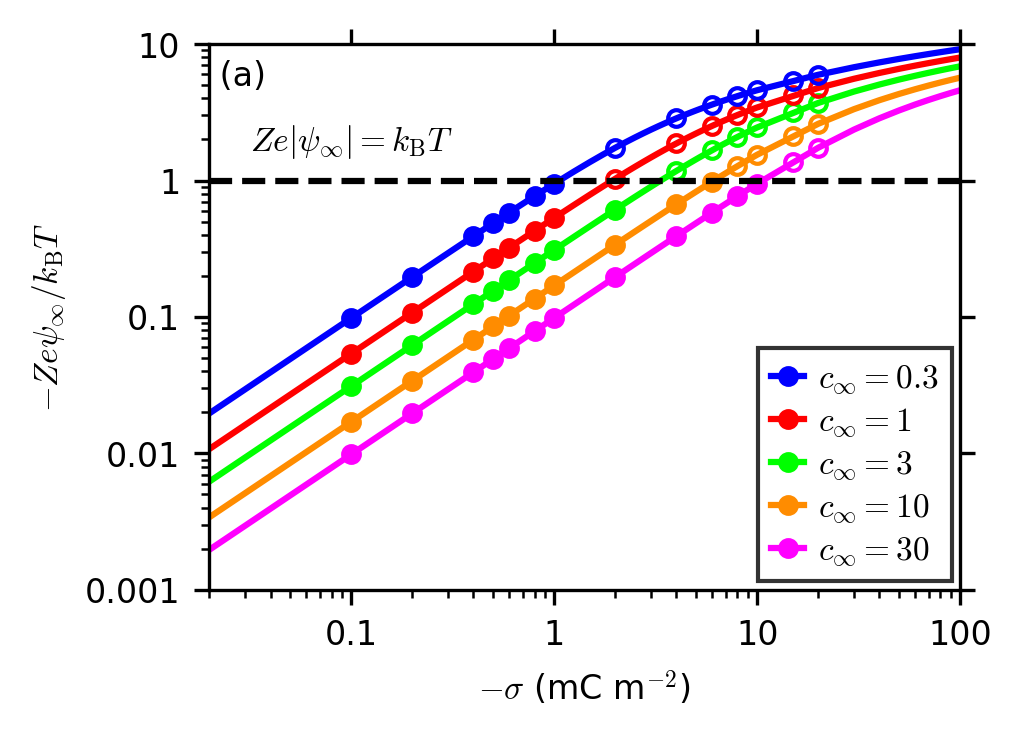}
\end{flushleft}
  \end{minipage}%
 \begin{minipage}{.5\textwidth}
   \begin{flushright}
\includegraphics[scale = 1, trim={0.175cm 0.0875cm 0.175cm 0.0875cm},clip]{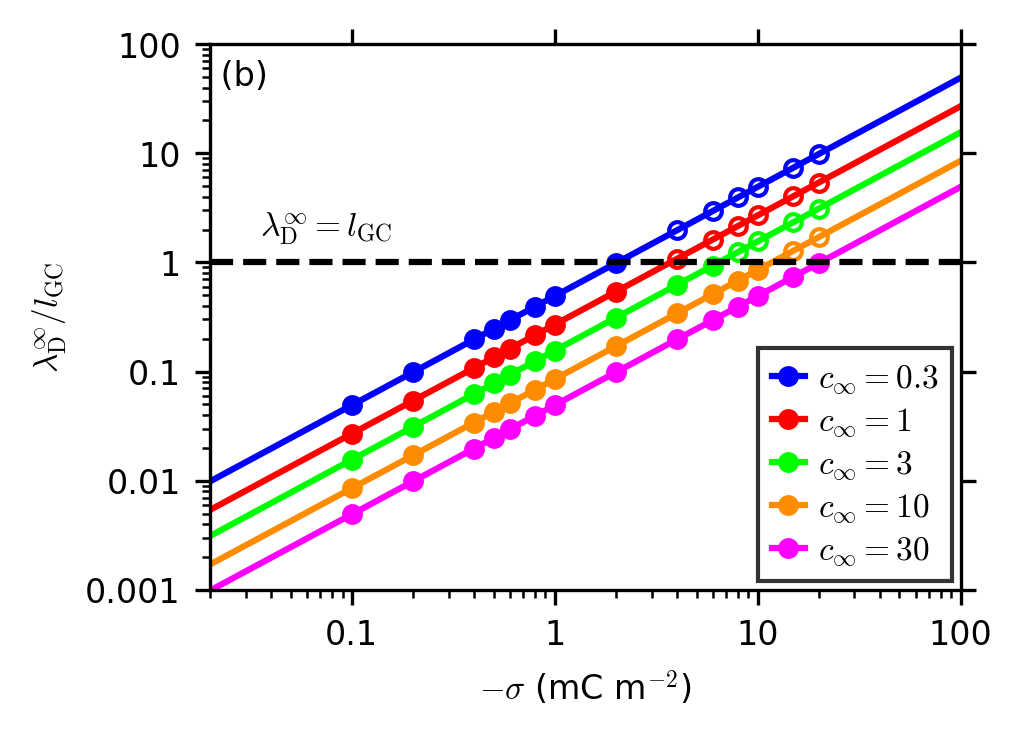}
  \end{flushright}
  \end{minipage}
    \caption{\label{sigma-potentials} (a) Magnitude of the surface potential energy of a plane, $Ze|\psiinf|$, relative to the thermal energy, $\kBT$, and (b) ratio of the Debye length $\lDinf$ to the Gouy--Chapman length $l_{\mathrm{GC}}$, vs surface charge density $\sigma$ for a pore radius of $5$~nm and a range of equilibrium solute concentrations $\cinf$ (mol~m$^{-3}$). Parameters used in FEM simulations that correspond to (a) $Ze|\psiinf|$ that are $< \kBT$ (filled symbols) and $> \kBT$ (empty symbols) far from the pore mouth, respectively, and (b) $\lDinf$ that are $< l_{\mathrm{GC}}$ (filled symbols) and $> l_{\mathrm{GC}}$ (empty symbols) are shown. (a) $Ze|\psiinf| = \kBT$ and (b) $\lDinf = l_{\mathrm{GC}}$ are also shown (dashed line).}
 \end{figure}

Instead of using the boundary conditions on the virtual concentration variable $c_{\mathrm{s}}$ defined in the main paper, let us write the concentration of the higher electrolyte concentration reservoir, $c_{\mathrm{H}}$, with respect to the electrochemical potential difference, $\Delta \psi_{\mathrm{s}}$;~\cite{fairReverseElectrodialysis1971} i.e.
\begin{equation}
     c_{\mathrm{H}} = c_{\mathrm{L}} + \Delc = c_{\mathrm{L}} \exp{\left(-\frac{Z_i e \Delta \psi_{\mathrm{s}} }{\kBT} \right)} ,
\end{equation}
such that
\begin{equation}
    -\Delta \psi_{\mathrm{s}}  = \frac{\kBT}{Z_i e}\ln{ \left(\frac{c_{\mathrm{L}} + \Delc}{c_{\mathrm{L}}} \right)}  =\frac{\kBT}{Z_i e} \Delta \ln{c} . 
\end{equation}
For flux contributions due to an induced electric field, $\kBT\Dellnc/(Z_i e)$ plays an analogous role to an electric field applied on species $i$ with ion valence $Z_i$. Assuming that the non-equilibrium bulk concentration profiles of all species are identical, we use the ion flux density in Eq.~\eqref{diff-advec2} of the main paper to write the total ion flux density for a $Z$:$Z$ electrolyte as
\begin{align}
    \label{total_ion_flux_density}
    \boldsymbol{j} & =  - D_+ \exp{\left(-\frac{Z _+ e \psi}{\kBT} \right)} \nabla c_{\mathrm{s}}^{+} + D_- \exp{\left(-\frac{Z _- e \psi}{\kBT} \right)} \nabla c_{\mathrm{s}}^{-} \nonumber \\
    & =  - \left[ (D_+ + D_-) \cosh{\left(\frac{Z e \psi}{\kBT} \right)} - (D_+ - D_-) \sinh{\left(\frac{Z e \psi}{\kBT} \right)} \right] \nabla c_{\mathrm{s}},
\end{align}
and the total electric current density as
\begin{align}
    \label{total_electric_current_density}
    \boldsymbol{j}_e & =  - e\left[ Z_+ D_+ \exp{\left(-\frac{Z _+ e \psi}{\kBT} \right)} \nabla c_{\mathrm{s}}^{+} + Z_- D_- \exp{\left(-\frac{Z _- e \psi}{\kBT} \right)} \nabla c_{\mathrm{s}}^{-} \right] \nonumber \\
    & =  - Ze \left[ (D_+ - D_-) \cosh{\left(\frac{Z e \psi}{\kBT} \right)} - (D_+ + D_-) \sinh{\left(\frac{Z e \psi}{\kBT} \right)} \right] \nabla c_{\mathrm{s}}.
\end{align}
As we have shown that an applied electric field gradient and applied concentration gradient differ by a factor of $Z_i$, the scaling of the concentration-gradient-driven total solute flux density and electric current density with the electric potential (derived from Eqs.~\eqref{total_ion_flux_density} and \eqref{total_electric_current_density}) in the linear-response regime are the same as the electric-field-driven total solute flux and electric current, respectively, in the same geometry. When the electric double layer is thin relative to the pore radius, we assume for a $Z$:$Z$ electrolyte that the concentration-gradient-driven total solute flux and electric-field-driven electric current exhibit the same scaling with the potential at a planar surface. Thus, we assume that $\delJ$ varies with the surface conductivity near a planar wall, $\kappa _{\mathrm{s}} ^{\infty}$, as with the surface contribution to the electric-field-driven electric current in Ref.~\citenum{leeLargeElectricSizeSurfaceCondudction2012}. Moreover, we can write the surface conductivity near a planar wall for $D_+ \approx D_-$ (Eq.~{\eqref{conductance}} in the main paper) in terms of hyperbolic functions as
\begin{align}
    \kappa_\mathrm{s} ^{\infty} &= \frac{Ze(\Dp) |\sigma|}{2 \kBT} \frac{l_{\mathrm{GC}}}{\lDinf} \left[\sqrt{\left( \frac{\lDinf}{l_{\mathrm{GC}}} \right)^2 + 1} - 1\right] \nonumber\\
    \label{ks_hyperbolic}
    &= \frac{\eps (\Dp)}{\lDinf} \left[\cosh{\left(\sinh^{-1}{\left( \frac{\lDinf}{l_{\mathrm{GC}}} \right)} \right)} - 1\right].
\end{align}
such that
\begin{equation}
    \label{ks_hyperbolic2}
     \kappa_\mathrm{s} ^{\infty}= \frac{\epsilon \epsilon _0 (\Dp)}{\lDinf}
     \left[ \cosh{\left( \frac{Ze |\psiinf|}{2\kBT} \right)}- 1 \right] .
\end{equation}
In the Debye--H\"uckel regime ($Ze |\psiinf|\ll \kBT$), Eq.~\eqref{ks_hyperbolic2} reduces to
\begin{equation}
\label{delJ_scale2}
    \kappa_{\mathrm{s}}^{\infty} \approx \frac{(Ze)^2 \sigma^2 \lDinf (\Dp)}{8 \epsilon \epsilon _0 (\kBT)^2} .
\end{equation}
When the contribution from the difference in ion diffusivities can be ignored, we can rearrange the surface contribution to the total concentration-gradient-driven solute flux in the Debye--H\"uckel regime (Eq.~\eqref{solute_flux4} in the main paper) to write
\begin{align}
    \label{delJ_scaling_derive1}
    \delJ = -\frac{\kBT}{(Ze)^2}\frac{(Ze)^2 \lDinf \sigma^2 (\Dp)}{2\eps (\kBT)^2} \left[ \left(\frac{a}{\lDinf} \right)^3 \int ^1 _0 \mathrm{d} \zeta \ ( \epotscaled |_{\nu=0} )^ 2 \right] \Dellnc .
\end{align}
At fixed $\lDinf/a$, Eq.~\eqref{delJ_scaling_derive1} gives the scaling
\begin{align}
    \label{delJ_scaling_derive2}
    \delJ \propto \frac{\kBT}{(Ze)^2}\frac{(Ze)^2 \lDinf \sigma^2 (\Dp)}{\eps(\kBT)^2 }  \Dellnc.
\end{align}
Substituting Eq.~\eqref{delJ_scale2} into Eq.~\eqref{delJ_scaling_derive2} and assuming that the width of the electric double layer is smaller than the pore radius gives the scaling in Eq.~\eqref{Jpropto3} of the main paper. When $l_{\mathrm{GC}} \ll \lDinf$, $\sqrt{(l_{\mathrm{GC}}/\lDinf)^2 + 1} \approx 1$ and Eq.~{\eqref{conductance}} in the main paper can be approximated as
\begin{equation}
\label{conductance2}
    \kappa_\mathrm{s} ^{\infty} \approx \frac{\eps (\Dp)}{\lDinf} \left( \frac{\lDinf}{l_{\mathrm{GC}}} - 1\right) \approx \frac{ \eps (\Dp)}{l_{\mathrm{GC}}},
\end{equation}
such that $\delJ$ is proportional to $\sigma$ when $l_{\mathrm{GC}} \ll \lDinf$ and the Debye length is smaller than the pore radius. When $l_{\mathrm{GC}} \gg \lDinf$, $\sinh^{-1}\left( \lDinf / l_{\mathrm{GC}} \right) \approx  \lDinf / l_{\mathrm{GC}}$ and Eq.~\eqref{ks_hyperbolic} reduces to
\begin{equation}
\label{conductance4}
    \kappa_\mathrm{s} ^{\infty} \approx \frac{\eps (\Dp)}{\lDinf} \left[ \cosh{\left(\frac{\lDinf}{l_{\mathrm{GC}}} \right)}  - 1 \right] \approx \frac{\eps \lDinf (\Dp) }{2 (l_{\mathrm{GC}})^2},
\end{equation}
which indicates that $\delJ$ is proportional to $\sigma^2$ when $l_{\mathrm{GC}} \gg \lDinf$, rather than strictly within the Debye--H\"uckel regime, when the width of the electric double layer is smaller than the pore radius. With reference to Eq.~\eqref{ks_hyperbolic2}, we can use the scaling relationships for $\delJ$ in Eq.~\eqref{Jpropto3} of the main paper and the relationship between Eqs.~\eqref{total_ion_flux_density} and \eqref{total_electric_current_density} to infer that 
\begin{align}
    \delI & \propto \left( \frac{\kBT}{Ze} \right) \frac{\epsilon \epsilon _0 (D_+ + D_-)}{\lDinf}\left[ \sinh{\left( \frac{Ze \psiinf}{2\kBT} \right)} \right] \Dellnc \nonumber \\
    & = 2 \sigma (\Dp) \Dellnc \nonumber \\
    & \propto \sigma (\Dp) \Dellnc,
\end{align}
which gives the scaling relationships for $\delI$ at fixed $\lDinf/a$ when the width of the electric double layer is smaller than the pore radius (Eq.~\eqref{Ipropto3} of the main paper). Using Eqs.~\eqref{total_ion_flux_density}, we can also consider the contribution from the difference in ion diffusivities to the scaling relationships for $\delJ$ (at fixed $\lDinf/a$ and when the width of the electric double is smaller than the pore radius) to write
\begin{equation}
\label{Jpropto4}
    \delJ \propto \frac{1}{Ze}\left[ \frac{\kBT}{Ze} \kappa_{\mathrm{s}} ^{\infty} - \alpha_1 (D_+ - D_-) \sigma \right] \Dellnc,
\end{equation}
where $\alpha_1$ is a fitting parameter that considers different contributions to $\delJ$ (from functions of $\lDinf/a$) due to the sum of and difference in ion diffusivities (as seen in Eqs.~\eqref{epot_integral_shortDebye2} and \eqref{epot_integral_shortDebye3} of the main paper). Similarly, we can use Eq.~\eqref{total_ion_flux_density} to consider the contribution from the difference in ion diffusivities to the scaling relationships of $\delI$ at fixed $\lDinf/a$ and when the width of the electric double is smaller than the pore radius to write
\begin{equation}
    \label{Ipropto4}
    \delI \propto \left[(D_+ + D_-) \sigma - \alpha_2 \left(\frac{\kBT}{Ze} \right) \left( \frac{D_+ - D_-}{D_+ + D_-} \right) \kappa_{\mathrm{s}}^{\infty} \right] \Dellnc,
\end{equation}
where $\alpha_2$ is a fitting parameter. Figure~\ref{sigma-potentials} shows the values of $Ze|\psiinf|/\kBT$, where $\psiinf$ is the potential at a planar surface, and $\lDinf/l_{\mathrm{GC}}$ as functions of the surface charge density for the parameters used in the simulations. We show Fig.~\ref{sigma-potentials} as a reference for the scaling relationships outside the Debye--H\"uckel regime.

\begin{figure}[!h]
    \includegraphics[scale = 1, trim={0.175cm 0.0875cm 0.175cm 0.0875cm},clip]{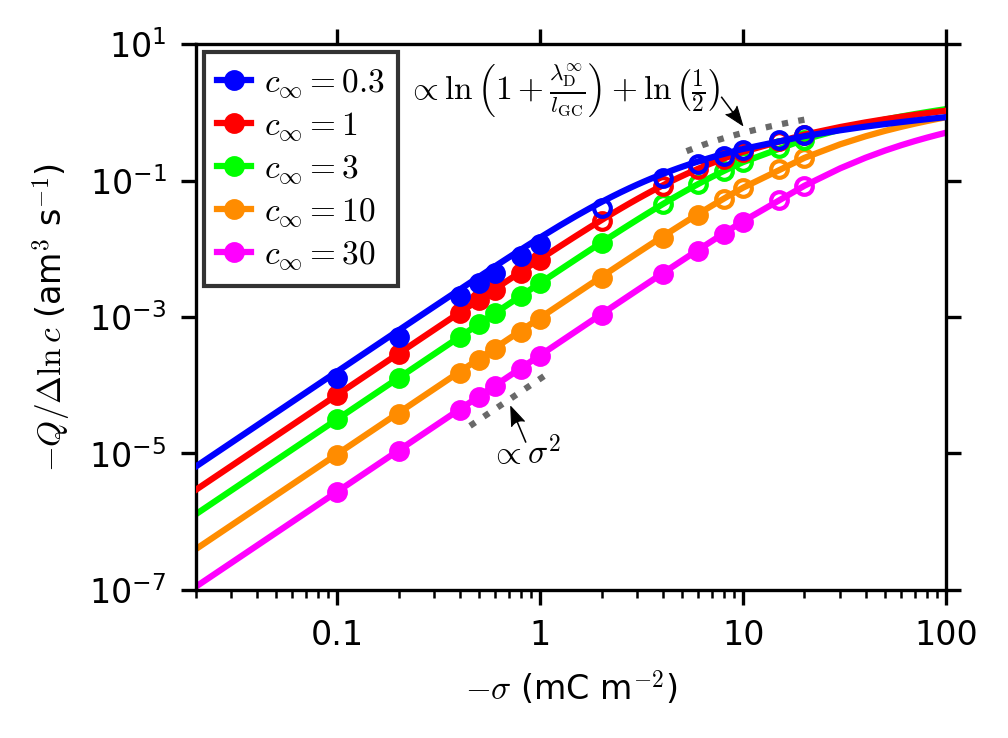}
    \caption{\label{Q_largepsi1} Flow rate $Q$ over $\Dellnc$ vs surface charge density $\sigma$ for a pore radius of $5$ nm and a range of equilibrium solute concentrations $\cinf$ (mol~m$^{-3}$) from FEM simulations (symbols). Simulations are shown for surface potential energies that are $< \kBT$ (filled symbols) and $> \kBT$ (empty symbols) far from the pore mouth, respectively. Fits of the scaling in Eq.~\eqref{other_Scaling_Q} in the main paper to the simulations (solid lines) are shown, along with scaling in limiting cases of $l_{\mathrm{GC}}/\lDinf$ (dotted lines). The scaling at high $|\sigma|$ was calculated for $\cinf = 0.3$~mol~m$^{-3}$.}
 \end{figure}

\begin{figure}[!h]
 \centering
 \begin{minipage}{.5\textwidth}
\begin{flushleft}
    \includegraphics[scale = 1, trim={0.175cm 0.0875cm 0.175cm 0.0875cm},clip]{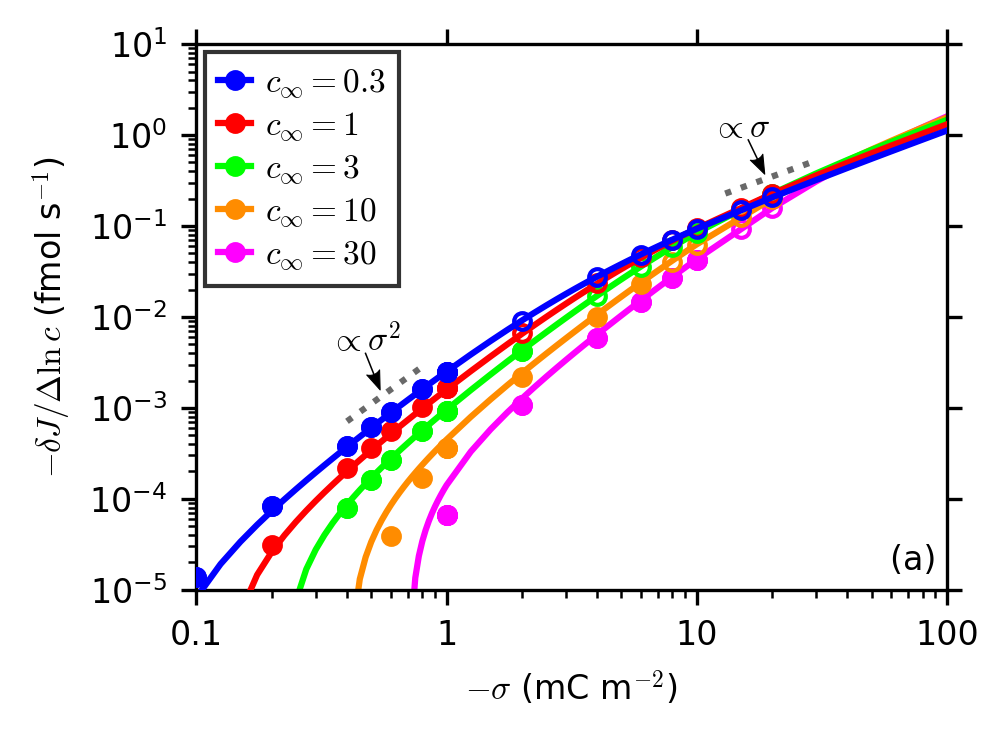}
\end{flushleft}
  \end{minipage}%
 \begin{minipage}{.5\textwidth}
   \begin{flushright}
\includegraphics[scale = 1, trim={0.175cm 0.0875cm 0.175cm 0.0875cm},clip]{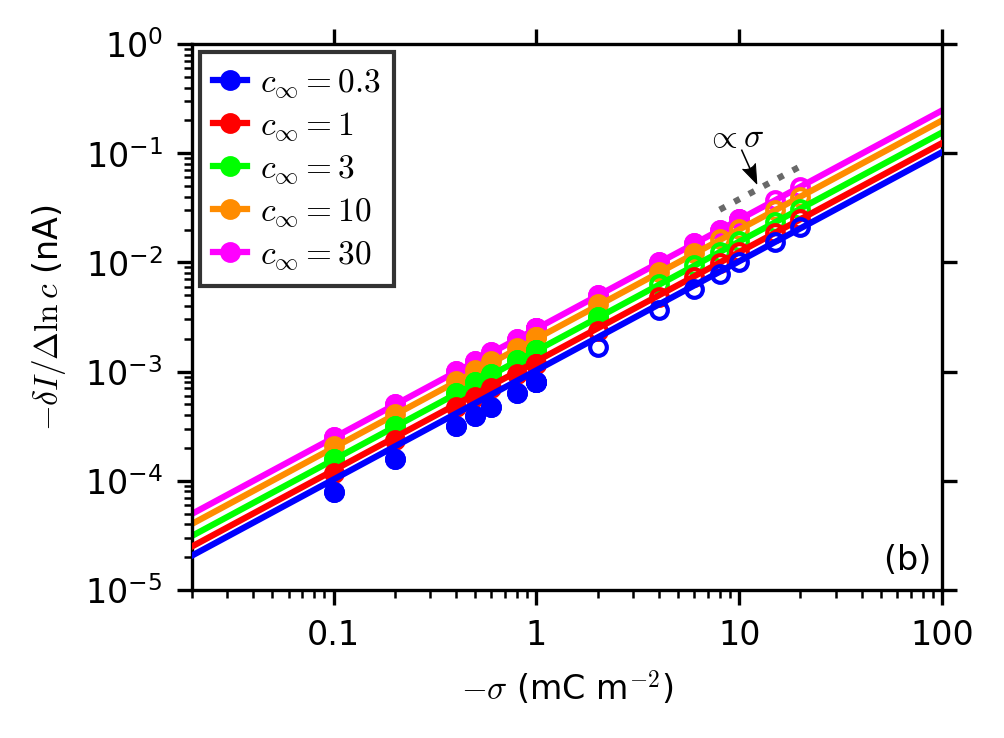}
  \end{flushright}
  \end{minipage}
    \caption{\label{fig:J-large_sigma} Surface contribution to the (a) total solute flux, $\delta J$, and (b) electric current $\delta I$, over $\Dellnc$ vs surface charge density $\sigma$ for a pore radius of $5$~nm and a range of equilibrium solute concentrations $\cinf$ (mol~m$^{-3}$) from FEM simulations (symbols). Simulations are shown for surface potential energies that are $< \kBT$ (filled symbols) and $> \kBT$ (empty symbols) far from the pore mouth, respectively. Fits of the scaling in (a) Eq.~\eqref{Jpropto4} for $\alpha_1 = 1/2$ and (b) Eq.~\eqref{Ipropto3} in the main paper to the simulations are shown (solid lines), along with scaling for various powers of $\sigma$ (dotted lines).}
 \end{figure}

In Figs.~\ref{Q_largepsi1} and \ref{fig:J-large_sigma}, we verify the scaling relationships for $Q$, $\delJ$ and $\delI$ at fixed $\lDinf/a$ for non-overlapping electric double layers via fits of the scaling relationships in Eq.~\eqref{other_Scaling_Q} in the main paper, Eq.~\eqref{Jpropto4} for $\alpha_1 = 1/2$, and Eq.~\eqref{Ipropto3} in the main paper to the simulation data when $\sigma$ is varied. For $\delI$, we neglected the contribution from the difference in ion diffusivities to the scaling in Eq.~\eqref{Ipropto4} as it is negligibly small for KCl and the parameters used in the simulations. For simulations where $a=5$ nm, $\cinf \geq 3$ mol m$^{-3}$ correspond to non-overlapping electric double layers, while $\lDinf/a = 1.64$ and $\lDinf/a = 2.85$ for $\cinf = 1$ and $\cinf = 0.3$ mol m$^{-3}$, respectively. The heuristic scaling relationships for $Q$ at arbitrary strengths of the electric potential and fixed $\lDinf/a$ seem to hold for $\cinf \geq 1$ mol m$^{-3}$ for the $\sigma$ used in the simulations, while those for $\delJ$ and $\delI$ seem to hold for all $\cinf$ in the simulations and $\cinf \geq 3$ mol$^{-3}$, respectively.

\clearpage

\section{General expressions for the concentration-gradient-driven fluxes for thin electric double layers (relative to the pore radius)} 
\label{sec:kappa*a2}

\begin{figure}[!h]
    \centering
    \includegraphics[scale = 1, trim={0.125cm 0.0875cm 0.125cm 0.0875cm},clip]{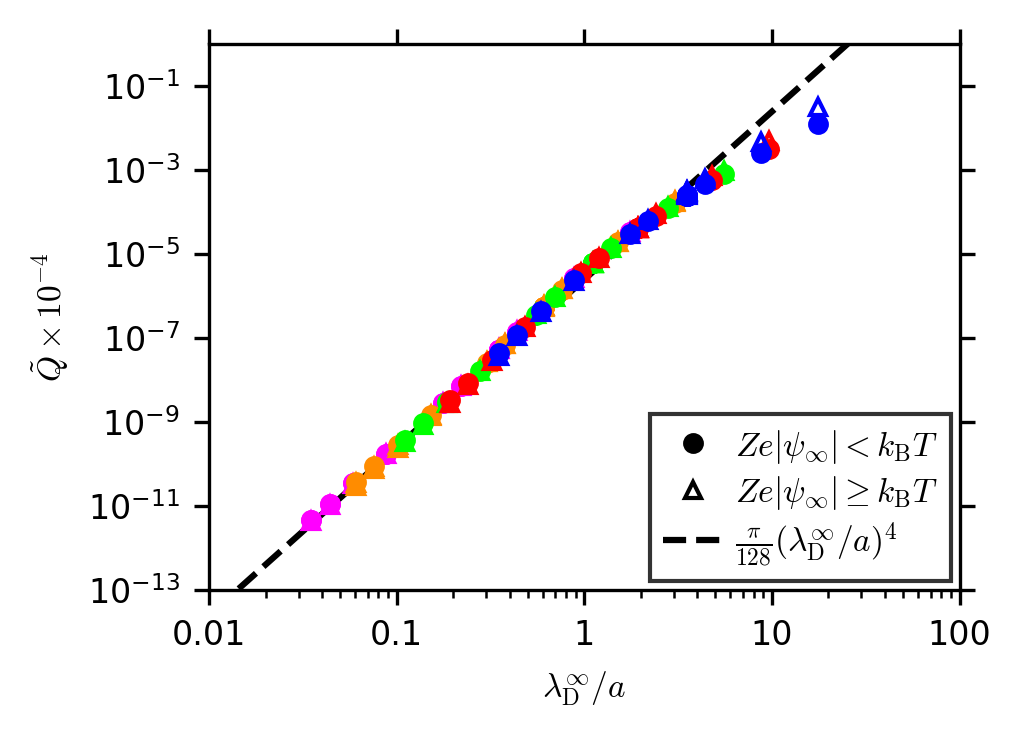}
    \caption{\label{final_eq1} Dimensionless flow rate $\Tilde{Q}$ vs $\lDinf / a$ from FEM simulations for electric potential energies far from the pore that are $< \kBT$ (filled circles) and $\geq \kBT$ (empty triangles) for equilibrium bulk concentrations $\cinf$ of $0.3$ (blue), $1$ (red), $3$ (green), $10$ (orange) and $30$ (magenta) mol~m$^3$. The approximate expression for the dimensionless flux for $\lDinf \ll a$ obtained with fits of $(\lDinf/a)^4$ to the simulation data for which $\lDinf < 0.5a$ are shown for all $\lDinf/a$ in the simulations (dashed lines).}
\end{figure}

\begin{figure}[!h]
 \centering
 \begin{minipage}{.5\textwidth}
\begin{flushleft}
    \includegraphics[scale = 1, trim={0.175cm 0.0875cm 0.175cm 0.0875cm},clip]{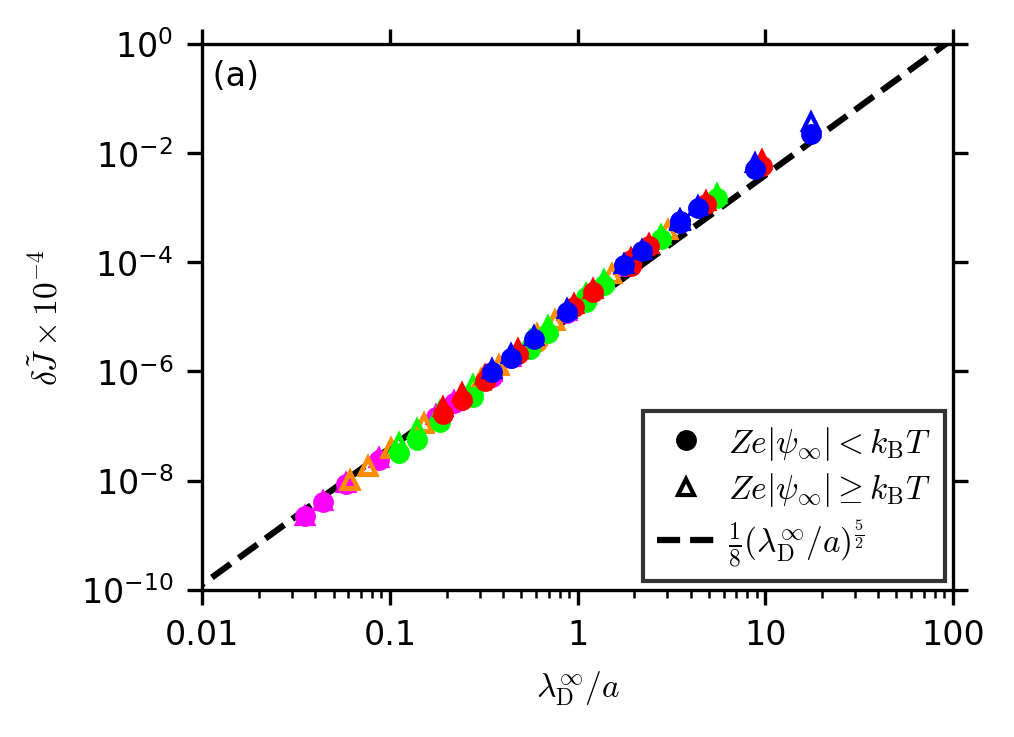}
\end{flushleft}
  \end{minipage}%
 \begin{minipage}{.5\textwidth}
\begin{flushright}
\includegraphics[scale = 1, trim={0.175cm 0.0875cm 0.175cm 0.0875cm},clip]{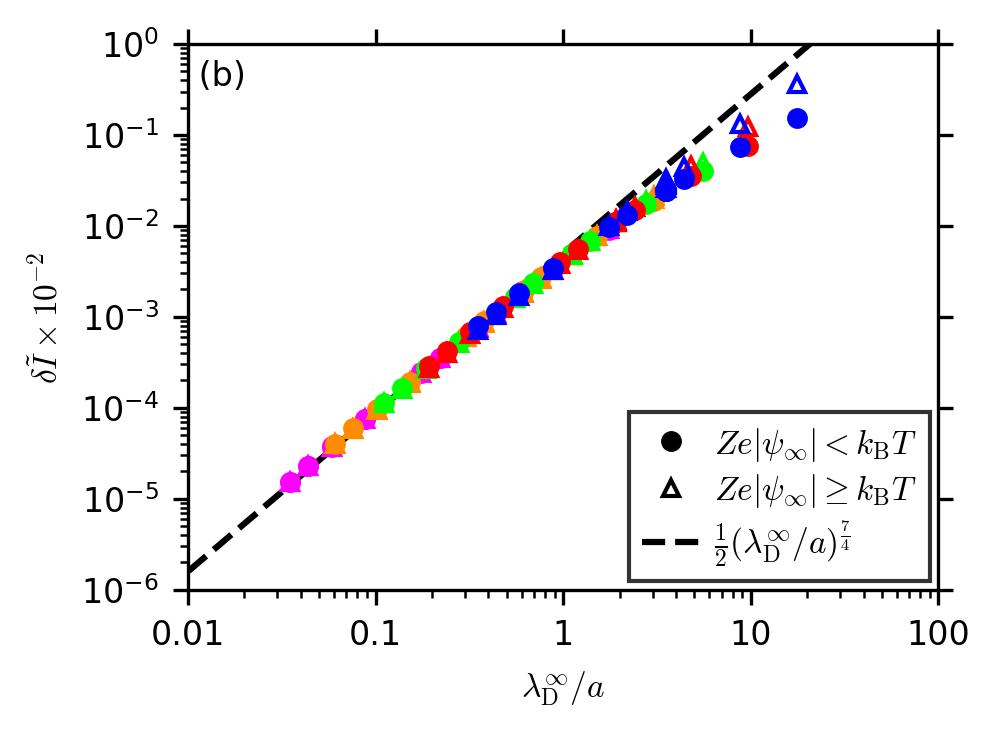}
\end{flushright}
  \end{minipage}
    \caption{\label{final_eq2} Dimensionless surface contribution to the (a) total solute flux, $\Jscaled$, and (b) total electric current, $\Iscaled$, vs $\lDinf / a$ from FEM simulations for electric potential energies far from the pore that are $< \kBT$ (filled circles) and $\geq \kBT$ (empty triangles) for equilibrium bulk concentrations $\cinf$ of $0.3$ (blue), $1$ (red), $3$ (green), $10$ (orange) and $30$ (magenta) mol~m$^3$. The approximate expressions for the dimensionless fluxes for $\lDinf \ll a$ obtained with fits of (a) $(\lDinf/a)^{\frac{5}{2}}$ and (b) $(\lDinf/a)^{\frac{7}{4}}$ to the simulation data for which $\lDinf < 0.5a$ are shown for all $\lDinf/a$ in the simulations (dashed lines).}
 \end{figure}

As all simulation data for which $\lDinf < 0.5a$ show the same power-law scaling for the dimensionless fluxes as determined in the thin electric-double-layer limit, we have considered simulations for which $\lDinf < 0.5a$ in obtaining empirical equations for the fluxes in this regime for any strength of the electric potential. We have restricted the analysis to $\lDinf/l_{\mathrm{GC}}< 10|\Dm|/(\Dp)$ for $\delJ$ and fixed the dimensionless concentration difference at $\Delta \hat{c} = 0.4$. We fitted $\Qscaled$ (from the simulations) to $(\lDinf/a)^4$, $\Jscaled$ to $(\lDinf/a)^{\frac{5}{2}}$ and $\Iscaled$ to $(\lDinf/a)^{\frac{7}{4}}$ (see Fig.~\ref{kappaa_scaling} in the main paper), which gave slopes of $0.0254$, $0.121$ and $0.514$ for $\Qscaled$, $\Jscaled$ and $\Iscaled$, respectively. Referring to the expressions for the dimensionless fluxes in Eqs.~\eqref{Qtilde2}--\eqref{Itilde2} in the main paper, we used these fits to propose that $\Qscaled \approx \frac{\pi}{128}(\lDinf/a)^4$ (where $0.0254 \times 32/\pi = 0.258 \approx 1/4$), $\Jscaled \approx \frac{1}{8}(\lDinf/a)^{\frac{5}{2}}$ and $\Iscaled \approx \frac{1}{2}(\lDinf/a)^{\frac{7}{4}}$ as shown in Figs.~\ref{final_eq1} and \ref{final_eq2}. Substituting these expressions into Eqs.~\eqref{solute_flux4}, \eqref{electric_current4} and \eqref{Qtilde2}--\eqref{Itilde2} in the main paper, we can write the approximate expressions for the fluxes when $D_+ \approx D_- = D$ and $\lDinf \ll a$ for all $Ze|\psiinf|/\kBT$ in the simulations as
\begin{align}
    \label{final1}
    Q & \approx -\left( \frac{\kBT}{Ze} \right)^2 \frac{\epsilon \epsilon_0a}{8 \eta} \ln{\left(1 + \frac{\lDu}{4 \lDinf} \right)} \Dellnc , \\
    \label{final2}
    J & \approx -4aD\Delc - \frac{\kBT}{2 (Ze)^2} \left(\frac{a}{\lDinf} \right)^{\frac{1}{2}} \kappa_{\mathrm{s}}^{\infty} \Dellnc, \\
    \label{final3}
    I & \approx \left( \frac{a}{\lDinf} \right) ^{\frac{1}{4}} \sigma D \Dellnc ,
\end{align}
where all parameters and length scales are as given in the main paper. When $\Delc/\cinf \ll 1$ and $\lDinf \ll a$, the bulk contribution to the solute flux in Eq.~\eqref{final2} can be ignored when
\begin{align}
    \label{condition1}
    \frac{\lDinf}{l_{\mathrm{GC}}} \gg 2\left( \frac{a}{\lDinf} \right)^{\frac{1}{2}}  \sqrt{1 + \left( \frac{\lDinf}{a} \right)^{\frac{1}{2}} },
\end{align}
which reduces to $\lDinf/l_{\mathrm{GC}} \gg 2\left( a / \lDinf \right)^{\frac{1}{2}}$ as $\lDinf/a \rightarrow 0$. As Eq.~\eqref{final2} requires that $l_{\mathrm{GC}} \ll \lDinf$, we can ignore the contribution to the total solute flux from the difference in ion diffusivities in this limiting case. Since $\kappa_{\mathrm{s}}^{\infty}$ is independent of $\lDinf$ when $l_{\mathrm{GC}} \ll \lDinf$ (see Eq.~\eqref{conductance2}), the total solute flux is proportional to $(\lDinf)^{-\frac{1}{2}}$ when $\lDinf \ll a$ and the bulk contribution can be ignored.

\clearpage

\section{Derivation of scaling relationships for the electroosmotic flow through a circular aperture in a 2D membrane in the thin-electric-double-layer regime}\label{sec:EDF}

Here we use the same approximation for the dimensionless equilibrium potential $\epotscaled$ applied in the main paper to derive the concentration-gradient-driven flow rate for $\lDinf \ll a$. For the first time, we obtain scaling laws for electroosmosis through a 2D membrane in the thin electric-double-layer regime ($\lambda_{\mathrm{D}} = \lDinf$ is the Debye screening length in the absence of an applied concentration gradient) using in existing analytical expression in Ref.~\citenum{maoElectroosmoticFlowNanopore2014} for the electroosmotic flow rate in the Debye--H\"uckel regime in this geometry. We can write the electrosmotic flow rate in Ref. \citenum{maoElectroosmoticFlowNanopore2014} in terms of the dimensionless equilibrium electric potential $\psi_0= \frac{\sigma a}{\epsilon \epsilon _0} \Tilde{\psi _0}$, Debye screening length $\lD$, and the coordinate system described in the main paper as
\begin{align}
    \label{psi-flowrate}
    Q = -\frac{2 a^4 \sigma \Delta \psi }{\pi \eta (\lD)^2 } \int ^1 _0 \mathrm{d}\zeta \, \zeta^2 \int ^\infty _0 \mathrm{d} \nu \, \frac{\Tilde{\psi_0}}{1 + \nu^2} ,
\end{align}
where $\Delta \psi$ is the applied potential difference and all other parameters and length scales are as described in the main paper. Taking $\epotscaled \approx \lDinf/a$ for $r > a$ and $d < \lDinf$ and $\approx 0$ otherwise, we can approximate the integral in Eq.~\eqref{psi-flowrate} for $\lDinf \ll a$ as
\begin{align}
    \int ^1 _0 \mathrm{d}\zeta \ \zeta^2 \int ^\infty _0 \mathrm{d} \nu \, \frac{\Tilde{\psi_0}}{1 + \nu^2} &\approx \left( \frac{\lD}{a} \right) \int ^1 _0 \mathrm{d}\zeta \ \zeta^2 \int ^\infty _\zeta \mathrm{d} \nu \, \frac{{H(\lDinf - a \nu \zeta)}}{1 + \nu^2} \nonumber \\
    \label{psi-flow_rate2}
     &\approx \frac{1}{2} \left( \frac{\lD}{a} \right)^3, 
\end{align}
where we have used the solution to an analogous integral derived in Ref. \citenum{rankinEntranceEffectsConcentrationgradientdriven2019}. Substituting Eq.~\eqref{psi-flow_rate2} into Eq.~\eqref{psi-flowrate} gives the scaling for the electroosmotic flow rate,
\begin{align}
    \label{Qpropto3}
    Q \propto a \lD \sigma \Delta \psi  ,
\end{align}
in the limit $Ze|\psi| \ll \kBT$ and $\lDinf \ll a$. Similarly to the case of concentration-gradient-driven flow in the main paper, we compare the approximate scaling shown in Eq.~\eqref{psi-flow_rate2} with the integral in Eq.~\eqref{psi-flowrate} directly, such that the electroosmotic flow rate in the simulations is subject to the non-dimensionalization
\begin{align}
    \label{psi-flowrate2}
    Q =  -\frac{2 a^4 \sigma \Delta \psi }{\pi \eta (\lD)^2 } \Qscaled .
\end{align}

\begin{figure}[!ht]
    \centering
    \includegraphics[scale = 1, trim={0.175cm 0.0875cm 0.175cm 0.0875cm},clip]{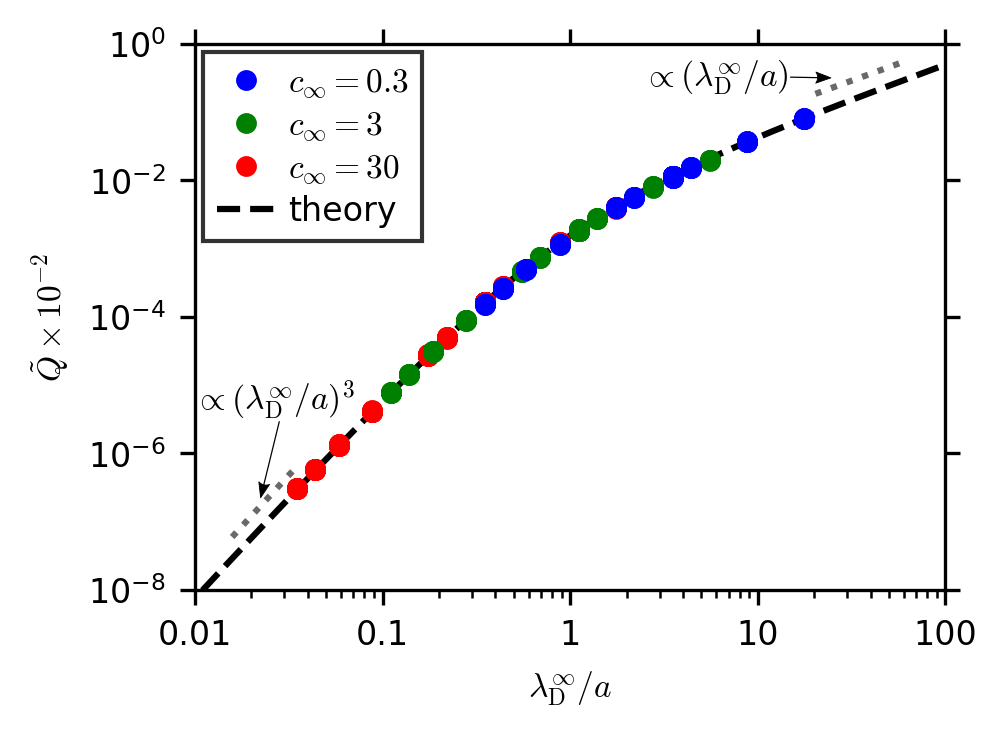}
    \caption{Dimensionless electroosmotic flow rate $\Qscaled$ vs $\lDinf / a$ from FEM simulations (symbols) and theory in Eq.~\eqref{psi-flowrate} (dashed line) for a range of average bulk concentrations $\cinf$ (mol~m$^{-3}$). The scaling is shown for various powers of $\lDinf / a$ for $\lDinf \ll a$ (Eq.~\eqref{psi-flow_rate2}) and $\lDinf \gg a$ (dotted lines).}
    \label{Q-dV}
\end{figure}

In Fig.~\ref{Q-dV}, we verify the scaling of $Q$ with $a$ and $\lDinf$ in the $\lDinf \ll a$ regime, as given in Eq.~\eqref{Qpropto3} and Table~\ref{scaling-flux} in the main paper, via FEM simulations of electric-field-driven flow for various $\cinf$, $a$ and $\sigma$ (for $Ze|\psi| < \kBT$) subject to the non-dimensionalization in Eq.~\eqref{psi-flowrate2}. Figure~\ref{Q-dV} includes all of the simulated surface charge densities that correspond to $Ze|\psi| < \kBT$ (see Sec.~\ref{sec:FEM} on finite-element method numerical simulations) and also shows the scaling for the electroosmotic flow rate for $\lDinf \gg a$ in Ref. \citenum{maoElectroosmoticFlowNanopore2014} subject to the non-dimensionalization in Eq.~\eqref{psi-flowrate2}. 

\clearpage

\section{Derivation of scaling relationships for concentration-gradient-driven flow in a long cylindrical pore (Debye--H\"uckel regime)} \label{sec:cylinder}

\begin{figure}[!h]
    \centering
    \includegraphics[scale = 1.5]{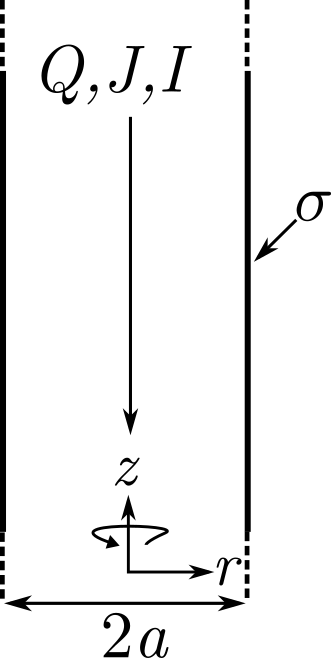}
    \caption{Schematic of flow of a solution through a cylindrical pore of radius $a$ in $(r,z)$ coordinates, which has surface charge density $\sigma$ and is symmetric about the $z$ axis. The pore has length $L$ and the effects of the pore ends are ignored ($L \gg a$).}
    \label{schematic-cylinder}
\end{figure}

Here we assume the same governing equations as used in the main paper (Eqs.~\eqref{poisson}--\eqref{newt}) to derive scaling relationships for the concentration-gradient-driven fluid fluxes as a function of the pore radius $a$, surface charge density $\sigma$, the equilibrium Debye screening length $\lDinf$ and pore length $L$ for a dilute solution in a long cylindrical pore (Fig.~\ref{schematic-cylinder}). We neglect entrance effects by assuming that the pore length is much larger than the pore radius ($L \gg a$).

This derivation begins with a similar approach to that in Ref.~\citenum{rankinEffectHydrodynamicSlip2016} for slip-dependent reverse electrodialysis (RED); however, we simplify the case significantly. We further assume small electric potential energies relative to $\kBT$ (Debye--H\"uckel regime), and implement a perturbation expansion, which allows us to ignore the axial-dependent term in the electric potential $\psi(r,z)$ (provided the concentration gradient is small) and only consider the equilibrium electric potential $\psi_0(r)$, which is only a function of the radial coordinate $r$. After considering a perturbation expansion, we use a similar approach to that in Ref.~\citenum{rankinEntranceEffectsConcentrationgradientdriven2019} for the concentration-gradient-driven flow of a dilute solution containing a neutral solute in a long cylindrical pore to determine the fluid fluxes for the electrolyte solution to first order. All parameters, fluxes and constants are the same as in the main paper unless otherwise specified.

Letting $\cs$ be the concentration of the electrolyte solution for a $Z$:$Z$ electrolyte where $\psi=0$, we can write the concentration of the ion species $i$ as \cite{rankinEffectHydrodynamicSlip2016}
\begin{align}
    \label{cs}
    c^{(i)}(r,z) = \cs(z) \exp{\left(-\frac{Z_i e \psi(r,z)}{\kBT}\right)} .
\end{align}
Assuming that $L \gg a$, the radial velocity $u_r$ will be negligible compared to the axial solution velocity $u_z$, and we can solve for the pressure using the $r$-component of Eq.~\eqref{stokes} in the main paper for a $Z$:$Z$ electrolyte by setting $u_r \approx 0$, which gives
\begin{align}
    \label{stokes1}
    \frac{\partial p}{\partial r} + Ze(c_+ - c_-) \frac{\partial \psi}{\partial r} = 0 .
\end{align}
Substituting Eq.~(\ref{cs}) into Eq.~\eqref{stokes1} and applying the product rule gives
\begin{equation}
\frac{\partial p}{\partial r} = 2 Ze \cs \sinh{\left(\frac{Ze \psi}{\kBT} \right)} \frac{\partial \psi}{\partial r} \\
\label{stokes2}
 = 2 \kBT \cs \frac{\partial}{\partial r}\left[\cosh{\left(\frac{Ze \psi}{\kBT} \right)}\right] .
\end{equation}
Assuming that $Ze|\psi| \ll \kBT$, Eq.~\eqref{stokes2} simplifies to
\begin{equation}
\label{stokes3}
\frac{\partial p}{\partial r}  \approx 2 \kBT \cs \frac{\partial}{\partial r} \left [ \frac{1}{2} \left(\frac{Ze \psi}{\kBT} \right) ^2 + 1 \right ]
= \frac{(Ze)^2}{\kBT} \cs \frac{\partial}{\partial r} \left ( \psi^2 \right ) .
\end{equation}

We assume that each system variable can be represented by a perturbation expansion with respect to the equilibrium value, where $\beta \ll 1$ is a dimensionless quantity that characterizes the perturbation to the system due to the applied concentration difference. Hence, to first order
\begin{align}
    \label{bu}
    \boldsymbol{u}(r,z) & = \beta \boldsymbol{u}_1 (r,z) , \\
    \label{bcs}
    \cs(z) & = \cinf + \beta c_{s_1}(z) , \\
    \label{bpsi}
    \psi(r,z) & = \psi _0 (r) + \beta \psi _1 (r,z) , \\
    \label{bp}
    p(r,z) &= p_0(r) + \beta p_1(r,z) ,
\end{align}
where the axial-dependent terms in the variables come from the applied concentration difference at the pore ends, and we can assume that these variables depend only on the radial coordinate at equilibrium when $L \gg a$. Substituting Eqs.~\eqref{bcs}--\eqref{bp} into Eq.~\eqref{stokes3} and expanding to $O(\beta)$ gives
\begin{equation}
\label{stokes4}
\frac{\partial  p_1(r,z)}{\partial r} \approx \frac{(Ze)^2}{\kBT} \bcs(z) \frac{\mathrm{d}}{\mathrm{d} r} \left ( \left( \psi_0 (r) \right) ^2 \right ) + \frac{2(Ze)^2}{\kBT} \cinf \frac{\partial}{\partial r} \left( \psi_0 (r) \psi_1(r,z) \right). 
\end{equation} 
We can integrate Eq.~\eqref{stokes4} to give the first-order component of the pressure,
\begin{equation}
\label{p1}
p_1(r,z) = p^{\infty}_1 + \frac{(Ze)^2 \cs(z)}{\kBT} \left ( \psi_0(r) \right)^2 + \frac{2(Ze)^2}{\kBT} \cinf \psi_0(r)  \psi_1(r,z) ,
\end{equation}
where $p_1^{\infty}(z)$ is the first-order component of the pressure at which $\psi_0=0$ with respect to $r$ and is constant. Taking the derivative of Eq.~\eqref{p1} with respect to $z$ gives
\begin{equation}
\label{dp1}
\frac{\partial p_1(r,z)}{\partial z} = \frac{(Ze)^2}{\kBT} \left ( \psi_0(r) \right)^2 \frac{\mathrm{d} \cs(z)}{\mathrm{d} z} + \frac{2(Ze)^2}{\kBT} \cinf \psi_0(r)  \frac{\partial \left ( \psi_1(r,z) \right )}{\partial z} .
\end{equation}

We can also assume that the axial gradient in the remaining axial component of the ﬂuid velocity can be ignored; hence, the axial component of Eq.~\eqref{stokes} in the main paper simpliﬁes to
\begin{equation}
\label{stokes5}
\frac{\eta}{r} \frac{\partial}{\partial r} \left(\frac{\partial u_z}{\partial r} \right) = \frac{\partial p}{\partial z} + Ze(c_+ - c_-) \frac{\partial \psi}{\partial z} .
\end{equation}
Substituting Eq.~\eqref{cs} into Eq.~\eqref{stokes5} and applying the product rule gives
\begin{equation}
\label{stokes8}
\frac{\eta}{r} \frac{\partial}{\partial r} \left(\frac{\partial u_z}{\partial r} \right)  = \frac{\partial p}{\partial z} - 2Ze\cs \sinh{\left( \frac{Ze \psi}{\kBT} \right)} \frac{\partial \psi}{\partial z}
 = \frac{\partial p}{\partial z} - 2\kBT \cs  \frac{\partial}{\partial z} \left [ \cosh{\left( \frac{Ze \psi}{\kBT} \right)} \right ].
\end{equation}
Assuming that $Ze|\psi| \ll \kBT$, Eq.~\eqref{stokes8} simplifies to
\begin{equation}
\label{stokes9}
\frac{\eta}{r} \frac{\partial}{\partial r} \left(\frac{\partial u_z}{\partial r} \right)  \approx \frac{\partial p}{\partial z} - 2\kBT \cs  \frac{\partial}{\partial z} \left [ \frac{1}{2} \left( \frac{Ze \psi}{\kBT} \right)^2 + 1 \right ]
 = \frac{\partial p}{\partial z} - \frac{(Ze)^2}{\kBT} \cs  \frac{\partial}{\partial z} \left ( \psi ^2 \right ).
\end{equation}
Substituting Eqs.~\eqref{bcs}--\eqref{bp} into Eq.~\eqref{stokes9}, expanding to $O(\beta)$, and substituting Eq.~\eqref{dp1} into this expression (noting that $\frac{\partial p _0}{\partial z}=0$) gives
\begin{equation}
\frac{\eta}{r} \frac{\partial}{\partial r} \left(\frac{\partial u_{z _1}}{\partial r} \right) = \frac{\partial p _1}{\partial z} - \frac{2 (Ze)^2}{\kBT} \cinf \psi_0(r)  \frac{\partial}{\partial z} \left ( \psi_1(r,z) \right )
\label{stokes7}
 = \frac{(Ze)^2}{\kBT} \frac{\mathrm{d} \bcs(z)}{\mathrm{d} z} (\psi_0(r))^2.
\end{equation}
Integrating Eq.~\eqref{stokes7} twice using the no-slip boundary condition ($\boldsymbol{u} = 0$) at the pore surface gives the axial velocity, $u_z = \beta u_{z_1}$, as
\begin{align}
\label{velocity}
u_{z}(r,z) = -\beta \frac{(Ze)^2}{\eta \kBT} \frac{\mathrm{d} \bcs}{\mathrm{d} z} \int ^a _{r'} \frac{\mathrm{d}r'}{r'} \int ^{r'} _0 \mathrm{d} r'' \, r'' \psi _0 ^2 .
\end{align}
The flow rate at any cross-section of the pore is
\begin{align}
\label{flow rate}
Q= \iint _{S} \mathrm{d}S \, \boldsymbol{u} \cdot \hat{\boldsymbol{n}} .
\end{align}
Substituting Eq.~\eqref{velocity} into Eq.~\eqref{flow rate}, where $\hat{\boldsymbol{n}}= \hat{\boldsymbol{z}}$ is the unit normal to the pore surface, gives 
\begin{align}
\label{flow_rate2}
Q= -\beta \frac{\pi}{2 \eta} \frac{(Ze)^2}{\kBT} \frac{\mathrm{d} \bcs}{\mathrm{d} z} \int^a _0 \mathrm{d} r \, r(a^2 - r^2)\psi _0^2.
\end{align}
Since the concentration-gradient-driven flow rate $Q$ is the same for any cross-section along the pore, we can assume that the derivative of $\bcs$ with respect to $z$ is uniform over the entire length of the pore, $L$, where $\Delc / \beta$ is the change in $\bcs$ over the length of the pore. Moreover, $\Dellnc \approx \Delc / \cinf$ for surface contributions to the fluxes in the linear-response regime, and $\lDinf$ is the same as given in Eq.~\eqref{Debye_length} of the main paper. Equation~\eqref{flow_rate2} can thus be given as 
\begin{align}
\label{flow_rate3}
Q= -\frac{\pi}{4 \eta} \frac{\epsilon \epsilon _0}{(\lDinf)^2} \frac{\Dellnc}{L} \int^a _0 \mathrm{d} r \,  r(a^2 - r^2)\psi _0^2  .
\end{align}

Using Eq.~\eqref{solute_flux_density} in the main paper, the total ion flux density for a long cylindrical pore can be given as
\begin{align}
\label{ionfluxdens}
\boldsymbol{j} = -\beta \frac{\mathrm{d} \bcs}{\mathrm{d}z} \left \{ (\Dp) \left[1 + \frac{1}{2} \left( \frac{Ze\psi_0}{\kBT} \right)^2 \right] - (\Dm) \left( \frac{Ze\psi_0}{\kBT} \right) \right \} .
\end{align}
Using Eq.~\eqref{electric-current-density2} in the main paper, the electric current density for a long cylindrical pore can be given as
\begin{align}
\label{elecfluxdens}
\boldsymbol{j}_{\mathrm{e}} = \beta Ze\frac{\mathrm{d} \bcs}{\mathrm{d}z} \left [ (\Dp) \left( \frac{Ze\psi_0}{\kBT} \right) - (\Dm) \right ] .
\end{align}
Substituting Eq.~\eqref{ionfluxdens} into Eq.~\eqref{solute_flux1} in the main paper, where $\hat{\boldsymbol{n}}= \hat{\boldsymbol{z}}$ is the unit normal to the pore surface, gives the total concentration-gradient-driven solute flux as
\begin{align}
\notag
J & = -2 \pi \beta \frac{\mathrm{d} \bcs}{\mathrm{d}z} \left \{ (\Dp) \left[\frac{a^2}{2} + \frac{1}{2} \left( \frac{Ze}{\kBT} \right)^2 \int^a _0 \mathrm{d}r \, r \psi_0^2 \right] - (\Dm) \frac{Ze}{\kBT} \int^a _0 \mathrm{d}r \, r \psi_0 \right \} \\
\label{soluteflux}
& = -\frac{\pi}{L}(\Dp) \left \{ a^2 \Delc + \frac{\epsilon \epsilon_0}{(\lDinf)^2} \Dellnc \left[ \int^a _0 \mathrm{d} r \, r \frac{\psi_0^2}{2\kBT} - \left(\frac{\Dm}{\Dp}\right) \int^a _0 \mathrm{d}r \, r \frac{\psi_0}{Ze} \right] \right \} ,
\end{align}
where
\begin{align}
\label{Jbulk}
\Jb & = - \pi a^2 (\Dp) \frac{\Delc}{L}
\end{align}
is the bulk contribution and 
\begin{align}
\label{delJ}
\delJ & = - \frac{\pi \epsilon \epsilon_0}{(\lDinf)^2} \frac{\Dellnc}{L} \left[ \frac{\Dp}{2 \kBT} \int^a _0 \mathrm{d}r \, r \psi_0^2 - \frac{\Dm}{Ze} \int^a _0 \mathrm{d}r \, r \psi_0 \right]
\end{align}
is the surface contribution to the total solute flux. 

Substituting Eq.~\eqref{elecfluxdens} into Eq.~\eqref{electric_current1} from the main paper gives the concentration-gradient-driven electric current as
\begin{align}
\notag
I & = 2 \pi \beta Ze \frac{\mathrm{d} \cs}{\mathrm{d} z} \left[ -\frac{a^2}{2} (\Dm) + (\Dp) \frac{Ze}{\kBT} \int ^a_0 \mathrm{d}r \, r\psi_0  \right] \\
\label{electriccurrent}
& = \frac{\pi}{L} \left[-a^2 Ze (\Dm) \Delc + \frac{\epsilon \epsilon_0}{(\lDinf)^2} (\Dp) \Dellnc \int ^a_0 \mathrm{d}r \, r \psi_0  \right] ,
\end{align}
where
\begin{align}
\label{Ibulk}
    \Ib = -\pi a^2 Ze (\Dm) \frac{\Delc}{L}
\end{align}
is the bulk contribution and 
\begin{align}
\label{delI}
    \delI = \pi \frac{\epsilon \epsilon _0}{(\lDinf)^2}(D_+ + D_-) \frac{\Dellnc}{L} \int ^a _0 \mathrm{d}r \, r \psi_0
\end{align}
is the surface contribution to the electric current.

In the limit that the equilibrium Debye screening length is much larger than the pore radius ($\lDinf \gg a$), the potential at the membrane surface is approximately constant and equal to the surface potential everywhere. In this limit, we can use the electric potential at a planar surface and approximate $\psi_0 = \frac{\sigma \lDinf}{\epsilon \epsilon_0}$. Inserting this expression into Eqs.~\eqref{flow_rate2}, \eqref{delJ} and \eqref{delI} gives the approximate fluxes for $\lDinf \gg a$ as
\begin{align}
    Q & \approx -\frac{\pi}{16}\frac{a^4 \sigma ^2}{\epsilon \epsilon _0 \eta} \frac{\Dellnc}{L} , \\
    \label{delJ2}
    \delJ & \approx -\frac{\pi a^2}{2} \frac{|\sigma|}{\lDinf} \frac{(\Dp)}{Ze} \left(\frac{\lDinf}{l_{\mathrm{GC}}} - \mathrm{sgn}(\sigma) \frac{\Dm}{\Dp} \right) \frac{\Dellnc}{L} , \\
    \delI & \approx \frac{\pi a^2 \sigma}{2\lDinf}  \frac{\Dellnc}{L},
\end{align}
where the contribution to  Eq.~\eqref{delJ2} due to $\Dm$ is negligible when $\lDinf/l_{\mathrm{GC}} \gg \frac{|D_+ - D_-|}{\Dp}$. Thus, when $\lDinf \gg a$, the scaling relationships for the fluid fluxes with $a$, $\sigma$, $\lDinf$ and $L$ are
\begin{align}
Q & \propto \frac{a^4 \sigma ^2 \Dellnc}{L} , \\
\delJ & \propto \frac{a^2 \sigma ^2 \Dellnc}{L} , \\
\delI & \propto \frac{a^2 \sigma}{\lDinf} \frac{\Dellnc}{L} .
\end{align}

In the limit that the equilibrium Debye screening length is much smaller than the pore radius ($\lDinf \ll a$), we can approximate the equilibrium electric potential using the potential near a planar surface in Eq.~\eqref{epot3} for $d = a-r$, where $\psi_0 = \frac{\sigma a}{\epsilon \epsilon_0} \Tilde{\psi_0}$. Equation~\eqref{flow_rate2} for $\lDinf \ll a$ can thus be approximated as
\begin{align}
\notag
    Q &\approx -\frac{\pi}{4}\frac{\sigma^2 \Dellnc}{\epsilon \epsilon _0 \eta L} \int ^a _0 \mathrm{d}r \, r(a^2 - r^2) \exp{\left(-\frac{2(a - r)}{\lDinf} \right)} \\
    \notag
      &= -\frac{\pi}{4}\frac{\sigma^2 \Dellnc}{\epsilon \epsilon _0 \eta L} \left[ \frac{(a\lDinf)^2}{2} - \frac{3a(\lDinf)^3}{2} + \frac{3(\lDinf)^4}{8} + \left(\frac{(a \lDinf)^2}{4} - \frac{3(\lDinf)^4}{8} \right) \exp{\left( -\frac{2a}{\lDinf} \right) } \right] \\
      &\approx -\frac{\pi}{8}\frac{(a \lDinf \sigma)^2}{\epsilon \epsilon _0 \eta} \frac{\Dellnc}{L} ,
\end{align}
which gives the scaling relationship
\begin{align}
    Q \propto \frac{(a \lDinf \sigma)^2 \Dellnc}{L}
\end{align}
in the limit $\lDinf \ll a$. Noting that
\begin{align}
    \int ^a _0 \mathrm{d}r \, {r \psi_0} & \approx \frac{\sigma \lDinf}{\epsilon \epsilon_0} \int ^a _0 \mathrm{d}r \, r \exp{\left(-\frac{a - r}{\lDinf} \right)} \nonumber \\
     & = \frac{\sigma \lDinf}{\epsilon \epsilon_0} \left[ a\lDinf - (\lDinf)^2 + (\lDinf)^2\exp{\left(-\frac{a}{\lDinf} \right)} \right] \nonumber \\
     & \approx \frac{a (\lDinf)^2 \sigma}{\epsilon \epsilon _0}
\end{align}
and
\begin{align}
    \int ^a _0 \mathrm{d}r \, {r \psi_0} & \approx \left( \frac{\sigma \lDinf}{\epsilon \epsilon_0} \right) ^2 \int ^a _0 \mathrm{d}r \, r \exp{\left(-\frac{2(a - r)}{\lDinf} \right)} \nonumber \\
     & = \left( \frac{\sigma \lDinf}{\epsilon \epsilon_0} \right) ^2 \left[ \frac{a\lDinf}{2} - \frac{(\lDinf)^2}{4} + \frac{(\lDinf)^2}{4} \exp{\left(-\frac{a}{\lDinf} \right)} \right] \nonumber \\
     & \approx \frac{a (\lDinf)^3 \sigma^2}{2 (\epsilon \epsilon _0)^2},
\end{align}
Eqs.~\eqref{delJ} and \eqref{delI} can thus be approximated in the $\lDinf \ll a$ regime as
\begin{align}
    \delJ & \approx -\frac{ \pi a |\sigma|(\Dp)}{2Ze} \frac{\Dellnc}{L}\left[ \frac{\lDinf}{l_{\mathrm{GC}}} - 2\mathrm{sgn}(\sigma)\frac{\Dm}{\Dp}\right ]  , \\
    \delI & \approx \frac{\pi a \sigma (\Dp) \Dellnc}{L}  ,
\end{align}
where the contribution to  Eq.~\eqref{delJ2} from the difference in ion diffusivities is negligible when $\lDinf/l_{\mathrm{GC}} \gg 2 \frac{|D_+ - D_-|}{\Dp}$. Thus, in the thin electric-double-layer limit, the scaling relationships for $\delJ$ and $\delI$ with $a$, $\sigma$, $\lDinf$ and $L$ are
\begin{align}
\label{cyl_scaling_delJ}
    \delJ & \propto \frac{a \lDinf \sigma^2 \Dellnc}{L} , \\
\label{cyl_scaling_delI}
     \delI & \propto \frac{a \sigma \Dellnc}{L} .
\end{align}

\clearpage

\section{Finite-element method numerical simulations} \label{sec:FEM}

\begin{figure}[!h]
    \centering
    \includegraphics[scale = 3]{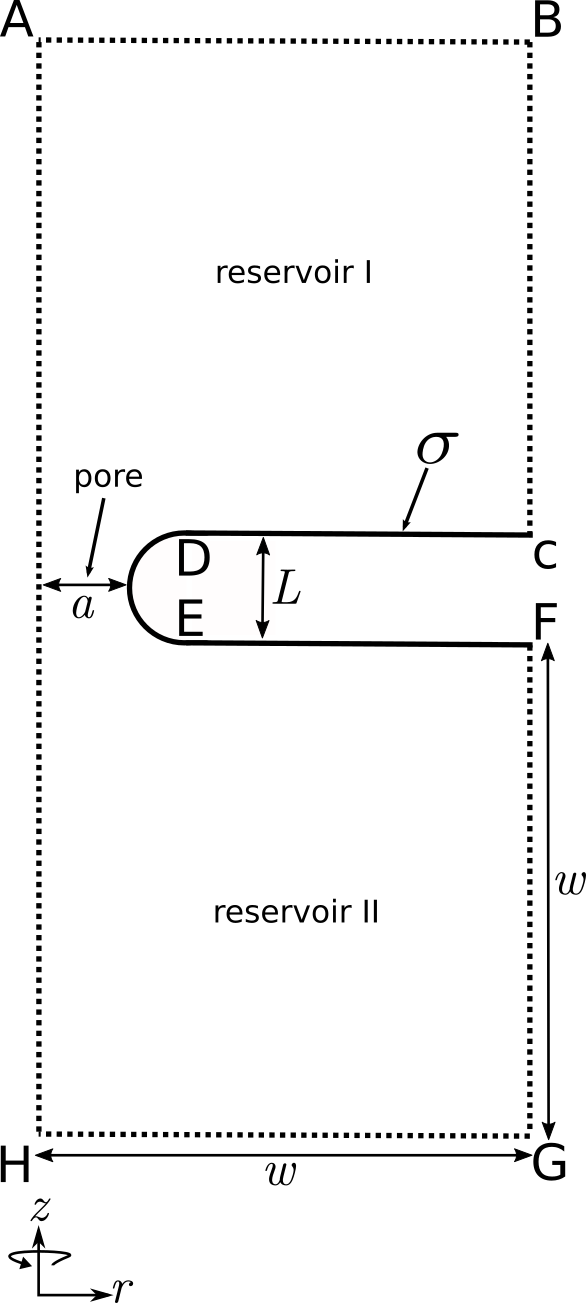}
    \caption{Schematic of the two-dimensional axisymmetric computational domain used in the FEM simulations (not to scale). The geometry has rotational symmetry about the boundary AH, where the solid lines denote solid–liquid boundaries and the dashed lines denote liquid boundaries.}
    \label{Schematic-FEM}
\end{figure}

The continuum hydrodynamic flow equations, given in Eqs.~\eqref{poisson}--\eqref{newt} in the main paper, were solved using finite-element method (FEM) simulations with COMSOL Multiphysics 4.3a \cite{comsol4.3a} for a thin planar membrane of thickness $L$ containing a circular aperture of radius $a$ connecting two large cylindrical fluid reservoirs (Fig.~\ref{Schematic-FEM}). A fully coupled solver, which is a damped version of Newton's method, and the PARDISO direct solver were used to solve the equations, where the damping option used to achieve convergence was "Automatic highly nonlinear (Newton)". 

\begin{figure}[!h]
    \centering
    \includegraphics[scale = 1, trim={0.175cm 0.0875cm 0.175cm 0.0875cm},clip]{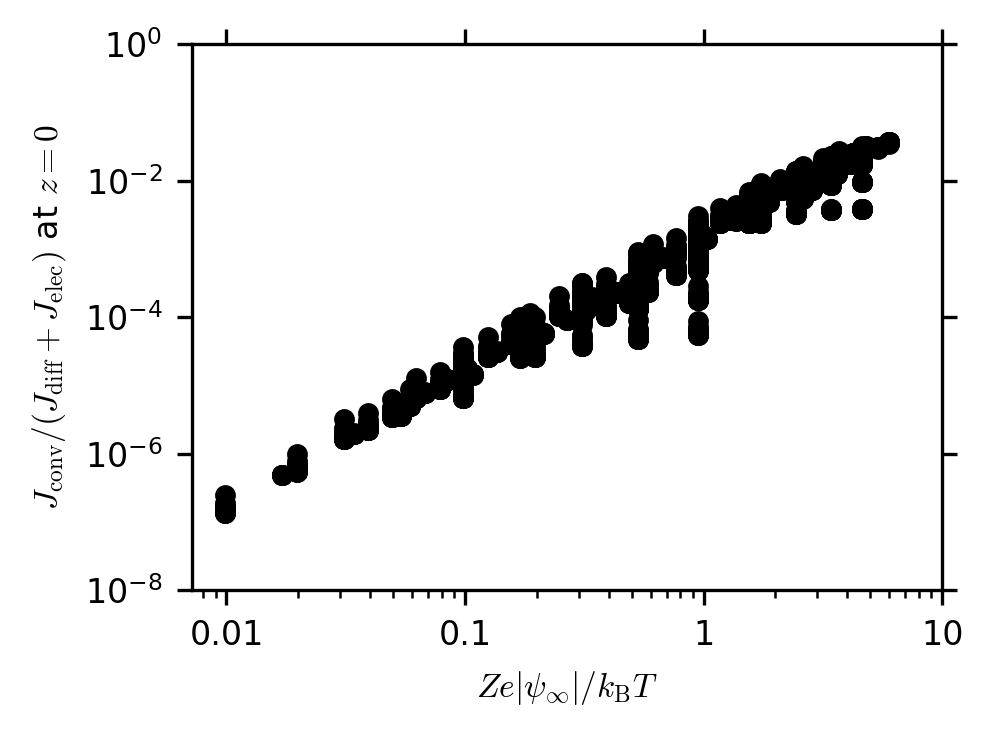}
    \caption{\label{peclet_verify} Total convective solute flux, $J_{\mathrm{conv}}$, over the sum of the total diffusive and electrophoretic solute fluxes, $J_{\mathrm{diff}}+J_{\mathrm{elec}}$, at $z = 0$ vs the magnitude of the potential energy of a planar surface, $Ze|\psiinf|$, relative to the thermal energy, $\kBT$, for all simulations of concentration-gradient-driven flow and non-zero surface charge (symbols).}
\end{figure}

A surface charge density $\sigma$ was applied to the boundaries DC, EF and DE. The boundaries AB, BC, FG and GH were of width $\mathrm{max}(12(l_{\mathrm{Du}_{\mathrm{L}}} + 2a), 75\lambda_{\mathrm{L}})$, where $l_{\mathrm{Du}_{\mathrm{L}}}$ and $\lambda_{\mathrm{L}}$ are the Dukhin and Debye screening lengths, respectively, far from the surface for the reservoir of lower solute concentration, and $a$ is the pore radius. The membrane thickness $L$ was no larger than $1/5$ of the pore radius and less than $1/5$ of the Debye screening length at the higher concentration reservoir in all cases, and the membrane surface between points D and E in Fig.~\ref{Schematic-FEM} was given a finite radius of curvature of $L/2$. A boundary layer mesh was used at all solid--liquid interface, with 5 boundary layers and a boundary layer stretching factor (mesh element growth rate) of $1.2$ and thickness adjustment factor of $10$. A predefined ("Normal") element size was used in the simulation domain and the maximum element size at the boundary between points D and E in Fig.~\ref{Schematic-FEM} was $2L$. The maximum element size between the points D and C was $\lambda_{\mathrm{H}}/5$ and that between the points E and F was $\lambda_{\mathrm{L}}/5$, where $\lambda_{\mathrm{H}}$ is the Debye screening length far from the surface for higher solute-concentration reservoir. Cubic and quadratic discretizations were used for the solute concentration and electric potential, respectively, while second-order elements were used for the velocity components and linear elements were used for the pressure field in the Stokes equation. We verified that the measured solution and solute fluxes did not change significantly with a finer mesh, a larger reservoir size or higher discretization orders. Table \ref{boundaries} lists the boundary conditions used to solve the equations and Table \ref{variables} lists the parameters used in the simulations. We verified that all of the FEM simulations of concentration-gradient-driven flow carried out in this work were within the low Péclet-number regime by comparing the convective solute flux to the diffusive and electrophoretic solute fluxes  in Fig.~\ref{peclet_verify}.

\begin{table}[!h]
\caption{\label{boundaries} Boundary conditions used to solve the continuum hydrodynamic flow equations in the FEM simulations, where $\boldsymbol{\hat{n}}$ is the unit normal to the surface and $i$ labels the ion type.}
\begin{tabular*}{.75\columnwidth}{@{\extracolsep{\fill}} l l}
\hline
boundary & conditions \\
\hline
AH      & $\boldsymbol{\hat{n}} \cdot \nabla c^{(i)} = \boldsymbol{\hat{n}} \cdot \boldsymbol{u} = \boldsymbol{\hat{n}} \cdot \nabla \boldsymbol{u} = 0$ \\
AB      & $c^{(i)} = c^{(i)}_{\mathrm{H}} = \cinf + \Delc/2$, $p = p_{\infty} = 0$, $\psi = \psi_{\mathrm{s}} = 0$  \\
GH          & $c^{(i)} = c^{(i)}_{\mathrm{L}} = \cinf - \Delc/2$, $p = p_{\infty} = 0$, $\psi = \psi_{\mathrm{s}} = 0$ \\
BC and FG         &  $\boldsymbol{\hat{n}} \cdot \boldsymbol{j}_i = \boldsymbol{\hat{n}} \cdot \boldsymbol{u} = \boldsymbol{\hat{n}} \cdot \nabla \boldsymbol{u} = 0$ \\
CD, DE, and EF         & $\boldsymbol{\hat{n}} \cdot \boldsymbol{j}_i = \boldsymbol{u} = 0$ \\ 
\hline
\end{tabular*}
\end{table}

\begin{table}[!h]
\caption{\label{variables} Parameters used in the FEM simulations, where the ion mobilities were calculated using the Einstein relation $\mu _i = D_i /\kBT$ with ion diffusivities $D_i$ chosen to be those for KCl. \cite{grayAIPHandbook1972} Where a range of values is given, parameters were fixed at the values in parentheses unless otherwise indicated.}
\begin{tabular*}{0.9\columnwidth}{@{\extracolsep{\fill}} l c c c}
\hline
quantity & symbol & unit & value \\
\hline
ion diffusivity ($+$ve)    & $D_+$      & m$^2$ s$^{-1}$ & $1.960 \times 10^{-9}$ \\
ion diffusivity ($-$ve)    & $D_-$      & m$^2$ s$^{-1}$ & $2.030 \times 10^{-9}$ \\
ion valence ($+$ve)        & $Z_+$      & --              & $1$ \\
ion valence ($-$ve)        & $Z_-$      & --              & $-1$ \\
membrane thickness         & $L$        & nm              & $0.2$ \\ 
pore radius                & $a$        & nm              & $1$--$50$ ($5$) \\
average bulk concentration & $\cinf$    & mol~m$^{-3}$    & $0.3$--$30$ ($0.3$, $1$, $3$, $10$ or $30$) \\
surface charge density     & $\sigma$   & mC m$^{-2}$     & $-(0.1$--$20)$ ($-1$ or $-10$) \\
concentration difference   & $\Delc$    & mol~m$^{-3}$    & $(0.01$--$18/11) \cinf$ \\
solution density           & $\rho_{\mathrm{w}}$ & g m$^{-2}$ & 1 \\
dielectric constant        & $\epsilon$ & --              & $78.46$ \\ 
temperature                & $T$        & K               & $298$ \\
\hline
\end{tabular*}
\end{table}

We also carried out parameter sweeps over $a$ at fixed $\sigma = -20$~mC~m$^{-2}$ and $\cinf = 10,30$~mol~m$^{-3}$ to obtain simulations in the $\lDinf \ll a$ regime for $Ze|\psi|  = 2.6\kBT$ and  $Ze|\psi| =1.7 \kBT$, respectively. In addition, we carried out simulations of electric-field-driven flow using the same model and parameters as described above, where potential differences of $\Delta \psi = 0$--$70$~mV were applied instead of a concentration difference. 

\begin{figure}[!h]
\centering
\begin{subfigure}[b]{.495\textwidth}
\includegraphics[width=\textwidth]{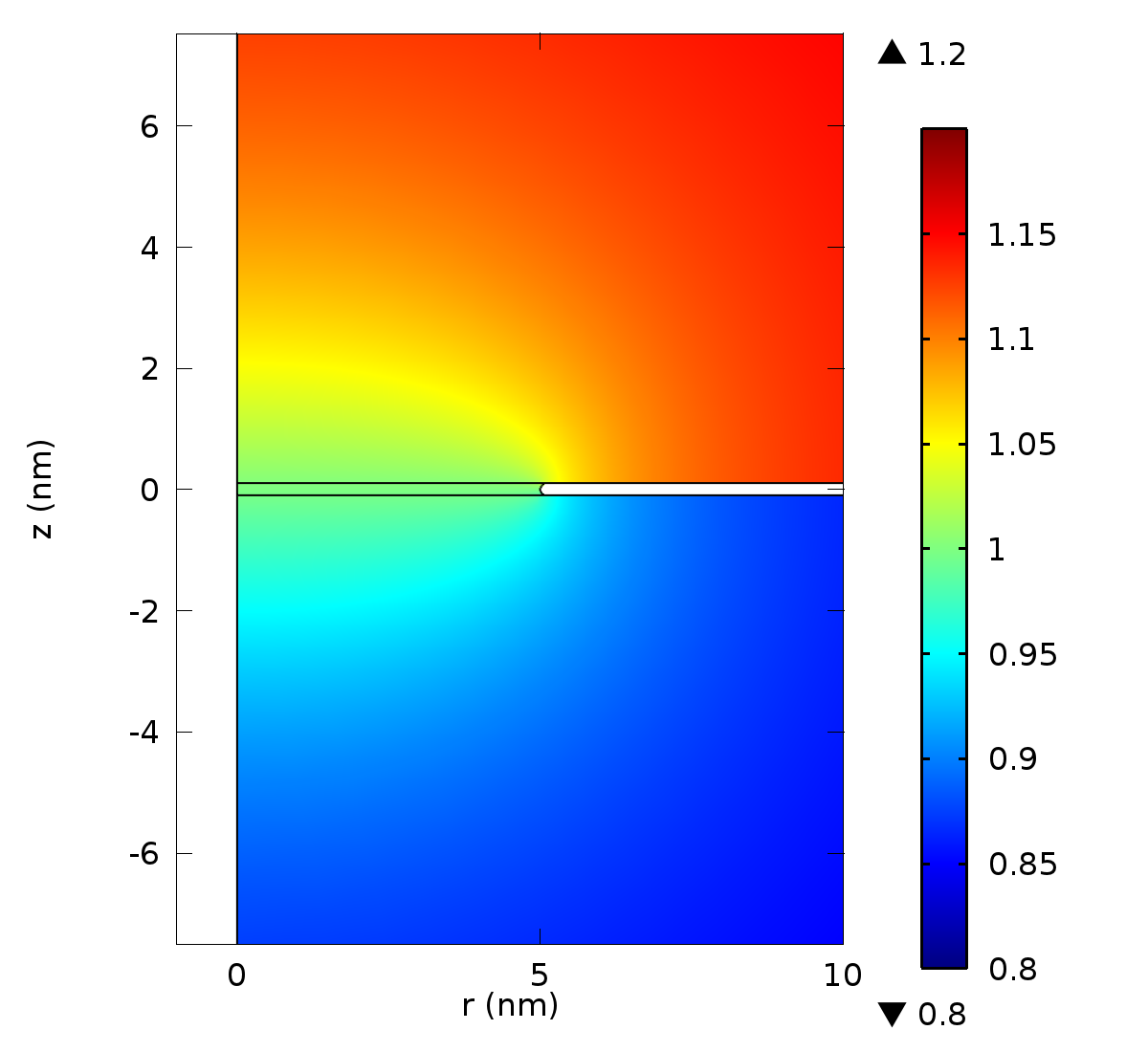}
\caption{}
\end{subfigure}
\begin{subfigure}[b]{.495\textwidth}
\includegraphics[width=\textwidth]{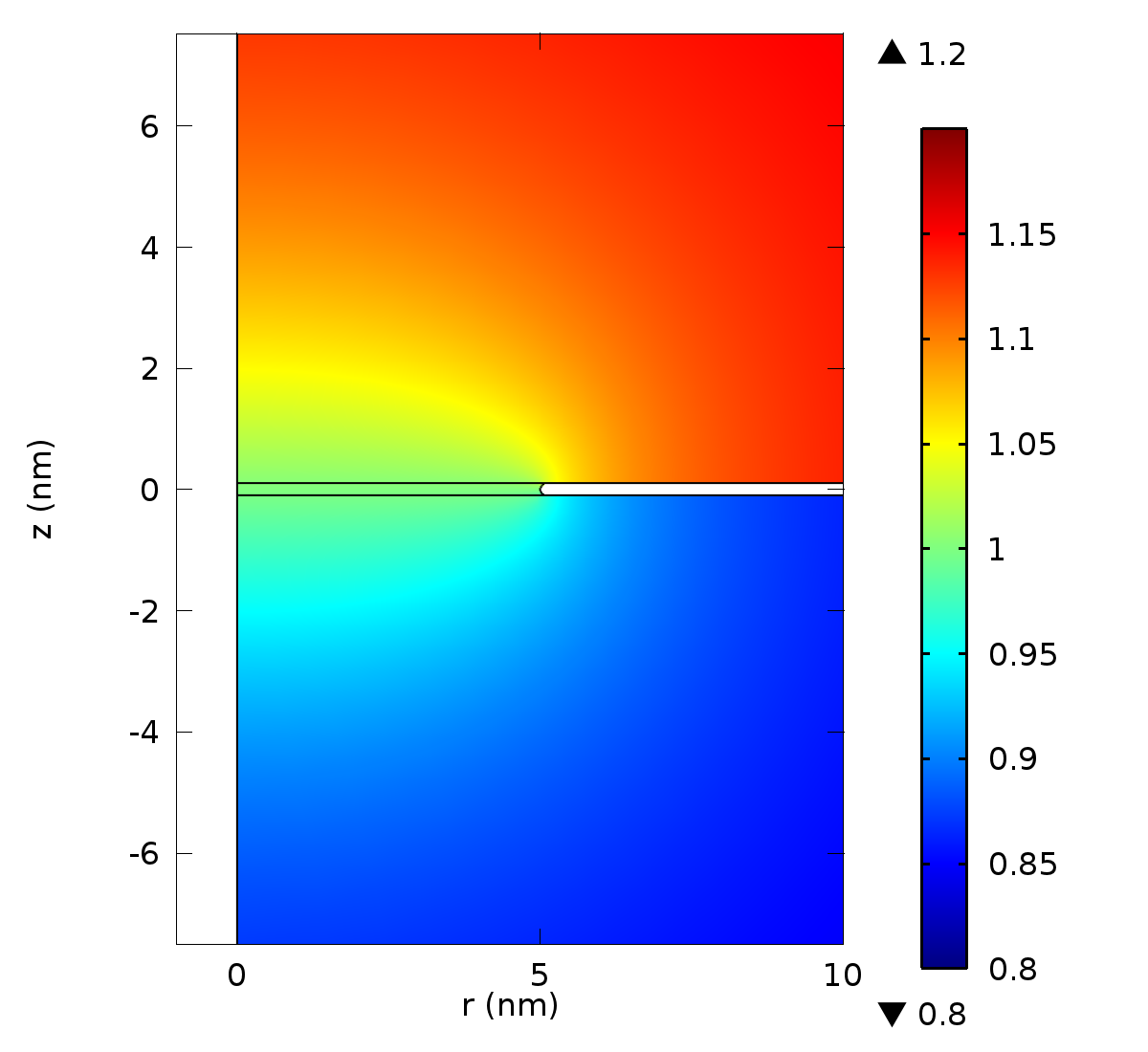}
\caption{}
\end{subfigure}

\begin{subfigure}[b]{.495\textwidth}
\includegraphics[width=\textwidth]{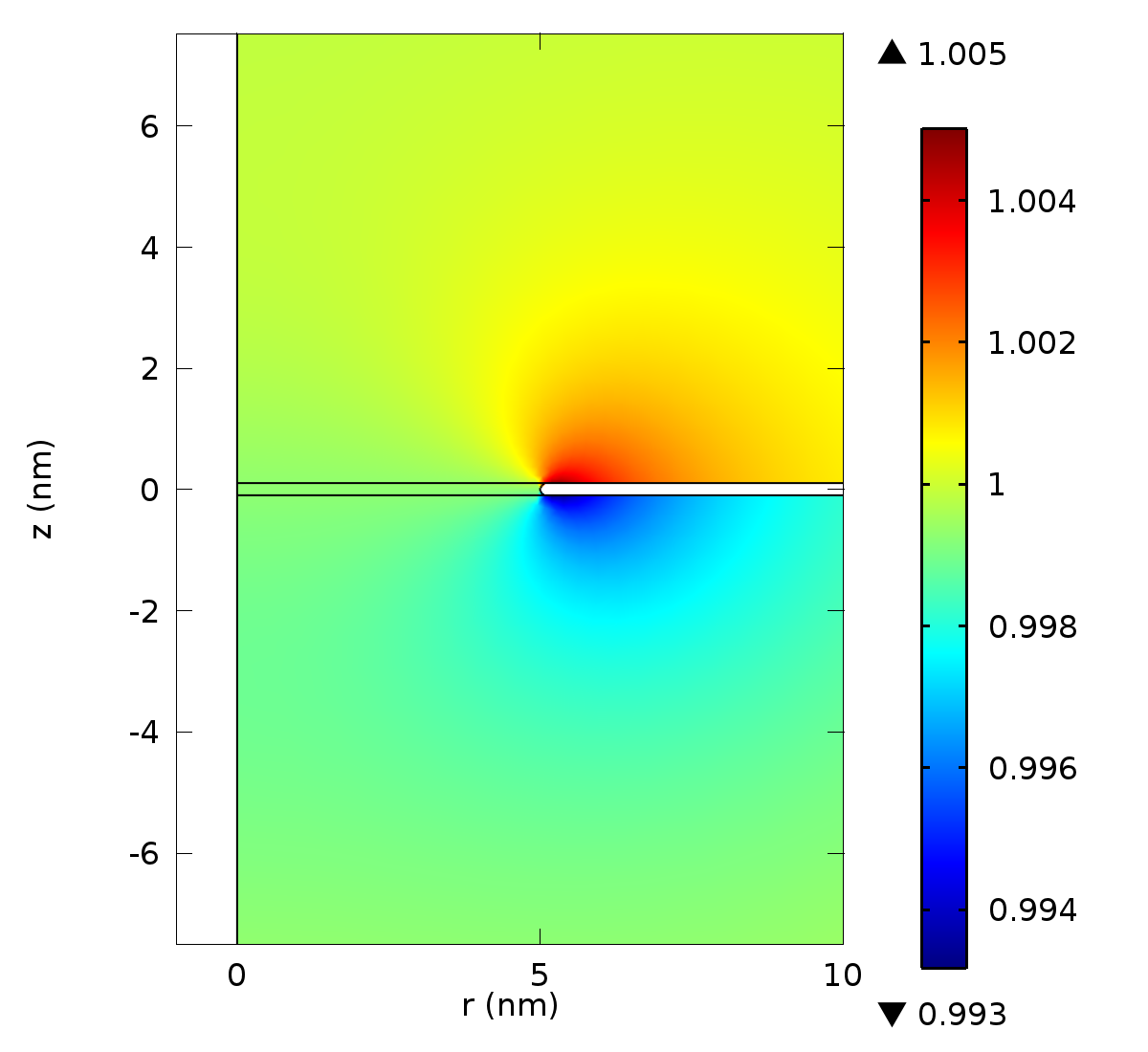}
\caption{}
\end{subfigure}
\begin{subfigure}[b]{.495\textwidth}
\includegraphics[width=\textwidth]{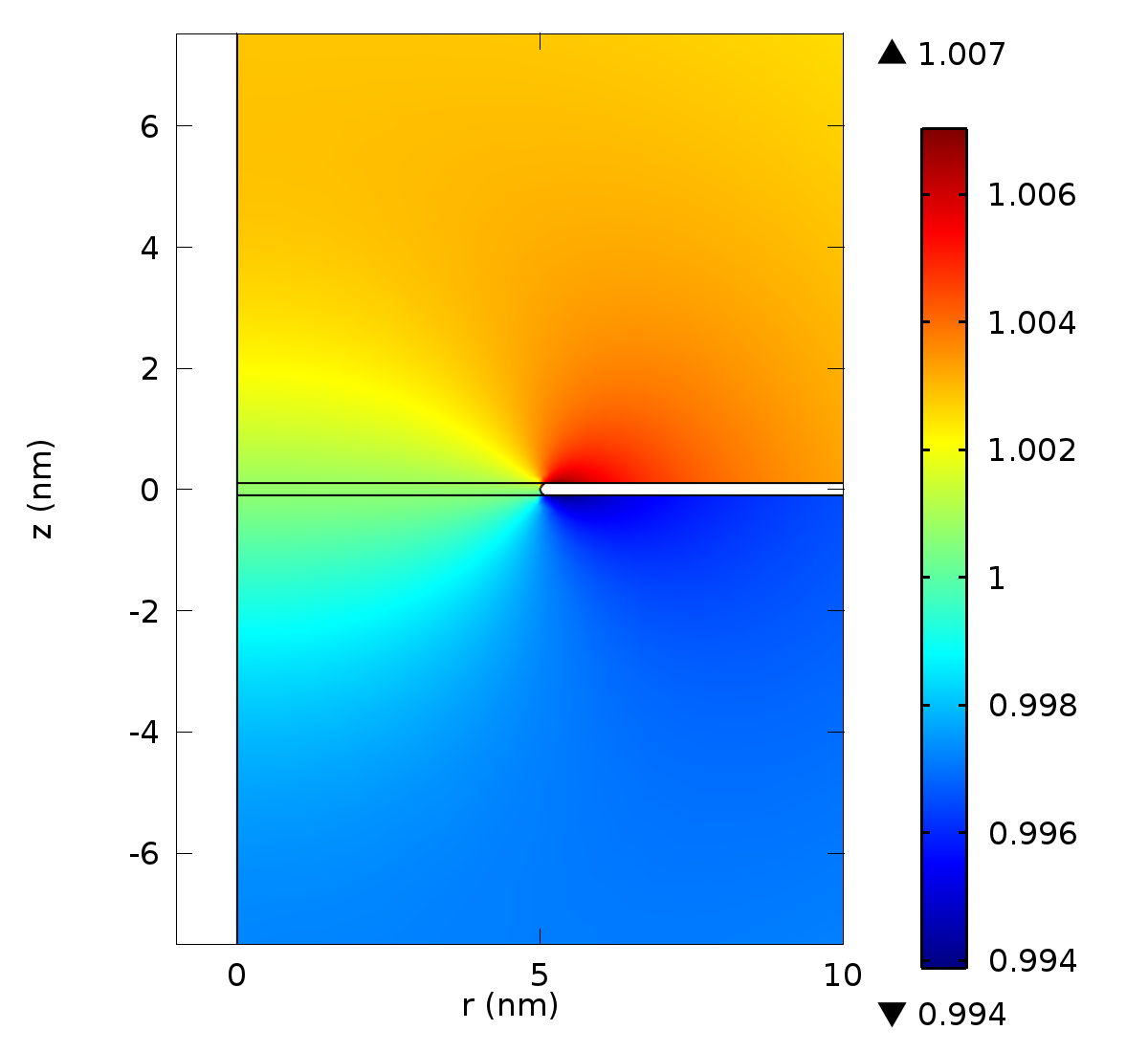}
\caption{}
\end{subfigure}
    \caption{\label{cs1} 2D plots of ((a),(b)) $\hat{c}_{\mathrm{s}}^{(i)}$ from the FEM simulations (Eq.~\eqref{cs_hat_simulations}) for the positive and negative ions, respectively, and ((c),(d)) $\hat{c}_{\mathrm{s}}^{(i)} / \hat{c}_{\mathrm{s}_{\mathrm{theory}}}$ for the positive and negative ions, respectively, for $\hat{c}_{\mathrm{s}_{\mathrm{theory}}}$ as derived in the theory (Eq.~\eqref{cs_hat}), as a function of the $r$ and $z$ coordinates near the pore entrance for a concentration difference $\Delc$ of $0.4 \cinf$ and a minimum value (maximum magnitude) of the electric potential of $-2.8$~mV near the membrane surface in the lower concentration reservoir ($Ze|\psi| \approx \kBT/10$).}
\end{figure}

\begin{figure}[!h]
\centering
\begin{subfigure}[b]{.495\textwidth}
\includegraphics[width=\textwidth]{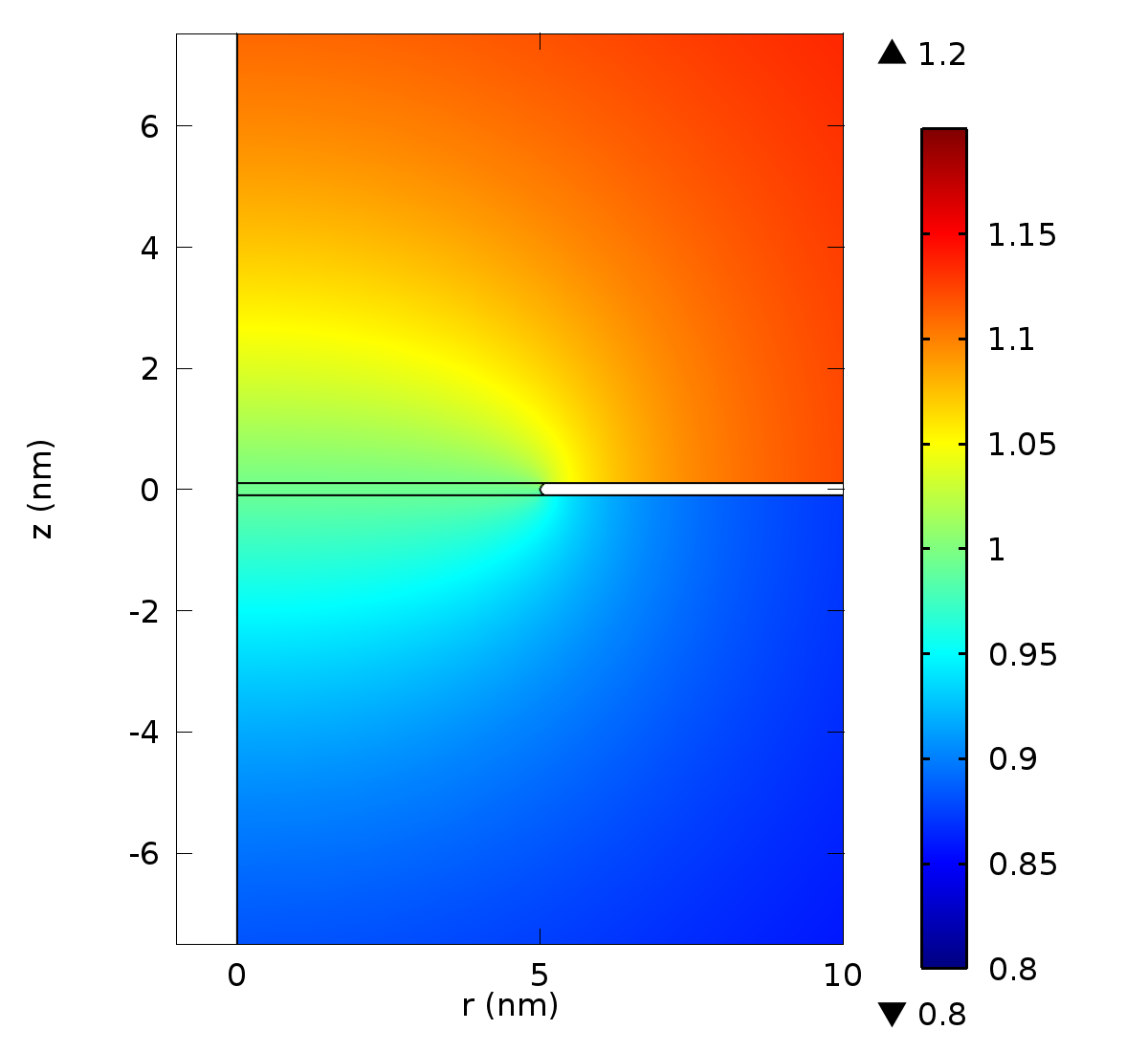}
\caption{}
\end{subfigure}
\begin{subfigure}[b]{.495\textwidth}
\includegraphics[width=\textwidth]{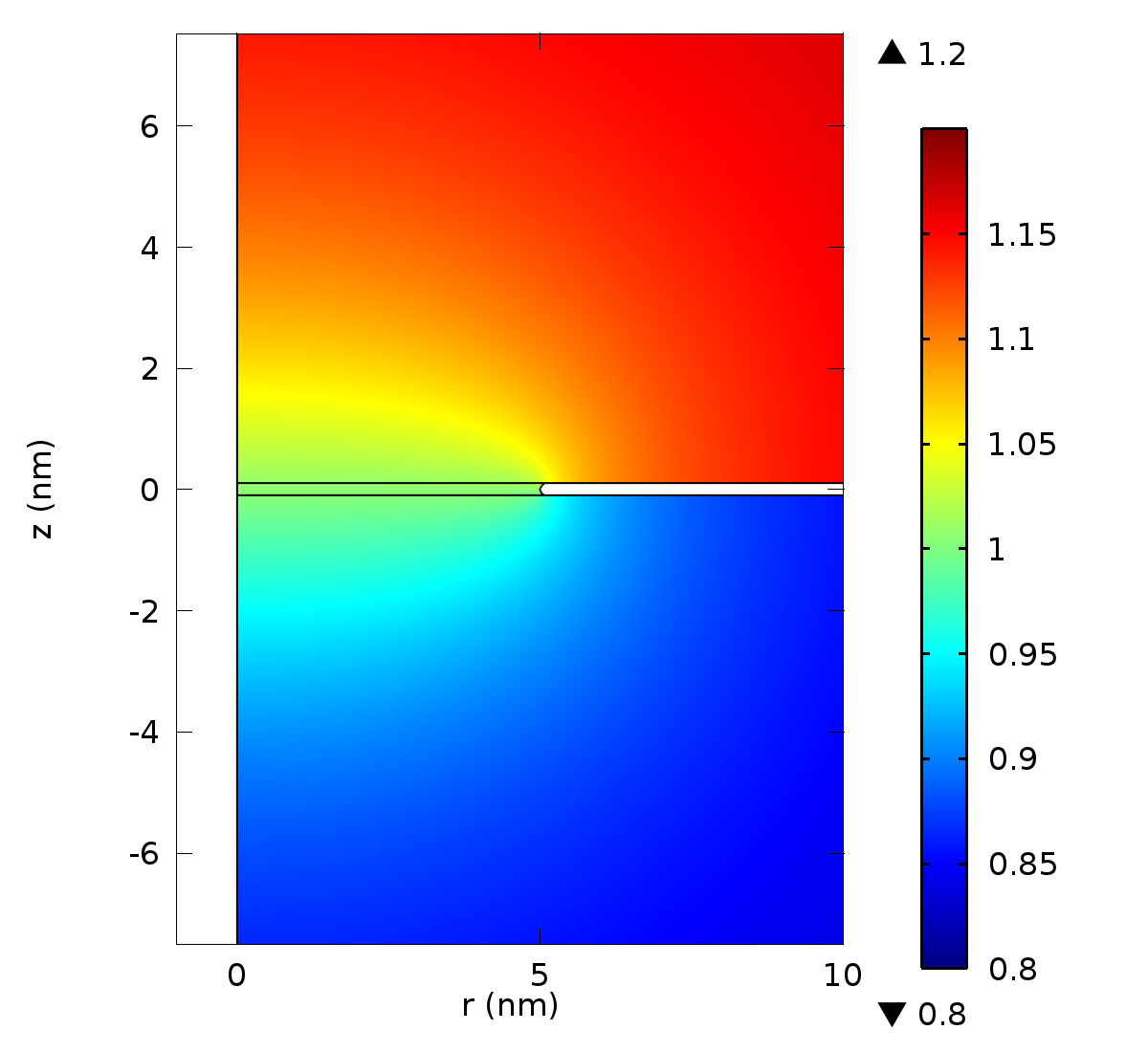}
\caption{}
\end{subfigure}

\begin{subfigure}[b]{.495\textwidth}
\includegraphics[width=\textwidth]{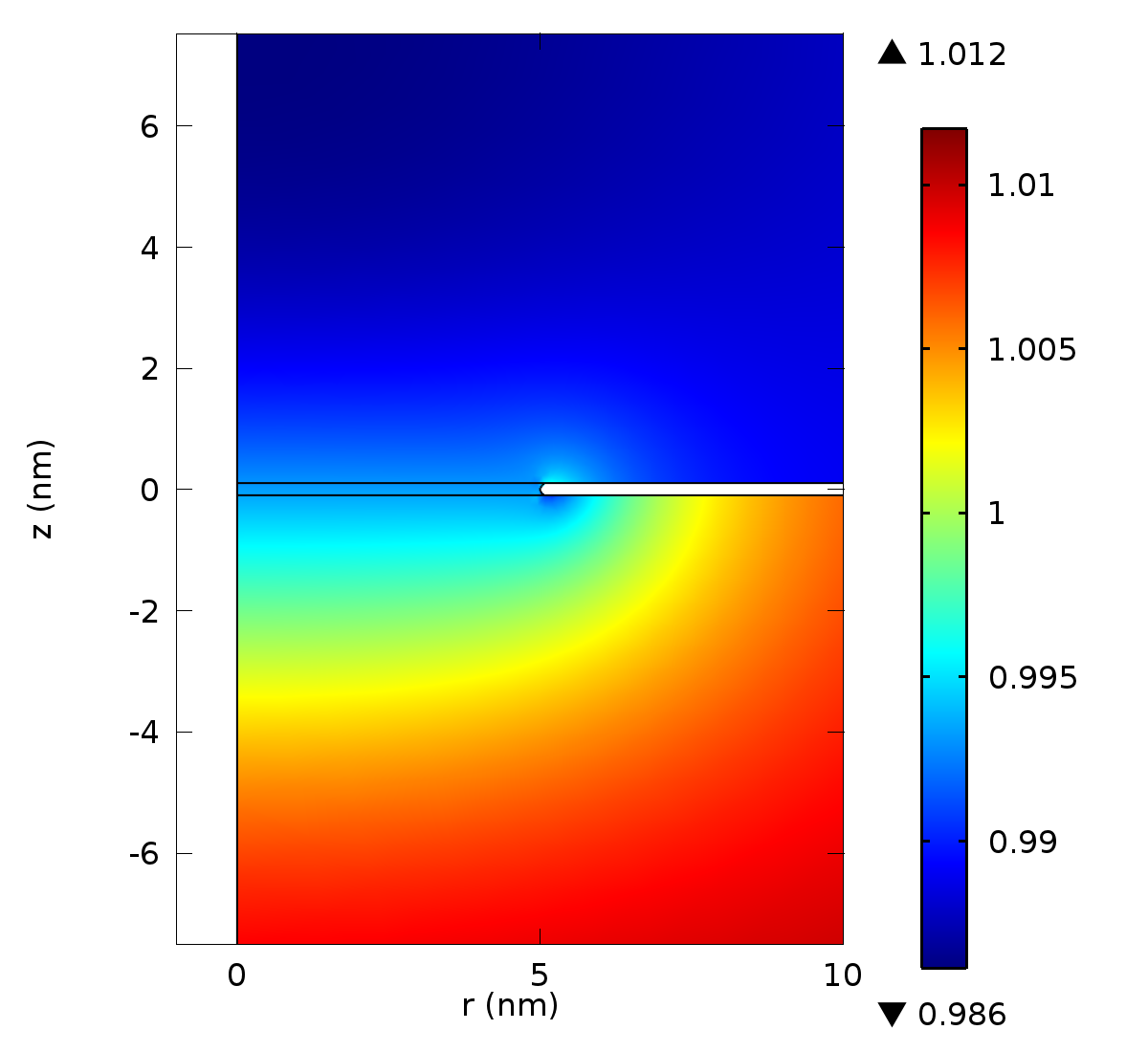}
\caption{}
\end{subfigure}
\begin{subfigure}[b]{.495\textwidth}
\includegraphics[width=\textwidth]{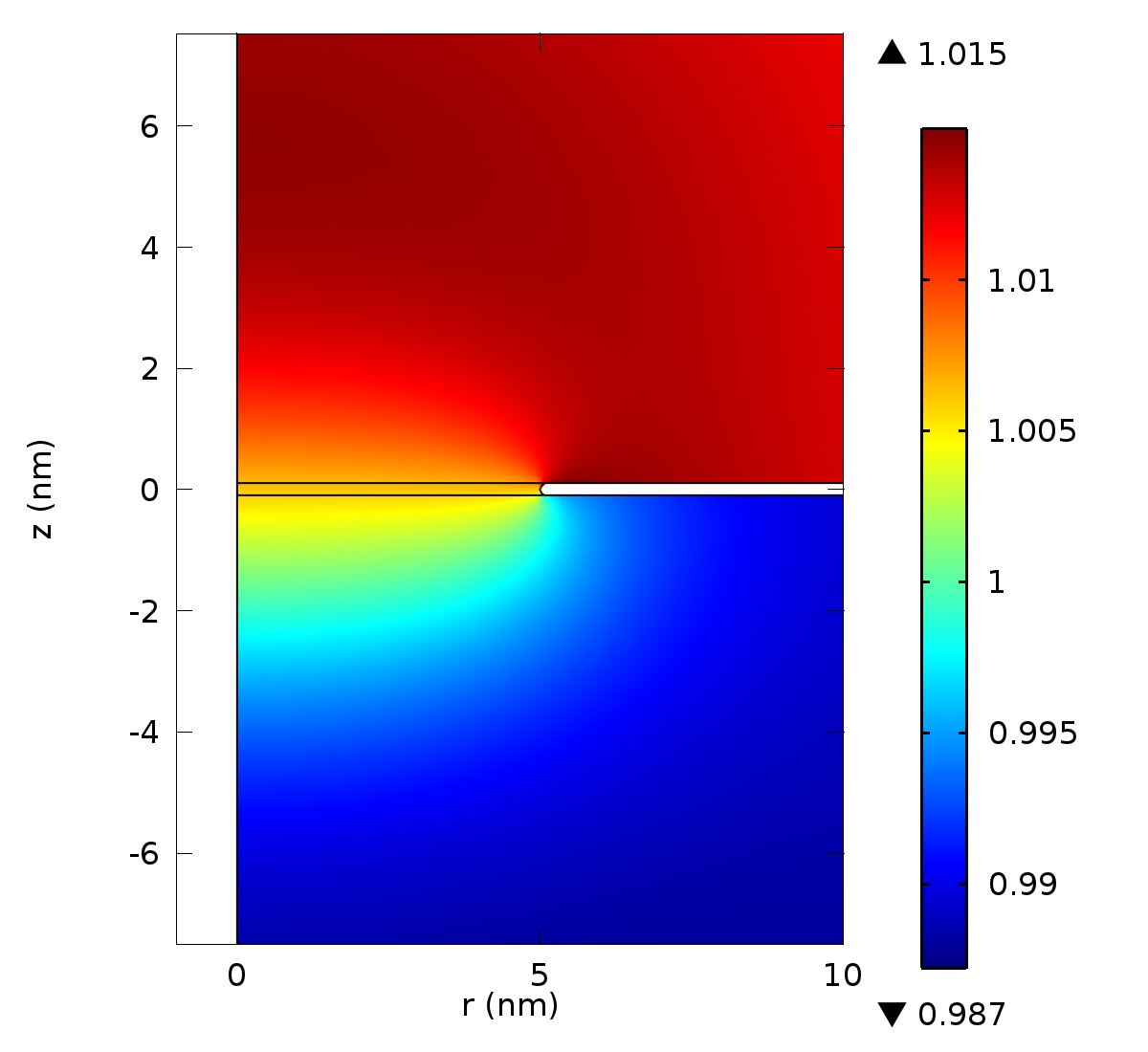}
\caption{}
\end{subfigure}
    \caption{\label{cs2} 2D plots of ((a),(b)) $\hat{c}_{\mathrm{s}}^{(i)}$ from the FEM simulations (Eq.~\eqref{cs_hat_simulations}) for the positive and negative ions, respectively, and ((c),(d)) $\hat{c}_{\mathrm{s}}^{(i)} / \hat{c}_{\mathrm{s}_{\mathrm{theory}}}$ for the positive and negative ions, respectively, for $\hat{c}_{\mathrm{s}_{\mathrm{theory}}}$ as derived in the theory (Eq.~\eqref{cs_hat}), as a function of the $r$ and $z$ coordinates near the pore entrance for a concentration difference $\Delc$ of $0.4 \cinf$ and  a minimum value (maximum magnitude) of the electric potential of $-27$ mV near the membrane surface in the lower concentration reservoir ($Ze|\psi| \approx \kBT$).}
\end{figure}

\begin{figure}[!h]
\centering
\begin{subfigure}[b]{.495\textwidth}
\includegraphics[width=\textwidth]{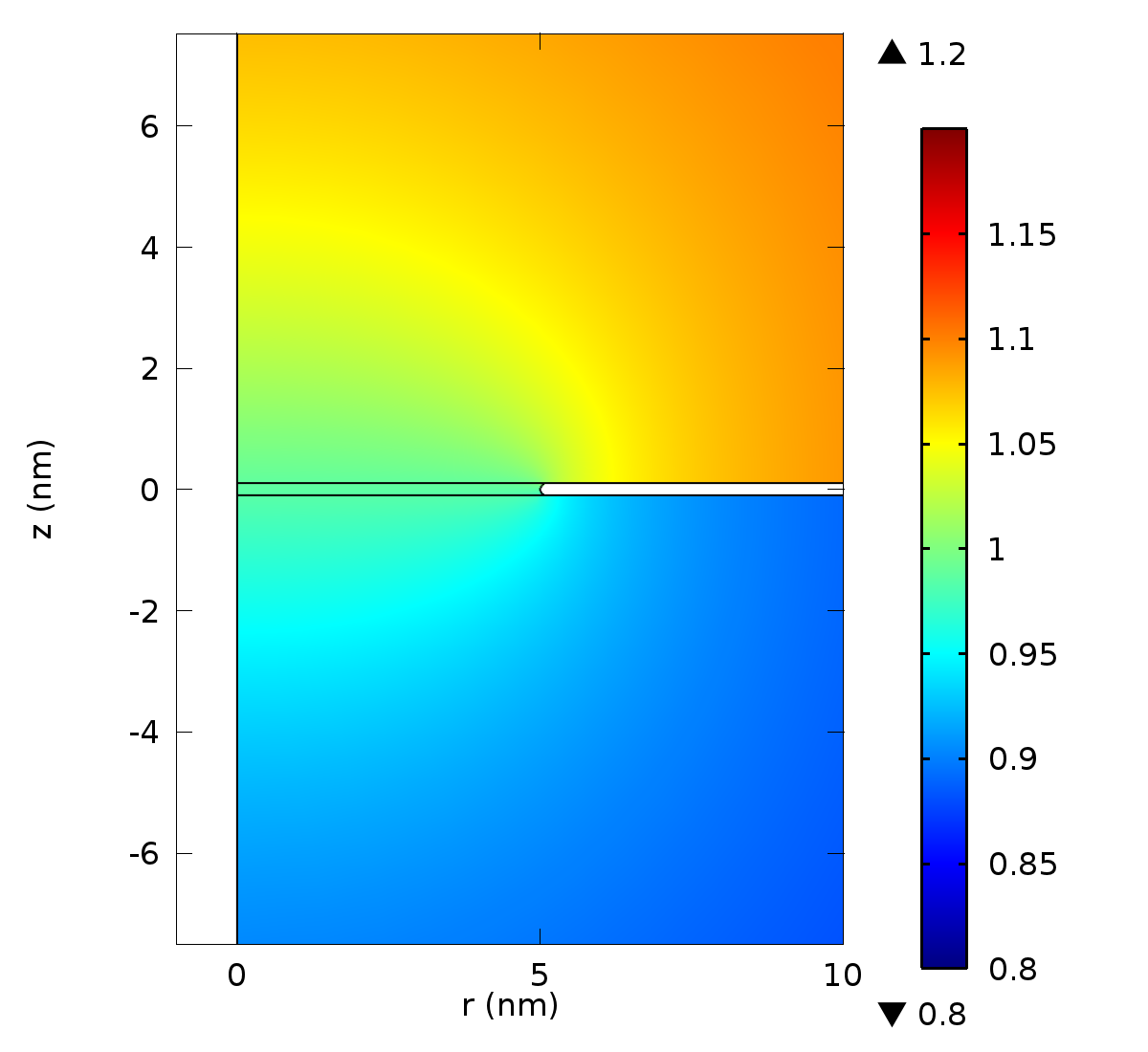}
\caption{}
\end{subfigure}
\begin{subfigure}[b]{.495\textwidth}
\includegraphics[width=\textwidth]{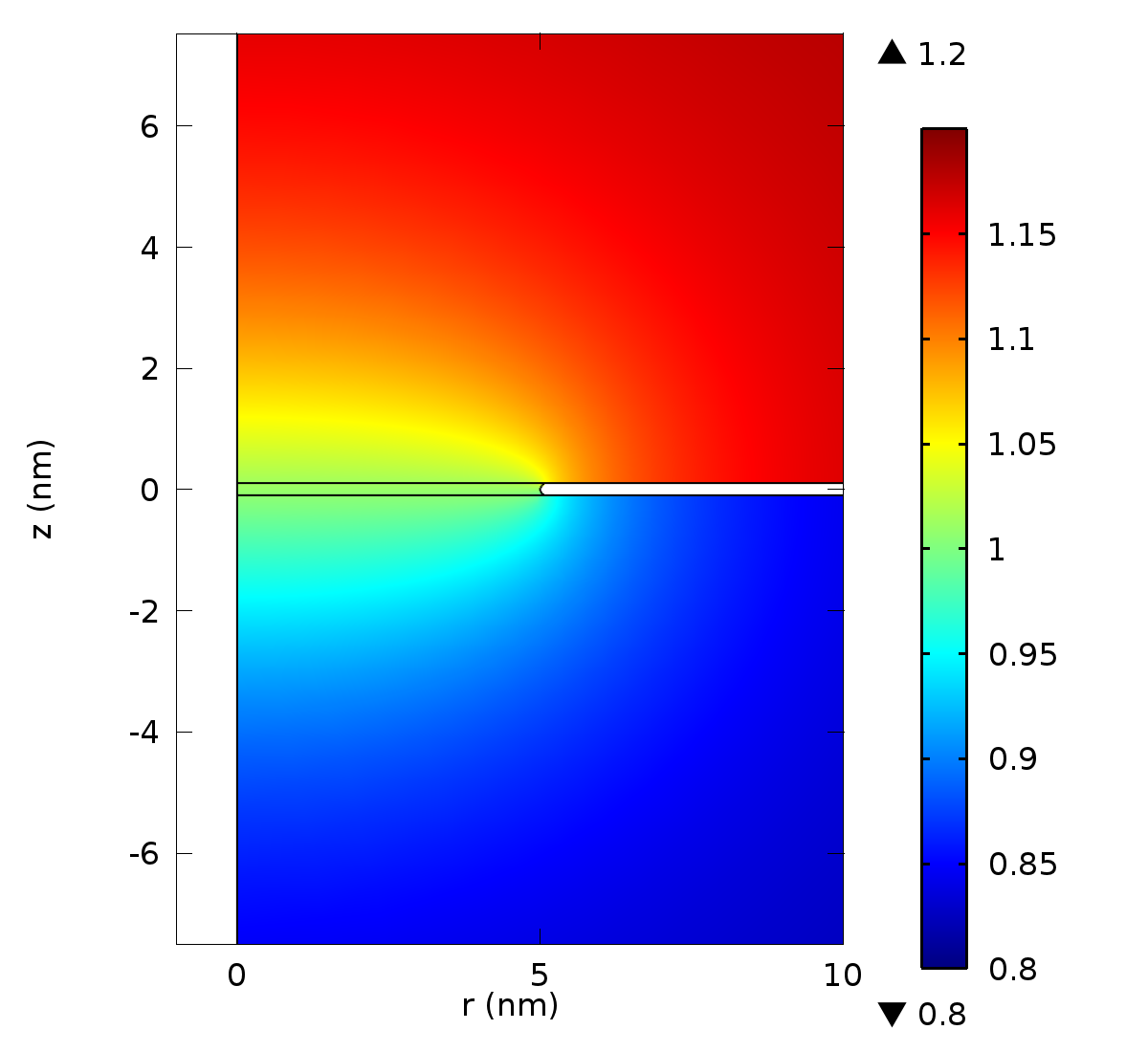}
\caption{}
\end{subfigure}

\begin{subfigure}[b]{.495\textwidth}
\includegraphics[width=\textwidth]{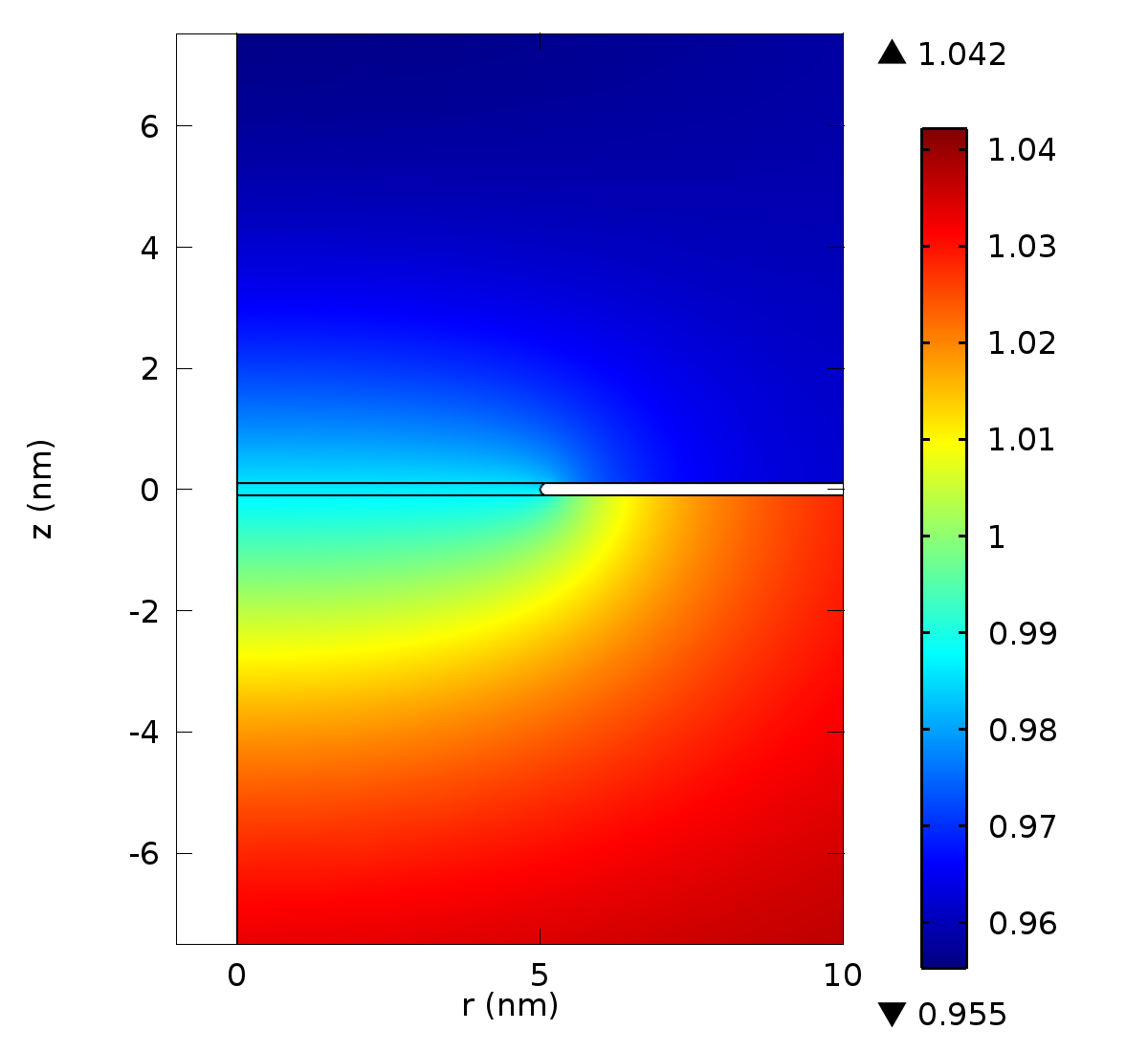}
\caption{}
\end{subfigure}
\begin{subfigure}[b]{.495\textwidth}
\includegraphics[width=\textwidth]{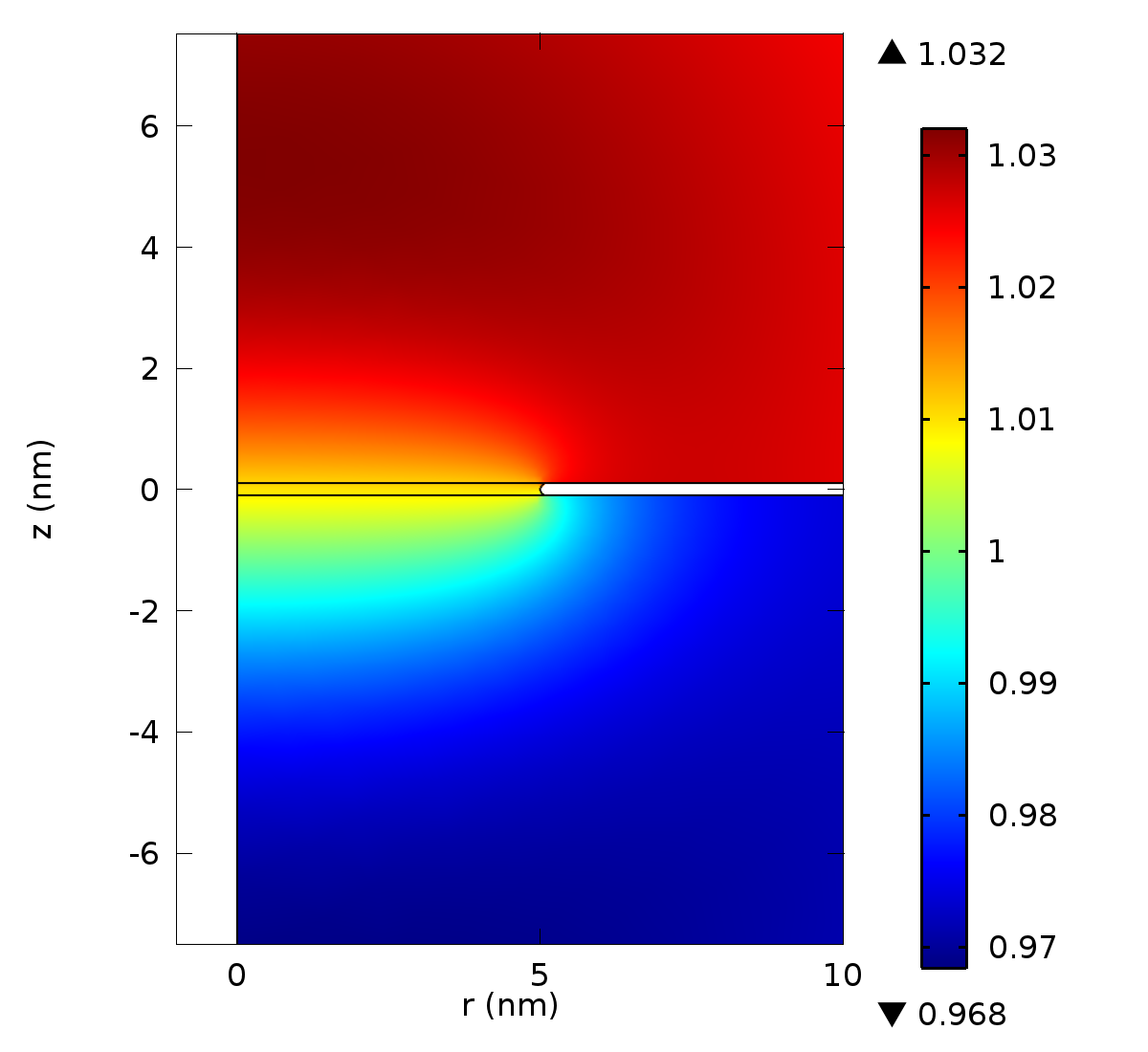}
\caption{}
\end{subfigure}
    \caption{\label{cs3} 2D plots of ((a),(b)) $\hat{c}_{\mathrm{s}}^{(i)}$ from the FEM simulations (Eq.~\eqref{cs_hat_simulations}) for the positive and negative ions, respectively, and ((c),(d)) $\hat{c}_{\mathrm{s}}^{(i)} / \hat{c}_{\mathrm{s}_{\mathrm{theory}}}$ for the positive and negative ions, respectively, for $\hat{c}_{\mathrm{s}_{\mathrm{theory}}}$ as derived in the theory (Eq.~\eqref{cs_hat}), as a function of the $r$ and $z$ coordinates near the pore entrance for a concentration difference $\Delc$ of $0.4 \cinf$ and  a minimum value (maximum magnitude) of the electric potential of $-78$ mV near the membrane surface in the lower concentration reservoir ($Ze|\psi| \approx 3\kBT$).}
\end{figure}

In Figs.~\ref{cs1}--\ref{cs3}, we verify the form of the electrolyte concentration distribution derived in the theory (Eqs.~\eqref{Boltzmann_dist} and \eqref{cs_solution} in the main paper) near the pore mouth for a range of surface charge densities $\sigma$ at a fixed pore radius of $5$~nm, solute concentration of $0.3$~mol~m$^{-3}$ and concentration difference of $\Delc = 0.4 \cinf$, which correspond to electric potential energies with magnitude up to $3 \kBT$ (potential for which the magnitude of the maximum deviation between $c_\mathrm{s}^{(i)}$ calculated from the simulations and $c_\mathrm{s}$ derived in the theory is less than 5\%). Similar results to those shown in Figs.~\ref{cs1}--\ref{cs3} were obtained for other parameter combinations that resulted in similar values for the electric potential.

\clearpage

\begin{figure}[!h]
    \centering
    \includegraphics[width=0.495\textwidth]{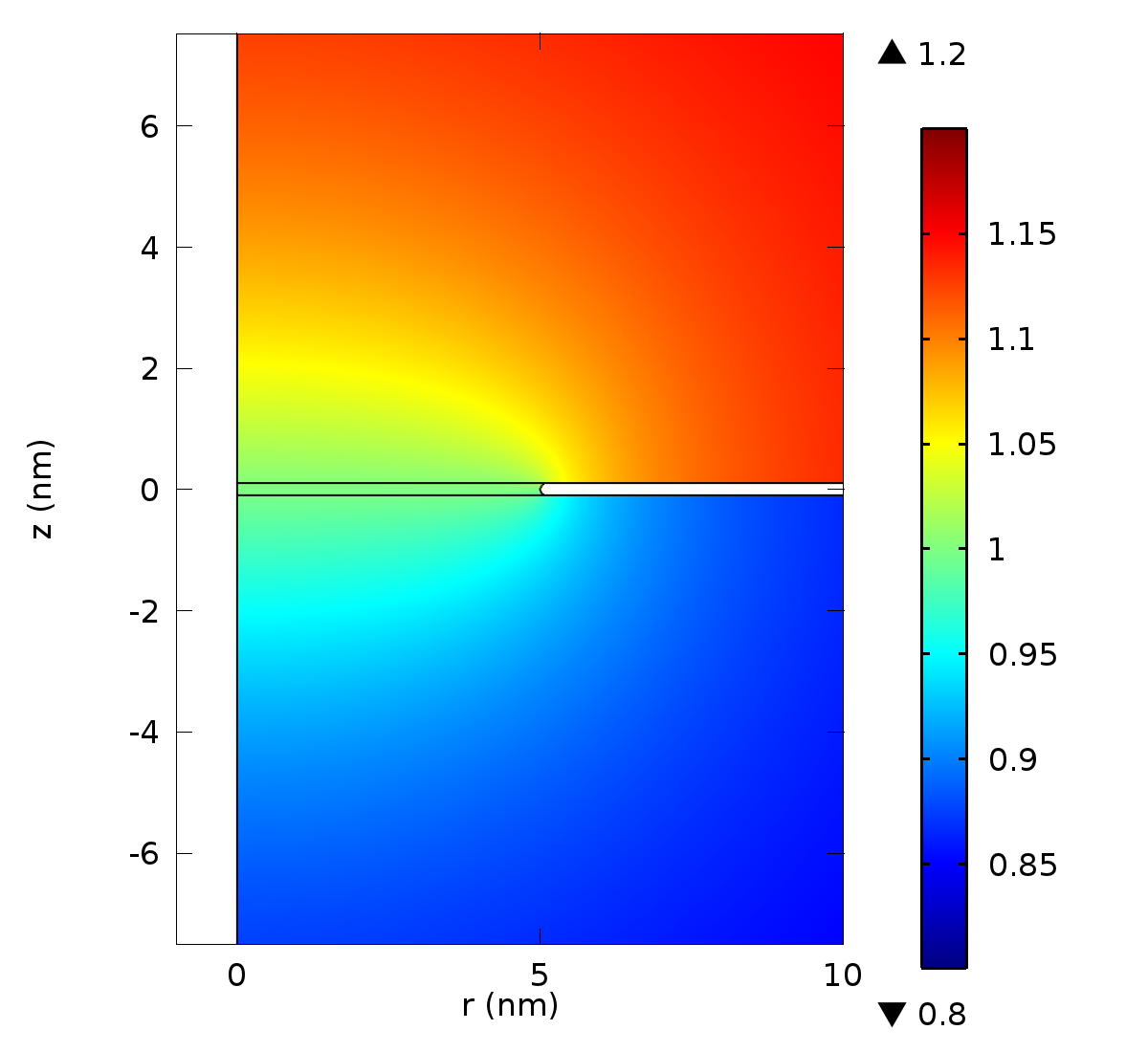}
    \caption{2D plot of the analytical expression for the dimensionless solute concentration of a $Z$:$Z$ electrolyte at an electric potential of zero, $\hat{c}_{\mathrm{s}}$, as a function of the $r$ and $z$ coordinates near the pore entrance for a concentration difference $\Delc$ of $0.4 \cinf$.}
    \label{cs4}
\end{figure}

From Eqs.~\eqref{conc_perturb} and \eqref{cs_solution} of the main paper, for a $Z$:$Z$ electrolyte (for which $c_\mathrm{s}^+ = c_\mathrm{s}^- = c_\mathrm{s}$, $c_\infty^+ = c_\infty^- = c_\infty$, and $\Delta c_+ = \Delta c_- = \Delta c)$,
\begin{align}
    \cs = \cinf + \frac{\Delc}{\pi}\tan^{-1}{\nu} ,
\end{align}
such that 
\begin{align}
\label{cs_hat}
    \hat{c}_{\mathrm{s}} = 1 + \frac{\Delta \hat{c}}{\pi}\tan^{-1}{\nu} ,
\end{align}
where $\Delta \hat{c} = \Delc / \cinf$, $\Delc$ is the applied concentration difference, and $\cinf$ is the average bulk solute concentration. Equation~\eqref{cs_hat} is plotted in Fig.~\ref{cs4} for a pore radius of $5$~nm.

From Eqs.~\eqref{Boltzmann_dist} and \eqref{conc_perturb} in the main paper,
\begin{align}
\label{cs_hat_simulations}
    \hat{c}_{\mathrm{s}}^{(i)} = \frac{c^{(i)}}{\cinf} \exp{\left(\frac{Z_i e \psi}{\kBT}\right)} 
\end{align}
can be calculated from the FEM simulations, where $c^{(i)}$ is the concentration of species $i$.

\clearpage

\clearpage

\clearpage

\clearpage

\subsection{Scaling of concentration-gradient-driven fluid fluxes with the applied concentration difference} \label{sec:Delc}

\begin{figure}[!h]
 \centering
 \begin{minipage}{.5\textwidth}
   \centering
    \includegraphics[scale = 1, trim={0.1cm 0.0875cm 0.1cm 0.0875cm},clip]{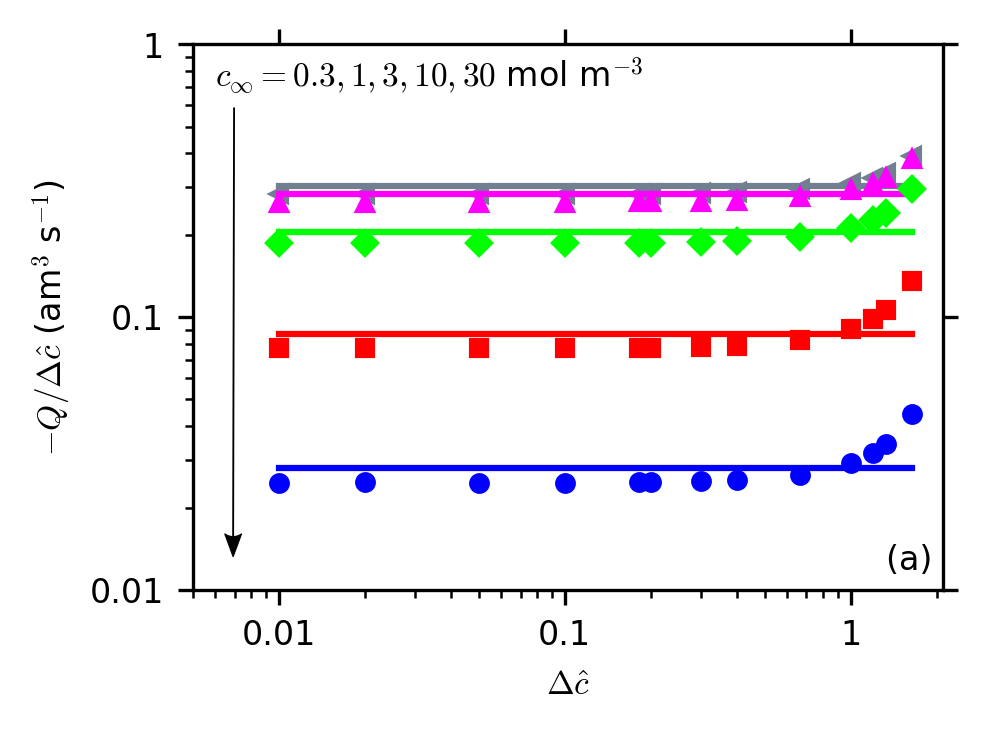}
    \includegraphics[scale = 1, trim={0.1cm 0.0875cm 0.1cm 0.0875cm},clip]{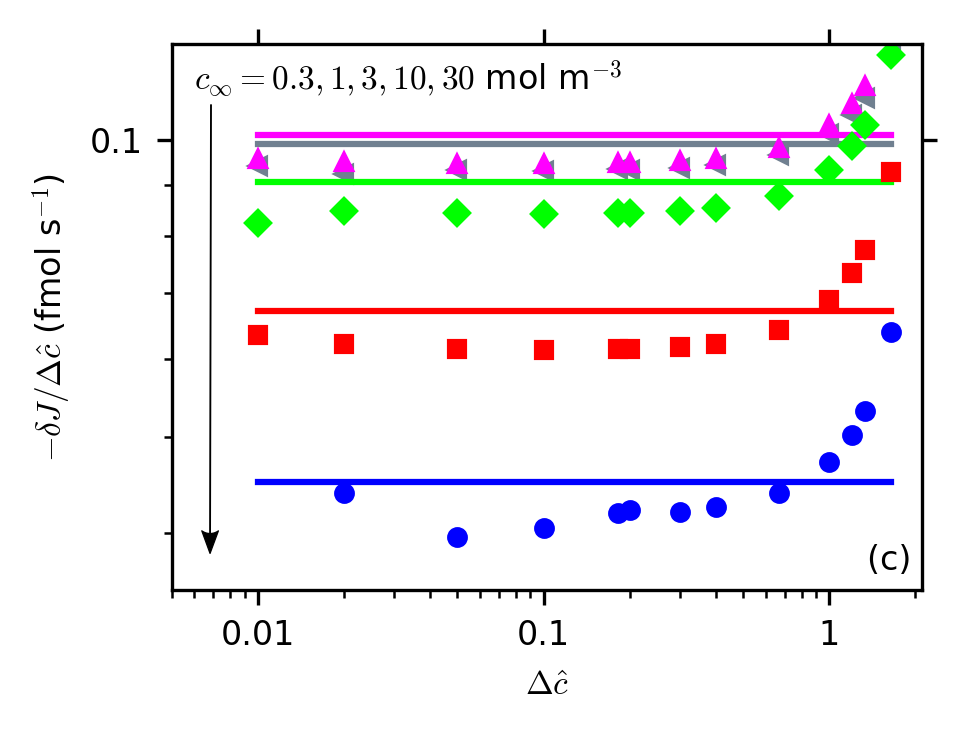}
    \includegraphics[scale = 1, trim={0.125cm 0.0875cm 0.125cm 0.0875cm},clip]{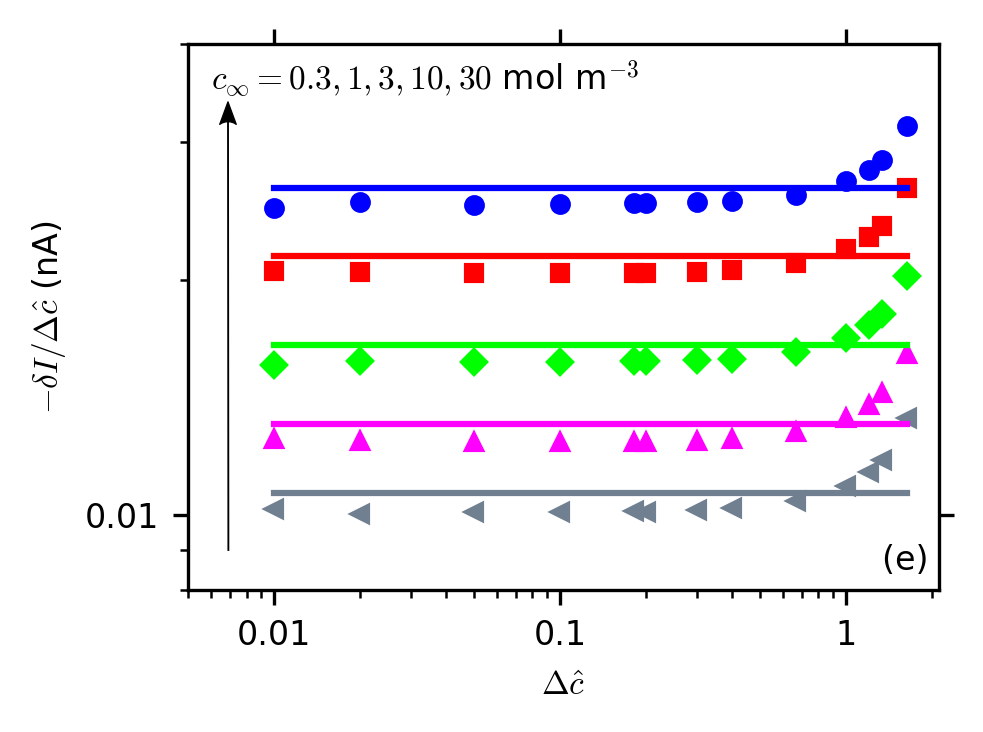}
  \end{minipage}%
 \begin{minipage}{.5\textwidth}
   \centering
    \includegraphics[scale = 1, trim={0.1cm 0.0875cm 0.1cm 0.0875cm},clip]{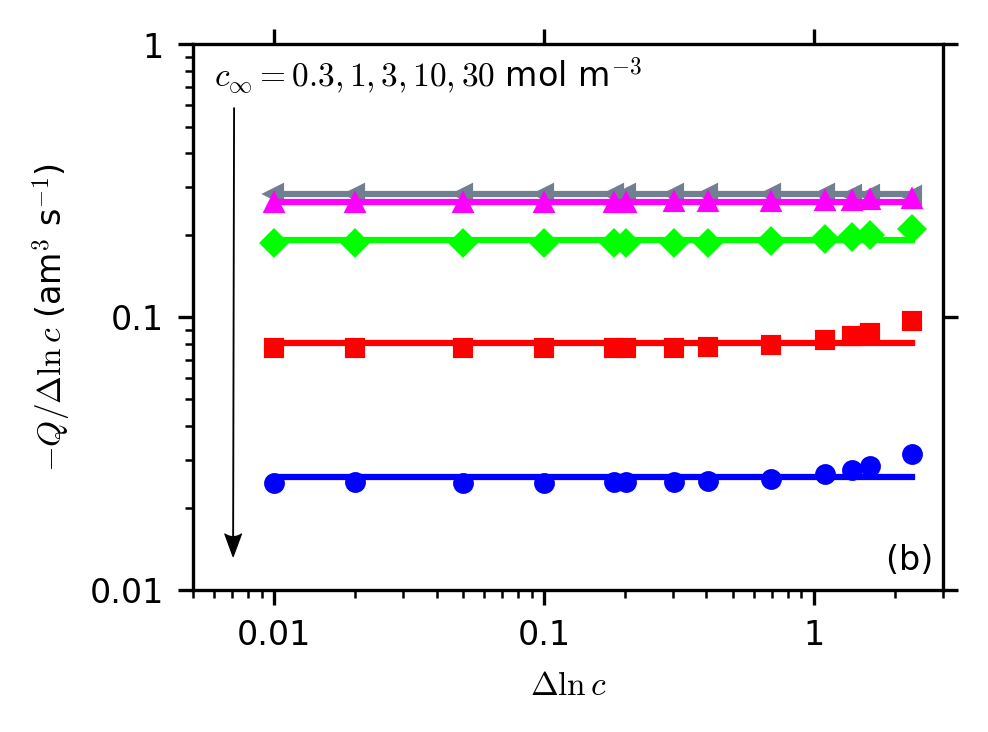}
    \includegraphics[scale = 1, trim={0.1cm 0.0875cm 0.1cm 0.0875cm},clip]{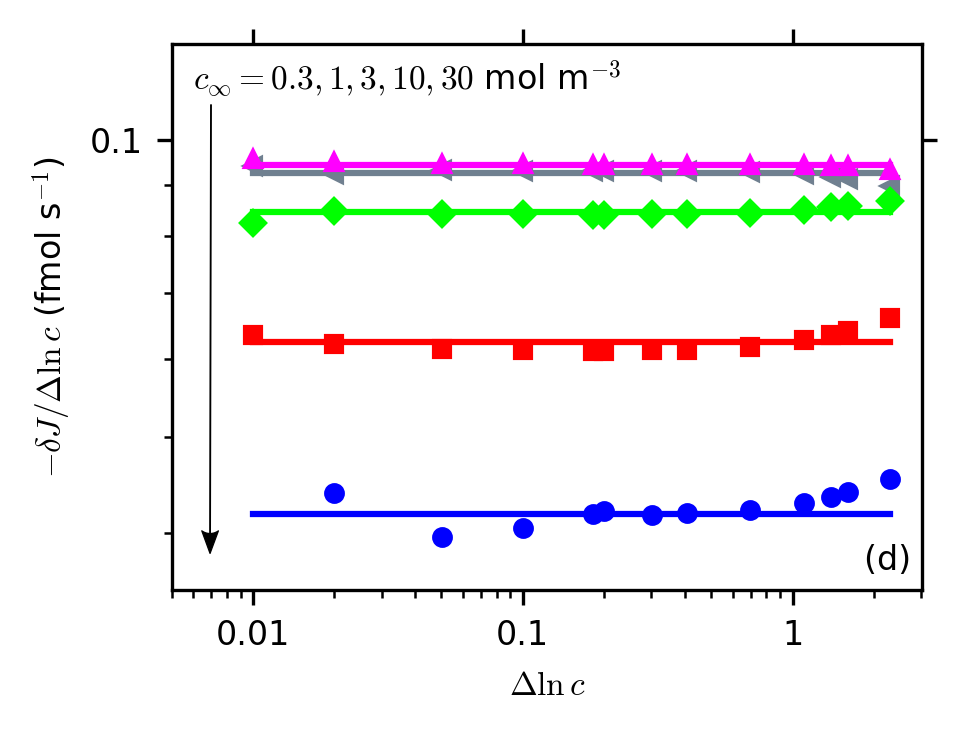}
    \includegraphics[scale = 1, trim={0.1cm 0.0875cm 0.1cm 0.0875cm},clip]{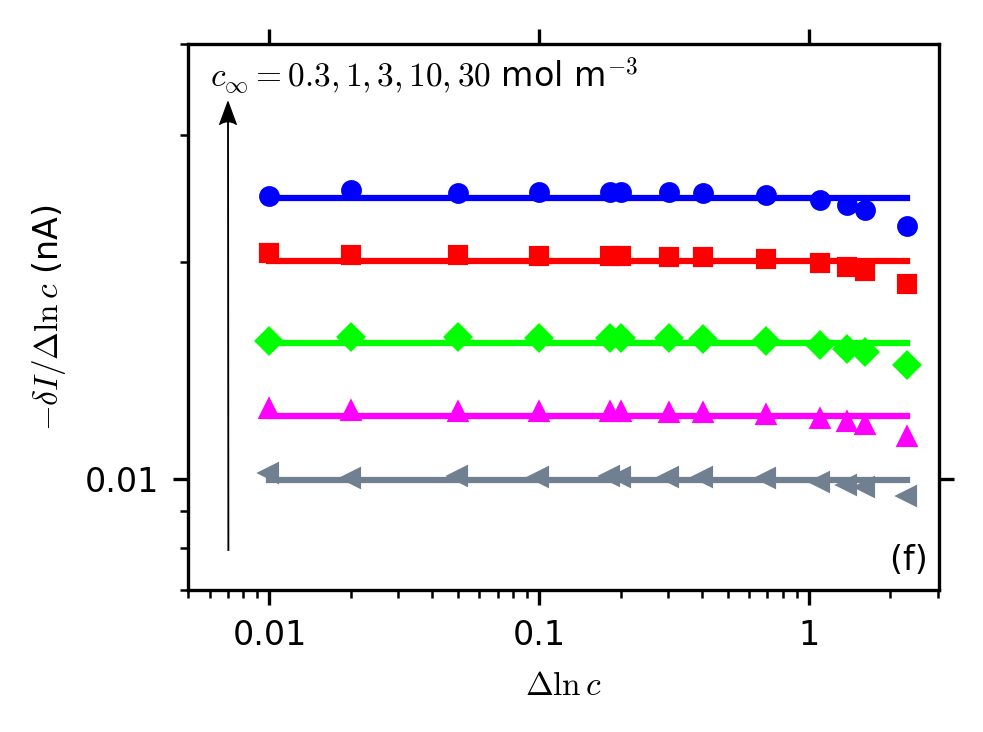}
  \end{minipage}
    \caption{\label{fig:Delc3} (a) $Q/\Delta \hat{c}$ vs $ \Delta \hat{c}$, (b) $Q/\Delta \ln{c}$ vs $\ln{c}$, (c) $\delJ/ \Delta \hat{c}$ vs $\Delta \hat{c}$, (d) $\delJ/ \Delta \ln{c}$ vs $\Delta \ln{c}$ and (d) $\delI/ \Delta \hat{c}$ vs $\Delta \hat{c}$, (d) $\delI/ \Delta \ln{c}$ vs $\Delta \ln{c}$. The points are the simulation data and the lines are horizontal fits to the points used to verify linear scaling with $\Delta c$ or $\Dellnc$ for various $\sigma$, where $1/100 \leq \Delta \hat{c} \leq 18/11$.  Arrows indicate the direction of increasing $\cinf$, where $a$ and $\sigma$ were fixed at $5$~nm and $-10$~mC~m$^{-2}$, respectively. Note there is numerical error in $\delJ$ for large $\cinf$ as $\Jb$ dominates. Fluxes and parameters are as defined in the main paper.}
\end{figure}

Note that we exclude simulations for $\Delc < 0.1\cinf$ from analyses in the main paper and the supplementary material due to the large numerical error in $\delJ$ (Fig.~\ref{fig:Delc3}).

\begin{figure}[!h]
 \centering
 \begin{minipage}{.5\textwidth}
   \centering
    \includegraphics[scale = 1, trim={0.1cm 0.0875cm 0.1cm 0.0875cm},clip]{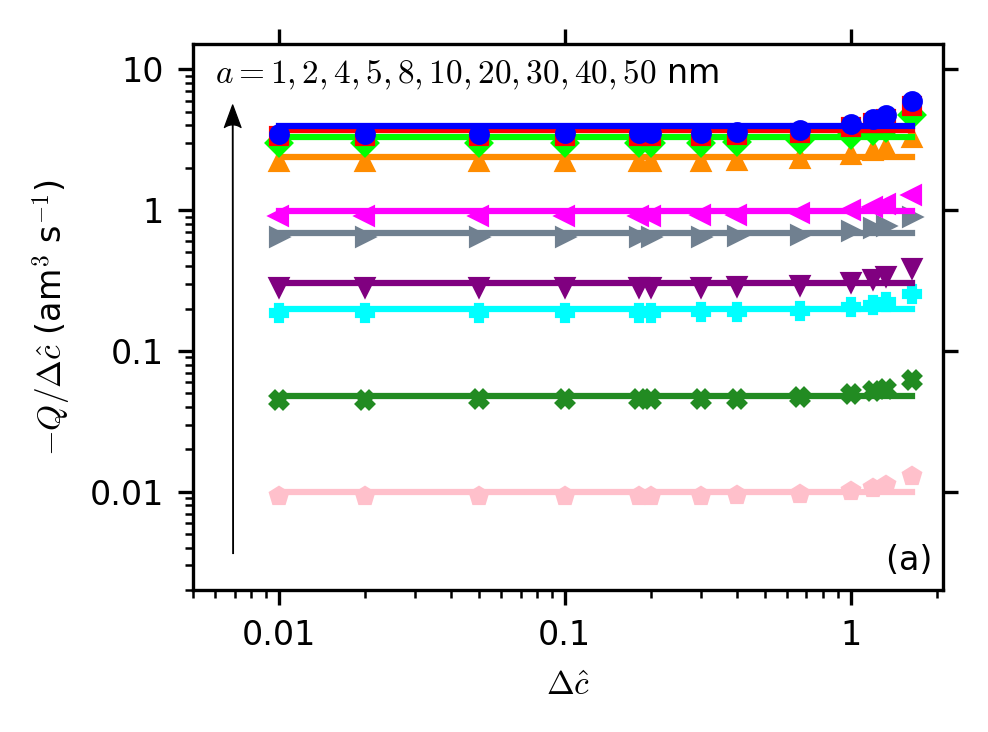}
    \includegraphics[scale = 1, trim={0.1cm 0.0875cm 0.1cm 0.0875cm},clip]{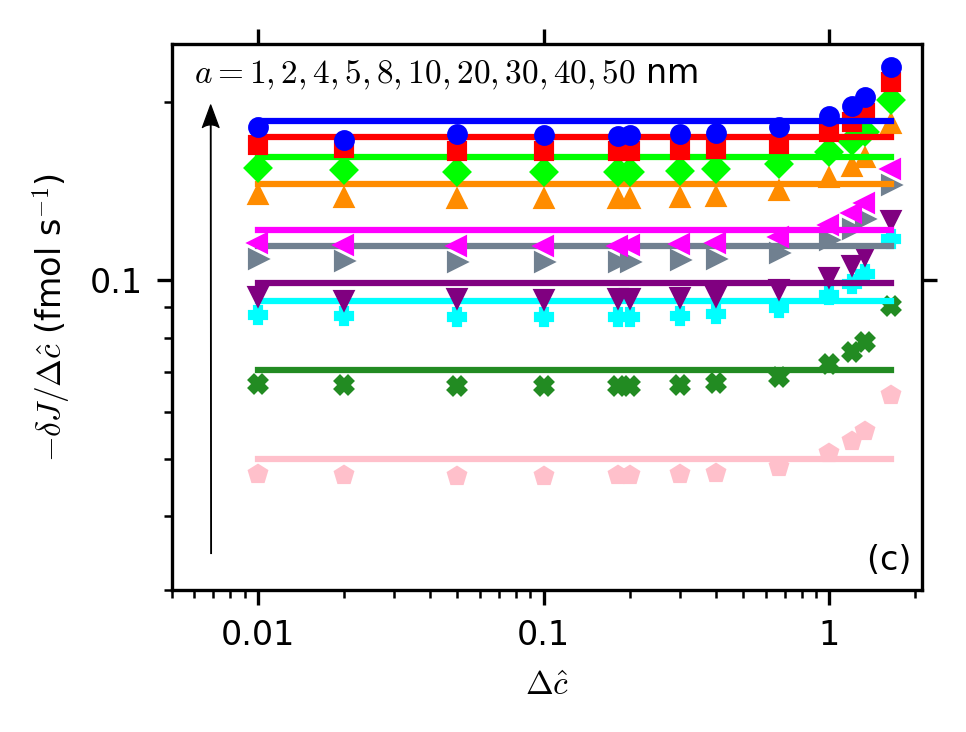}
    \includegraphics[scale = 1, trim={0.1cm 0.0875cm 0.1cm 0.0875cm},clip]{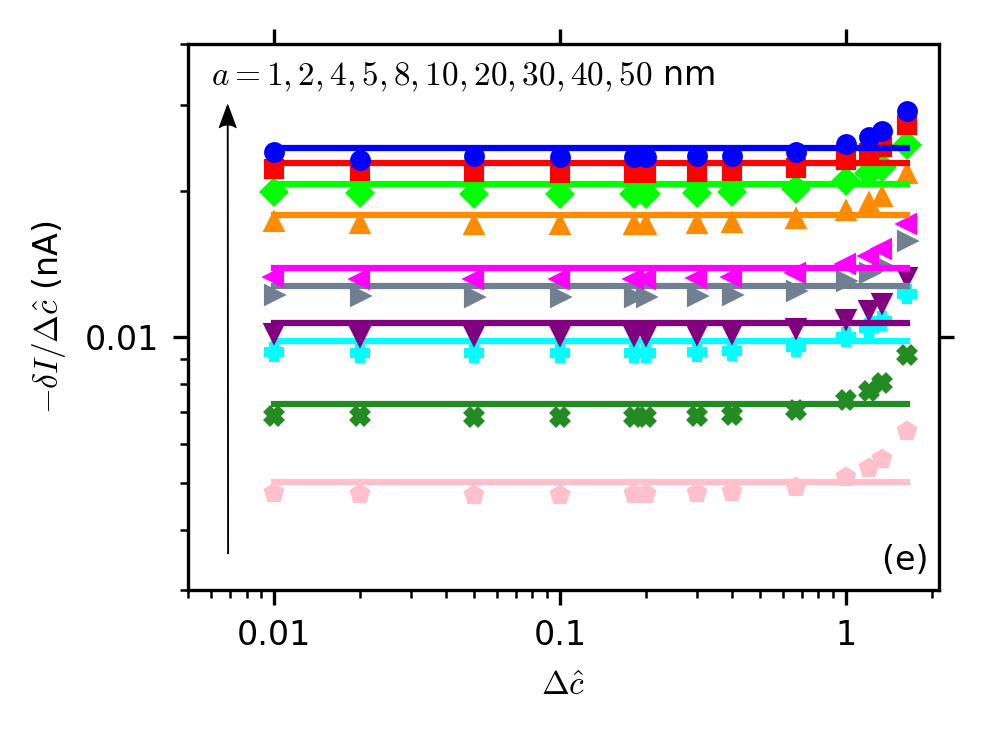}
    
  \end{minipage}%
 \begin{minipage}{.5\textwidth}
   \centering
   \centering
    \includegraphics[scale = 1, trim={0.1cm 0.0875cm 0.1cm 0.0875cm},clip]{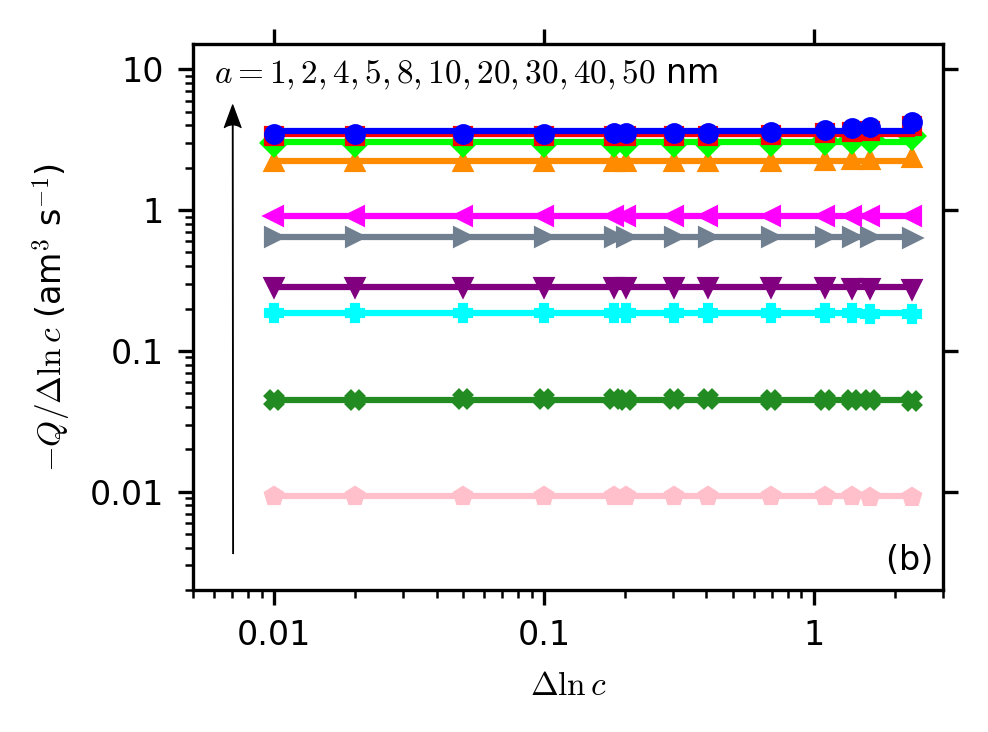}
    \includegraphics[scale = 1, trim={0.1cm 0.0875cm 0.1cm 0.0875cm},clip]{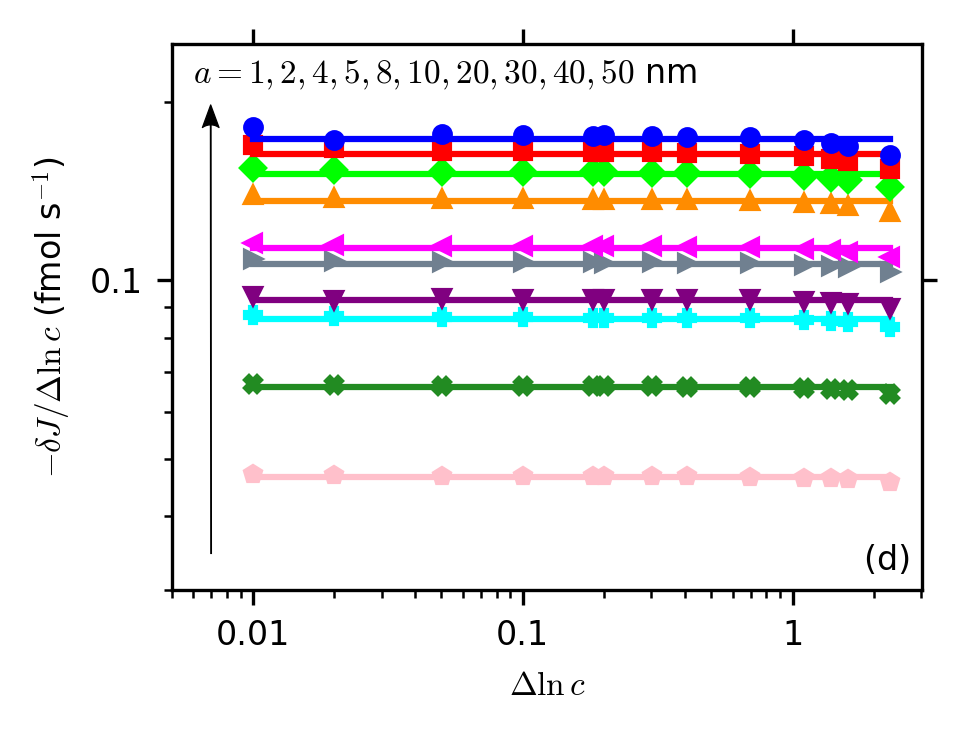}
    \includegraphics[scale = 1, trim={0.1cm 0.0875cm 0.1cm 0.0875cm},clip]{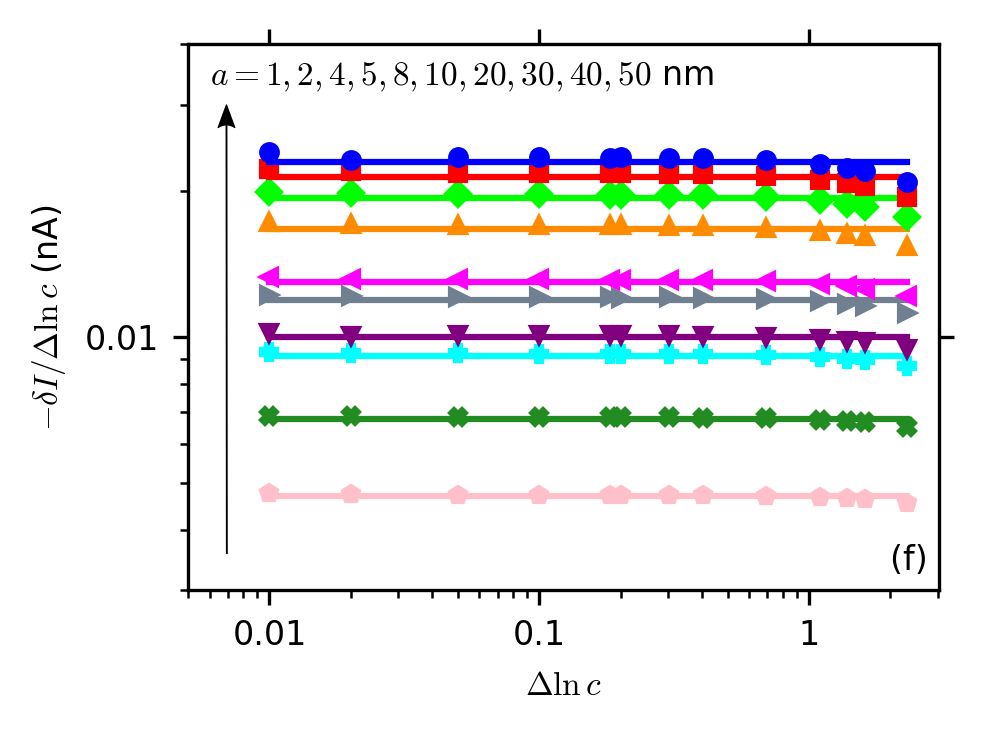}
    
  \end{minipage}
    \caption{\label{fig:Delc1} (a) $Q/\Delta \hat{c}$ vs $ \Delta \hat{c}$, (b) $Q/\Delta \ln{c}$ vs $\ln{c}$, (c) $\delJ/ \Delta \hat{c}$ vs $\Delta \hat{c}$, (d) $\delJ/ \Delta \ln{c}$ vs $\Delta \ln{c}$ and (d) $\delI/ \Delta \hat{c}$ vs $\Delta \hat{c}$, (d) $\delI/ \Delta \ln{c}$ vs $\Delta \ln{c}$. The points are the simulation data and the lines are horizontal fits to the points used to verify linear scaling with $\Delta c$ or $\Dellnc$ for various $a$, where $1/100 \leq \Delta \hat{c} \leq 18/11$.  Arrows indicate the direction of increasing $a$, where $\sigma$ and $\cinf$ were fixed at $-10$~mC~m$^{-2}$ and $0.3$~mol~m$^{-3}$, respectively. Fluxes and parameters are as defined in the main paper.}
\end{figure}

\clearpage

\begin{figure}[!h]
 \centering
 \begin{minipage}{.5\textwidth}
   \centering
    \includegraphics[scale = 1, trim={0.1cm 0.0875cm 0.1cm 0.0875cm},clip]{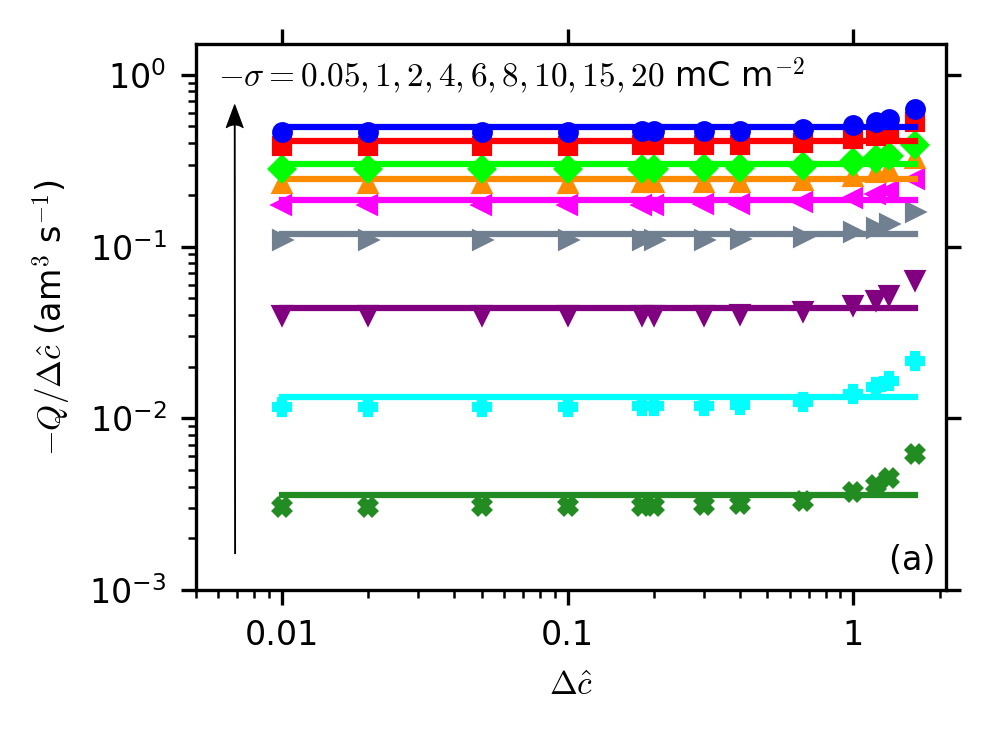}
    \includegraphics[scale = 1, trim={0.1cm 0.0875cm 0.1cm 0.0875cm},clip]{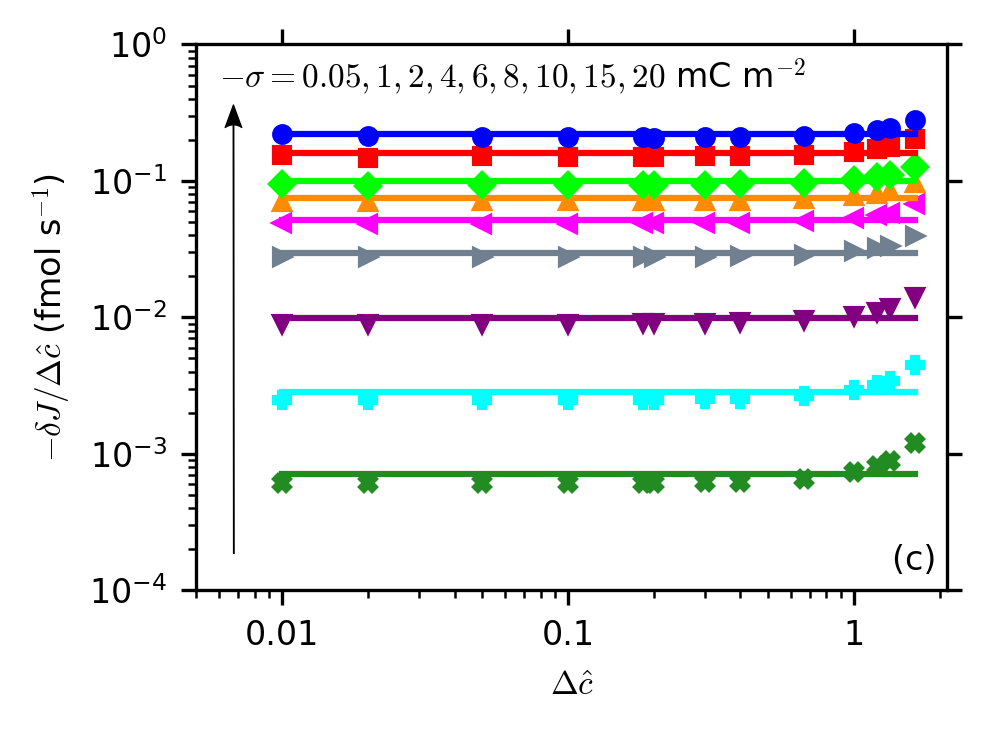}
    \includegraphics[scale = 1, trim={0.1cm 0.0875cm 0.1cm 0.0875cm},clip]{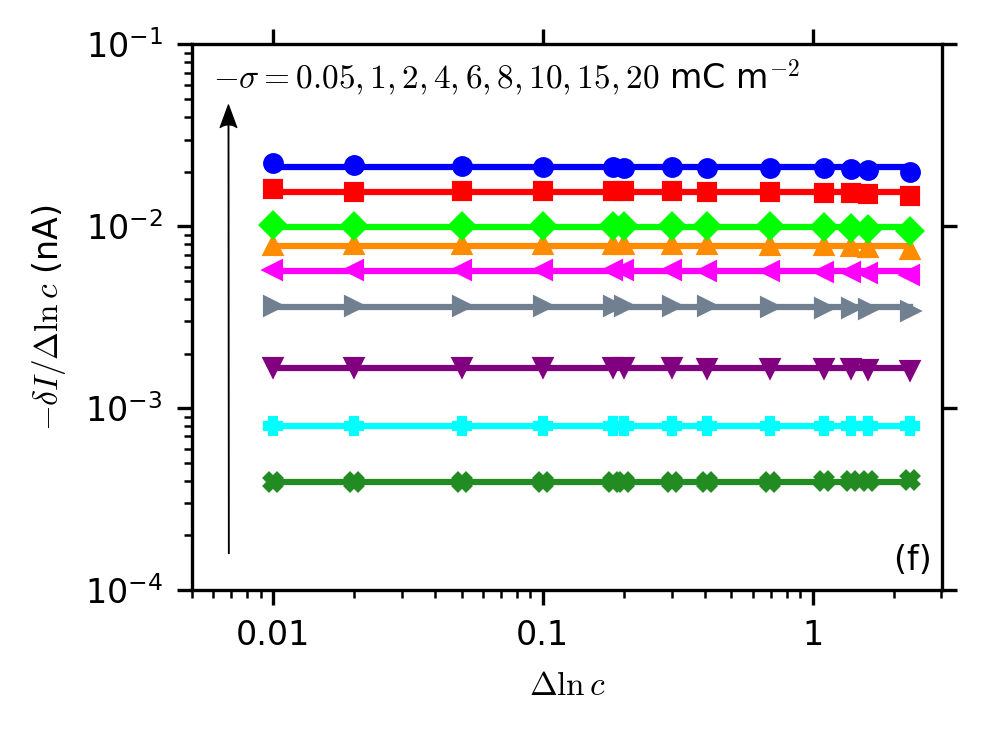}
    
  \end{minipage}%
 \begin{minipage}{.5\textwidth}
   \centering
   \centering
    \includegraphics[scale = 1, trim={0.1cm 0.0875cm 0.1cm 0.0875cm},clip]{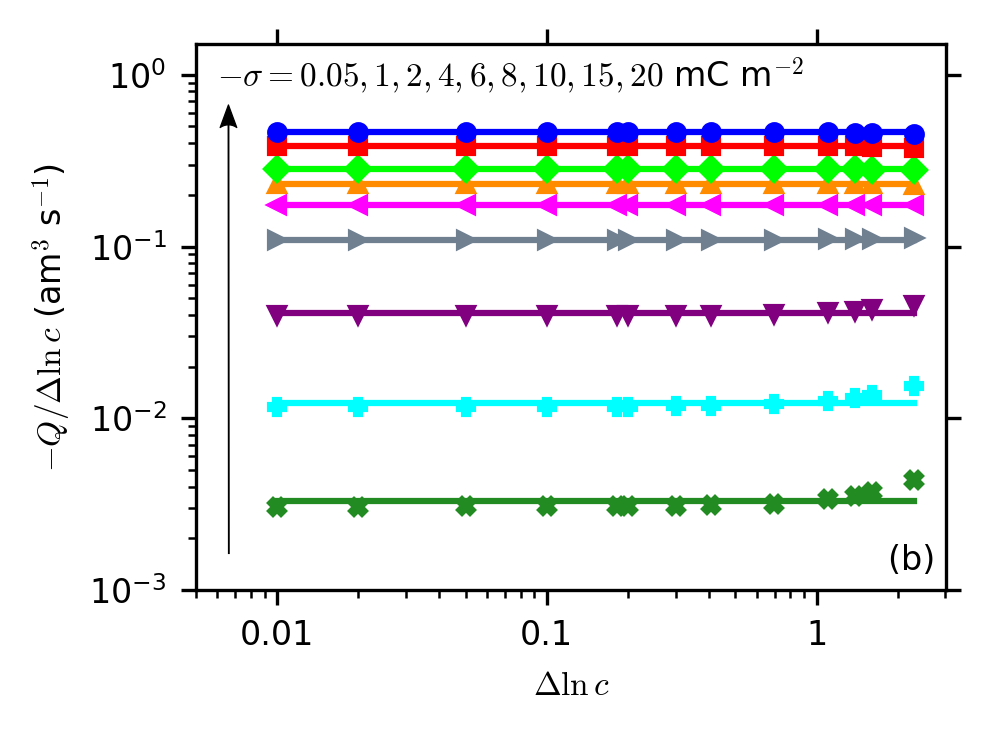}
    \includegraphics[scale = 1, trim={0.1cm 0.0875cm 0.1cm 0.0875cm},clip]{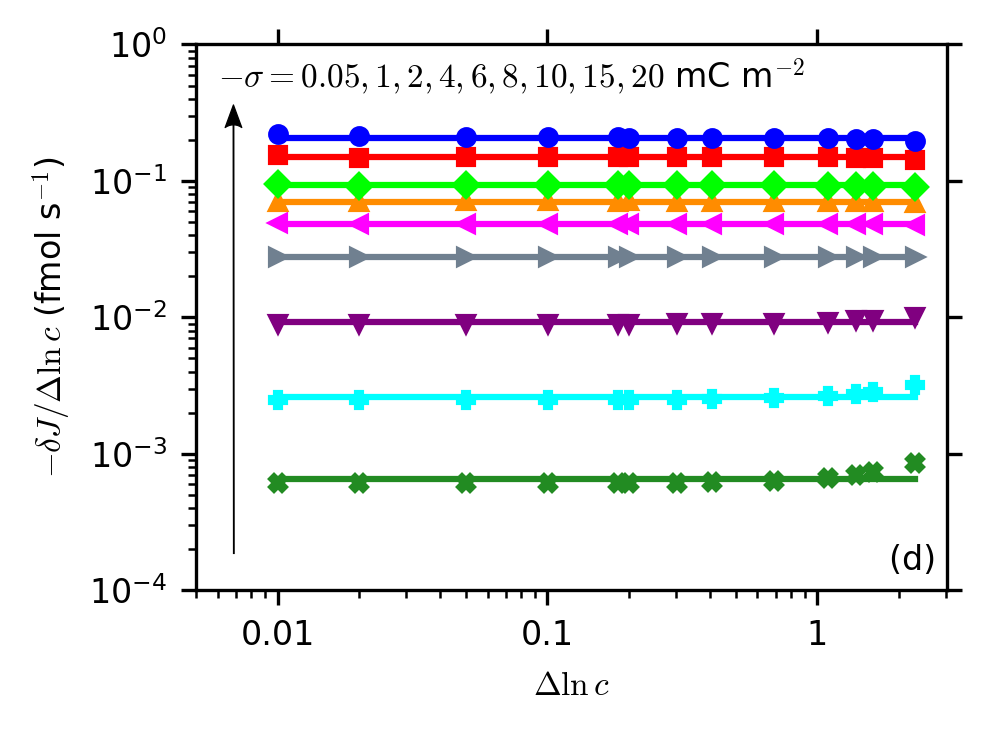}
    \includegraphics[scale = 1, trim={0.1cm 0.0875cm 0.1cm 0.0875cm},clip]{electrolyte-2d-membrane-KCl_H2O-L0.2-a5_c0.3_sig-dc-mesh9-res4-P2+P1-c3-V2-fluxes-paper6_annotated.png}
    
  \end{minipage}
    \caption{\label{fig:Delc2} (a) $Q/\Delta \hat{c}$ vs $ \Delta \hat{c}$, (b) $Q/\Delta \ln{c}$ vs $\ln{c}$, (c) $\delJ/ \Delta \hat{c}$ vs $\Delta \hat{c}$, (d) $\delJ/ \Delta \ln{c}$ vs $\Delta \ln{c}$ and (d) $\delI/ \Delta \hat{c}$ vs $\Delta \hat{c}$, (d) $\delI/ \Delta \ln{c}$ vs $\Delta \ln{c}$. The points are the simulation data and the lines are horizontal fits to the points used to verify linear scaling with $\Delta c$ or $\Dellnc$ for various $\sigma$, where $1/100 \leq \Delta \hat{c} \leq 18/11$.  Arrows indicate the direction of increasing $-\sigma$, where $a$ and $\cinf$ were fixed at $5$~nm and $0.3$~mol~m$^{-3}$, respectively. Fluxes and parameters are as defined in the main paper.}
\end{figure}

\clearpage

\clearpage

\begin{figure}[!h]
 \centering
 \begin{minipage}{.5\textwidth}
   \centering
    \includegraphics[scale = 1, trim={0.1cm 0.0875cm 0.1cm 0.0875cm},clip]{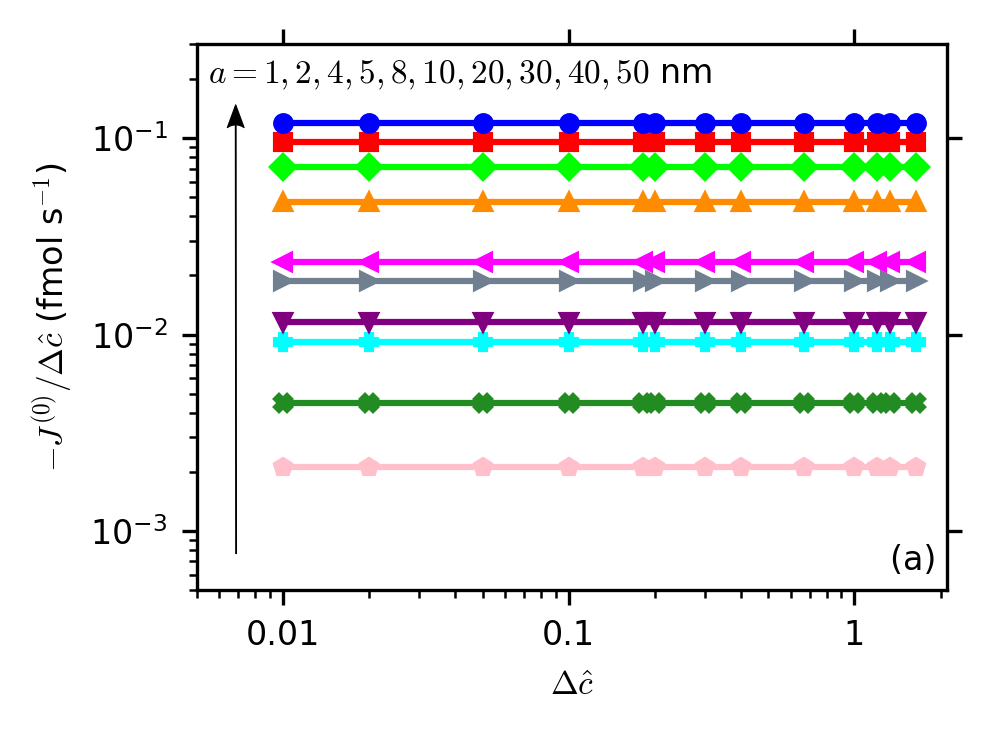}
    \includegraphics[scale = 1, trim={0.125cm 0.0875cm 0.125cm 0.0875cm},clip]{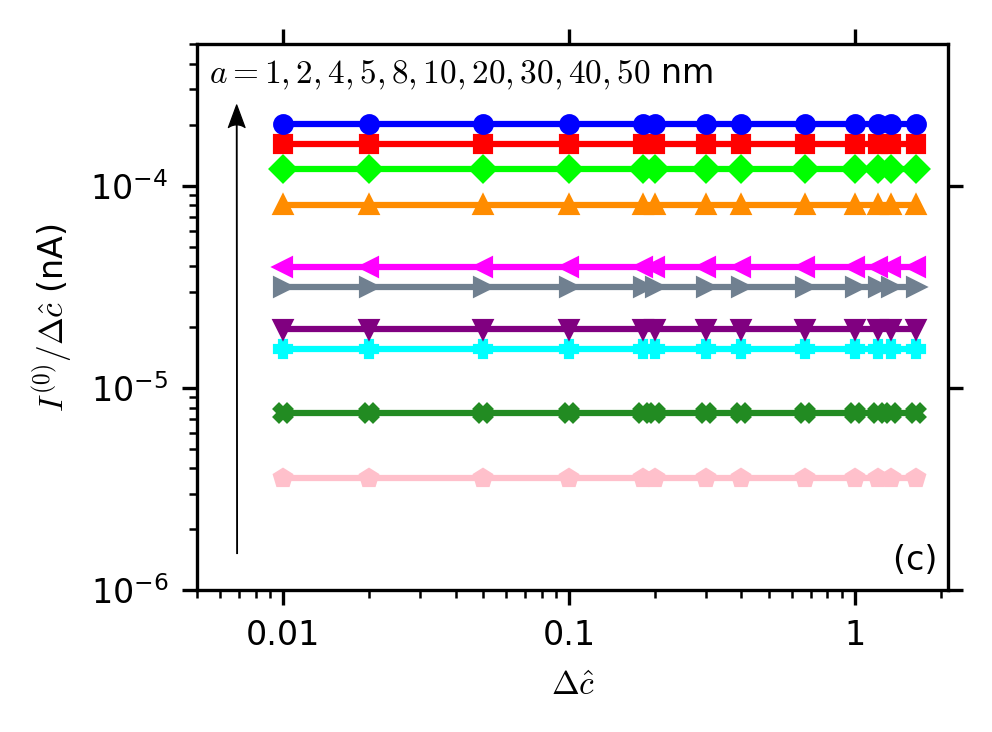}
  \end{minipage}%
 \begin{minipage}{.5\textwidth}
   \centering
    \includegraphics[scale = 1, trim={0.1cm 0.0875cm 0.1cm 0.0875cm},clip]{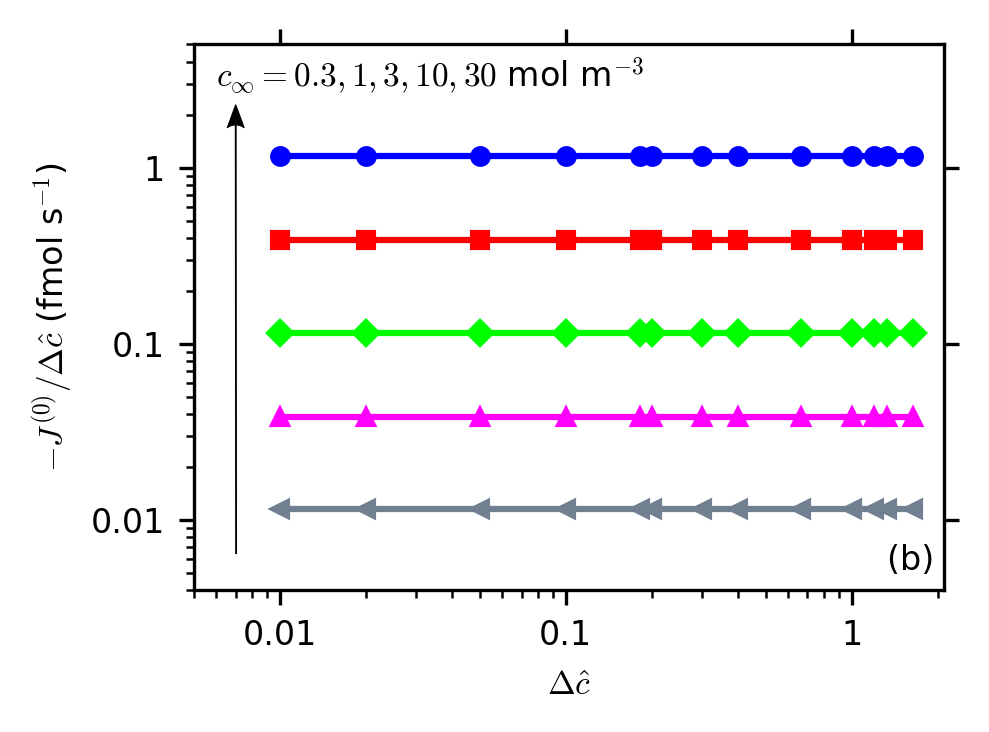}
    \includegraphics[scale = 1, trim={0.1cm 0.0875cm 0.1cm 0.0875cm},clip]{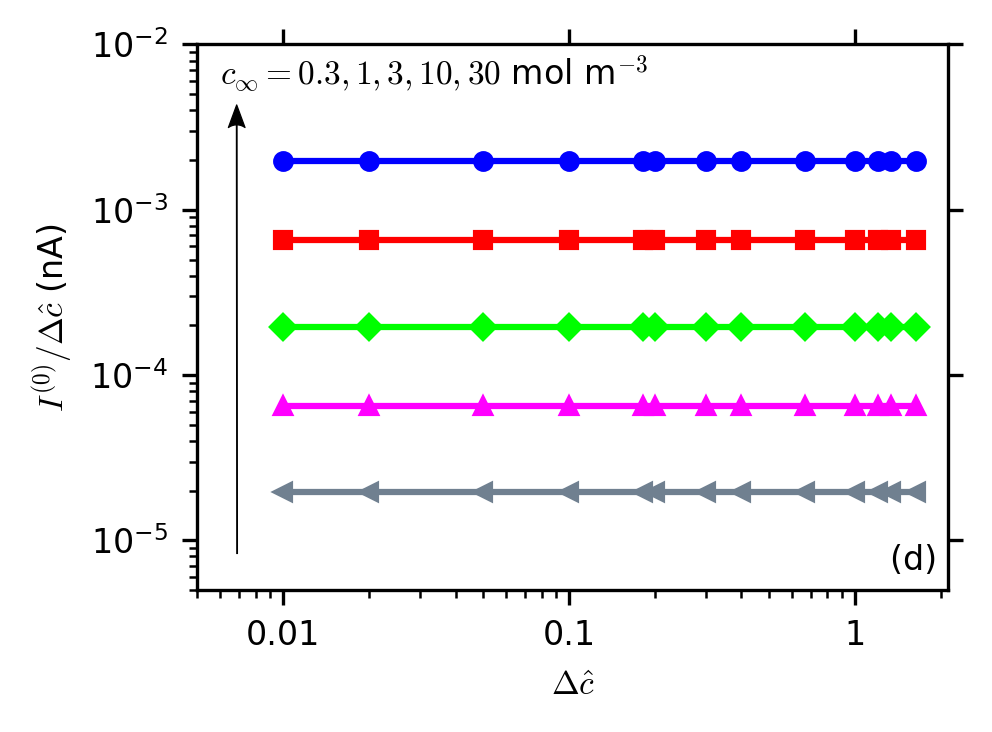}
  \end{minipage}
    \caption{\label{fig:Delc7} ((a),(b)) $\Jb/\Delta \hat{c}$ and ((c),(d)) $\Ib/\Delta \hat{c}$ vs $\Delta \hat{c}$ for $1/100 \leq \Delta \hat{c} \leq 18/11$. The points are the simulation data and the lines are horizontal fits to the points used to verify linear scaling with $\Delc$ for various ((a),(c))  $a$ for a fixed $\cinf = 0.3$ mol~m$^{-3}$, and ((b),(d)) $\cinf$ at fixed $a = 5$ nm. Arrows indicate the direction of increasing variables. Fluxes and parameters are as defined in the main paper.}
\end{figure}

\clearpage

\subsection{Bulk contribution to the solute flux and electric current} 
\label{sec:bulk}

\begin{figure}[!h]
 \centering
 \begin{minipage}{.5\textwidth}
   \begin{flushleft}
    \includegraphics[scale = 1, trim={0.175cm 0.0875cm 0.175cm 0.0875cm},clip]{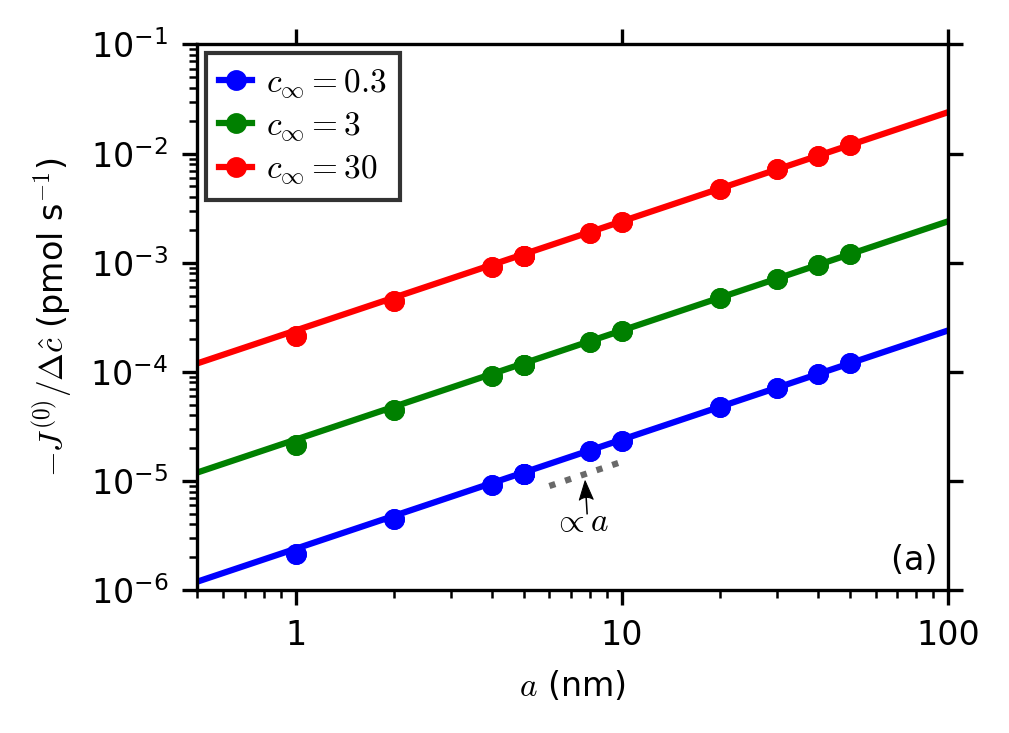}
    \end{flushleft}   
  \end{minipage}%
 \begin{minipage}{.5\textwidth}
   \begin{flushright}
\includegraphics[scale = 1, trim={0.175cm 0.0875cm 0.175cm 0.0875cm},clip]{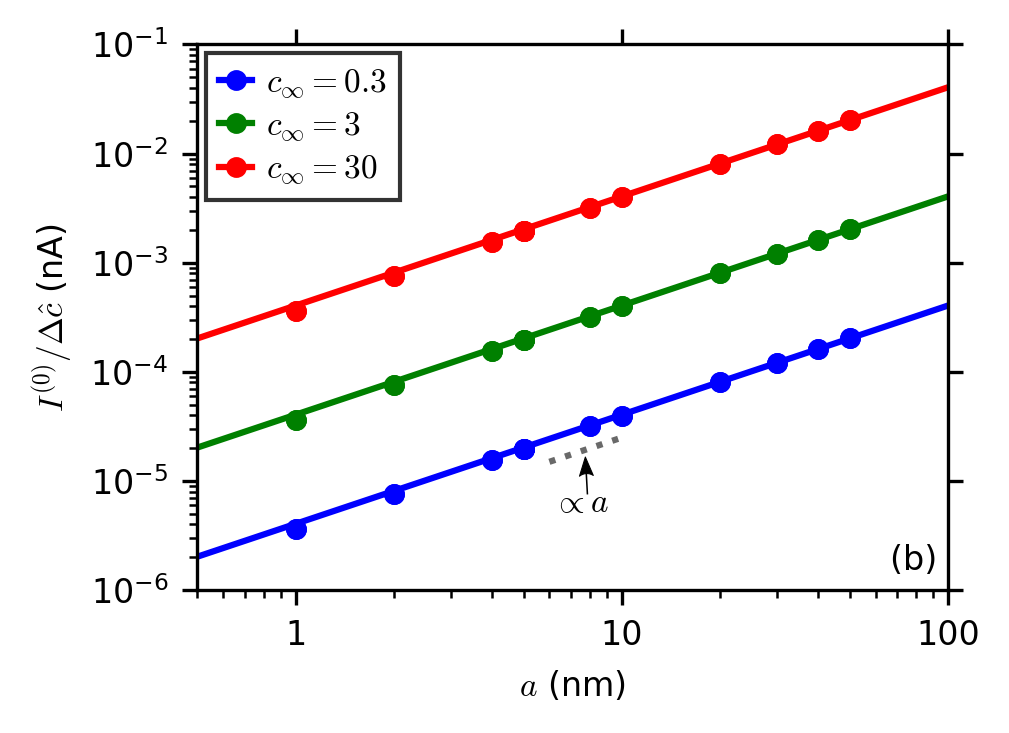}
   \end{flushright}
  \end{minipage}
    \caption{\label{fig:bulk} (a) Bulk contribution to the (a) total solute flux, $\Jb$, and (b) electric current, $\Ib$, over $\Delta \hat{c} = \Delc / \cinf$ vs pore radius $a$ from FEM simulations (symbols) and theory in Eq.~\eqref{solute_flux4} (solid lines) for a range of equilibrium solute concentrations $\cinf$ (mol~m$^{-3}$).}
 \end{figure}

The theory in the main paper predicts that the bulk contributions to the total solute flux, $\Jb$, and electric current, $\Ib$, are linearly related to the concentration difference $\Delc$, which can be seen in Fig.~\ref{fig:Delc7}. Thus, $\Jb$ scales linearly with $\cinf$ at fixed $c _{\mathrm{H}} / c _{\mathrm{L}}$, which is analogous to the case of a neutral solute\cite{rankinEntranceEffectsConcentrationgradientdriven2019} and the same as for thick membranes (see Sec.~\ref{sec:cylinder}). The theory also predicts that $\Jb$ and $\Ib$ are proportional to $a$, where Fig.~\ref{fig:bulk} shows good quantitative agreement between the theory and the simulations. This scaling of $\Jb$ with $a$ is the same as for a neutral solute, \cite{rankinEntranceEffectsConcentrationgradientdriven2019} whereas $\Jb \propto a^2$ (and $\Ib \propto a^2$) in thick membranes.

\clearpage

\bibliography{2D_membrane_CDF-references}

\makeatletter\@input{2D_membrane_CDF-paper-aux.tex}\makeatother